\documentclass[3p]{elsarticle} 
 \makeatletter
 \def\ps@pprintTitle{%
  \def\@oddhead{\footnotesize \centerline{\textcopyright 2022 This manuscript version is made available under the CC-BY-NC-ND 4.0 license \url{http://creativecommons.org/licenses/by-nc-nd/4.0/}}}
  \let\@evenhead\@empty
  \def\@oddfoot{\footnotesize DOI:\text{ }\url{https://doi.org/10.1016/j.eswa.2021.116304}}%
 \let\@evenfoot\@oddfoot}
\makeatother
\usepackage{comment}
\usepackage{soul}
\usepackage{amsmath}
\usepackage{algorithm}
\usepackage[noend]{algpseudocode}
\usepackage{multirow,tabularx}
\usepackage{color}
\usepackage{graphicx}
\usepackage{amsmath}
\usepackage{amsfonts}
\usepackage{lipsum}
\usepackage{multicol}
\usepackage{adjustbox}
\usepackage[graphicx]{realboxes}
\usepackage{subcaption} 
\usepackage{rotating} 
\usepackage{flushend}
\usepackage{natbib}
\usepackage{lineno}
\usepackage{adjustbox}
\usepackage[utf8]{inputenc}
\usepackage[ruled,lined, algo2e]{algorithm2e}
\usepackage{hyperref}
\hypersetup{colorlinks=true,linkcolor=blue,filecolor=magenta,urlcolor=cyan}

\usepackage{caption}
\usepackage{lineno} %count lines
\modulolinenumbers[5]
\usepackage{array}
\usepackage{amsthm}
\usepackage{amssymb}
\usepackage{threeparttable}
\usepackage{mathrsfs}
\usepackage{rotating}
\usepackage[graphicx]{realboxes}
\usepackage[normalem]{ulem}

\usepackage{geometry}
\usepackage{pdflscape}

\algdef{SE}[SUBALG]{Indent}{EndIndent}{}{\algorithmicend\ }%
\algtext*{Indent}
\algtext*{EndIndent}

\newtheorem{theorem}{Theorem}

\journal{Expert Systems with Applications}

\begin{document}
\begin{frontmatter}

\title{\LARGE \bf
Robust path-following control design of heavy vehicles based on multiobjective evolutionary optimization }

\author[]{Gustavo A. Prudencio de Morais$^{a}$}
\ead{gustavo\_{}gapm@usp.br}
\author[]{Lucas Barbosa Marcos$^{a}$}
\ead{lucasbmarcos@usp.br}
\author[]{Filipe Marques Barbosa$^{b}$}
\ead{filipe.barbosa@liu.se}
\author[]{Bruno H. G.~Barbosa$^{c}$}
\ead{brunohb@ufla.br}
\author[]{Marco Henrique Terra$^{a}$}
\ead{terra@sc.usp.br}
\author[]{Valdir Grassi Jr$^{a}$}
\ead{vgrassi@usp.br}

\address{$^{a}$Department of Electrical and Computer Engineering, São Carlos School of Engineering, University of São Paulo, São Carlos, Brazil}
\address{$^{b}$ Division of Automatic Control, Department of Electrical Engineering, Linköping University, Linköping, Sweden}
\address{$^{c}$Department of Automatics, Federal University of Lavras, Lavras, Brazil} 

% ABSTRACT
\begin{abstract}
The ability to deal with systems parametric uncertainties is an essential issue for heavy self-driving vehicles in unconfined environments. In this sense, robust controllers prove to be efficient for autonomous navigation. However, uncertainty matrices for this class of systems are usually defined by algebraic methods which demand prior knowledge of the system dynamics. In this case, the control system designer depends on the quality of the uncertain model to obtain an optimal control performance. This work proposes a robust recursive controller designed via multiobjective optimization to overcome these shortcomings. Furthermore, a local search approach for multiobjective optimization problems is presented. The proposed method applies to any multiobjective evolutionary algorithm already established in the literature. The results presented show that this combination of model-based controller and machine learning improves the effectiveness of the system in terms of robustness, stability and smoothness. 
\end{abstract}


\end{frontmatter}

% SECTION 1
\section{Introduction} 
\label{sec:1}

The past few years have seen a fast development of technologies regarding autonomous heavy vehicles. Applications of these technologies include mining, freight transportation \citep{alam2015heavy}, and urban driving \citep{held2018optimal}. In these contexts, such vehicles perform many tasks related to, for instance, path following \citep{barbosa2019robust}, path planning and optimization, trajectory generation \citep{oliveira2018combining}, speed control \citep{held2018optimal}, platooning \citep{alam2015heavy}, and fuel consumption optimization \citep{rodriguez2018speed}. Even though many advances concerning self-driving vehicles have recently emerged, the development of heavy autonomous vehicles remains a challenging subject. For instance, the payload of trucks may be much greater than the vehicle's weight itself \citep{kati2016robust}. Therefore, payload variations introduce severe changes in the vehicle dynamics and, consequently, strongly affect the accuracy of any modeling and control schemes.

Robust control methods are paramount to handle the extreme dynamic changes associated with time-varying vehicle mass. In this sense, a steering control for an autonomous vehicle subject to uncertain lateral disturbances, that combines fuzzy logic and model-based control to achieve robustness against external steering input was proposed in \cite{nguyen2020robust}. The authors solved the control problem as linear matrix inequalities, which have to be solved in a batch and, therefore, lack the properties of recursiveness. A robust event-triggered fault tolerant automatic steering control strategy for autonomous land vehicles was developed by  \cite{zhang2021novel}. The authors proposed an approach that encompasses uncertainties related to the parameters, time delay, and actuator fault in a robust controller. However, they suppose that the uncertainties' measures are contained in a certain polytopic region determined by known values of its vertices.

Furthermore,  a robust lateral control approach based on immersion and invariance control theorem was proposed in \cite{mohammadzadeh2020novel}. The robustness of the proposed control method was analyzed under parametric uncertainties, disturbances and weather-related conditions. However, these authors assumed that the uncertainties are associated to specific sources, such as cornering stiffness and forward speed, instead of a general-purpose uncertainty that affects the whole system without an identifiable source. Therefore, these considerations demonstrate that robust control techniques often require some modeling or estimation of the system's uncertainties, which is not an easy task. Consequently, the accurate estimation of these uncertainties relies on the expertise of the control system designer, which may result in suboptimal control system performance.

Thus, associating control design with optimization methods to avoid human dependence in the formulation of control plants is a possible solution for enhancing performance of robust control applications \citep{demorais2020vision}. Physical models, or white-box models, reflect the dynamics of the system and, in general, are represented by blocks that can be mathematically analyzed. However, in order to achieve such models, a deep knowledge of the system is necessary. On the other hand, machine learning methods are highly flexible and adaptable structures, which generally do not require prior knowledge (black-box approach) or mathematical analysis of the system \citep{moe2018machine,barbosaiet2019}. Therefore, although mathematical formulations of control systems and machine learning algorithms are uncorrelated, one can enhance performance in both methods by combining them.

Furthermore, robust control design can be commonly assigned as a multiobjective optimization problem (MOP). A MOP is defined as a problem involving two or more conflicting objectives to be optimized simultaneously, and engineering problems related to several objectives often appear in many real-world design applications \citep{zhou2011multiobjective}, such as data mining \citep{zhang2011generic,alatas2008modenar}, bioinformatics \citep{shin2005multiobjective,koduru2008multiobjective}, artificial neural networks \citep{delgado2008multiobjective,qasem2011radial}, manufacturing \citep{weinert2009use}, system identification \citep{barbosa2011,aguirre2017cep}, and pattern recognition \citep{GUEDES201665}. MOPs present a set of trade-off solutions, called Pareto optimal solutions, where an objective function cannot be optimized without decreasing performance in at least one other objective function \citep{he2017radial}. Therefore, multiobjective evolutionary algorithms (MOEAs) are promising optimizers for MOPs, due to their high capability of finding a set of trade-off solutions in a nonlinear search domain, especially considering MOPs with non-differentiable objective functions or without defined mathematical properties \citep{zhou2019decomposition}.

Since uncertainties values are constantly presented in a variety of problems, other  approaches for robust optimization and MOP have been presented in the literature for different research fields besides control applications. An optimization model for medical service performance in post-disasters rescue activities to decrease the effect of the uncertainties in the number of casualties was proposed by  \cite{sun2021robust}. Also, approaches for robust optimization for home health care, dealing with the uncertainties presented in service and travel time, in addition to conflict objectives, were proposed \citep{shahnejat2021robust, decerle2019memetic}. Other applications for robust optimization were also presented for data envelopment analysis with input and output data subject to uncertainty \citep{arabmaldar2021robust}, and cash logistics operations in bank branches \citep{lazaro2018improving}, showing that this research method can be applied to a variety of fields.

Furthermore, some authors have proposed promising works involving applications of MOEAs in MOPs represented by control system designs. For example, \cite{wu2018multiobjective} accomplished the multiobjective $\mathcal{H}_{2}/\mathcal{H}_{\infty}$ fuzzy control design for nonlinear mean-field jump diffusion systems, since the optimal $\mathcal{H}_{2}$ control design and optimal $\mathcal{H}_\infty$ robust control design performances are usually in conflict. Another application was proposed by \cite{malikopoulos2015multiobjective}, where the authors developed a multiobjective optimization framework to determine an optimal control policy for power management control problems within a stochastic formulation applied to hybrid electric vehicles. Besides, to design a robust Proportional-Integral-Derivative (PID) controller, a multiobjective optimization method was presented by \cite{zhao2011multi}, where the objectives were to minimize integral squared error and balance robust performance criteria.

In this sense, one motivation of this work is to design a robust control MOP for path-following and lateral control of a heavy vehicle subject to parametric uncertainties. Some works already proposed robust recursive methods for this task, as presented by \cite{barbosa2019robust}, where the authors obtained satisfactory results using algebraic manipulations. However, a major issue within this approach is to determine the matrices used to model the parametric uncertainties \citep{Cerri2014}. Thus, defining these parameters by means of MOEAs represents a promising solution.

The MOEAs are classified into three different categories, based on their selection strategies \citep{zhou2019decomposition}: (\textit{i}) Pareto dominance-based, where the algorithms modify the Pareto dominance relationship among the results, by dividing them in different fronts and modifying front dominance using diversity metrics \citep{deb2002fast};  (\textit{ii}) decomposition-based, represented by algorithms that convert a MOP into multiple single-objective optimization problems to solve them collaboratively \citep{deb2013evolutionary}; and (\textit{iii}) indicator-based, formed by algorithms that use performance indicator-based approaches to specify different solutions depending on their quality values \citep{hernandez2015improved,tian2016multi}. Also, many techniques can be applied to enhance performance of MOEAs in highly complex MOPs, as hybridizing these algorithms with global and local search methods.

As an improved local search technique for single-objective problems, \cite{de2019soft} proposed a new evolutionary algorithm, named Evolutionary Algorithm with Numerical Differentiation (EAND), for nonlinear single-objective optimization search domain. This method is composed of two main mutation processes: Global Search Procedure (GSP) and Local Search Procedure (LSP). The GSP drives the optimization process according to a dynamic parametrization based on the principles of numerical differentiation among the individuals. It intensifies the mutation process as the population becomes homogeneous, avoiding a premature convergence \citep{de2019soft}. Next, the LSP performs a search around the 20\%  fittest individuals of the population obtained during the GSP, and a performance improvement of the algorithm is achieved by applying this second mutation process.

This work proposes an adaptation of the LSP for multiobjective optimization environments, named Multiobjective Local Search Procedure (MO-LSP), that can be applied to any MOEA already established in the literature with few changes to its structure. The MO-LSP algorithm is applied at the end of the MOEA, performing a local search around the best values obtained so far, and outperforming the results for the robust control applications discussed. To evaluate the method's efficiency, the MO-LSP was compared to ten MOEAs within two robust control MOPs, namely, NSGA-II \citep{deb2002fast}, NSGA-III \citep{deb2013evolutionary}, $\theta$-DEA \citep{yuan2015new}, MOMBI-II \citep{hernandez2015improved}, MOEA/IGD-NS \citep{tian2016multi}, EFR-RR \citep{yuan2015balancing}, MaOEA-DDFC \citep{cheng2015many}, SPEA/R \citep{jiang2017strength}, SPEA2+SDE \citep{li2013shift} and BiGE \citep{li2015bi}. 

The main contributions of this paper are summarized as:
\begin{itemize}
    \item A robust path-following and lateral control is here designed as a MOP to improve different objective functions of tracking performance, while maintaining robustness for different series of time-varying loads.
    
    \item A case study for control modeling without mathematical formulation of the  parametric uncertainties is presented. A multiobjective optimization approach to adjust the uncertainty variables is used in order to reduce the vehicle’s lateral displacement and yaw rate, lateral velocity and displacement errors.
    
    \item The model's parametric uncertainties are estimated by MOEAs. The proposed approach is adaptable to different performance functions, as well as to different scenarios involving autonomous vehicles.
    
    \item Ultimately, this work proposes a novel adaptation of the LSP for multiobjective optimization environments, named Multiobjective Local Search Procedure (MO-LSP), which can be applied to any MOEA already established in the literature with minor adjustments to its structure. 
\end{itemize}

  To verify the effectiveness of this method, two experiments representing a generic and an applied robust control design were carried out. The MO-LSP algorithm was applied to ten MOEAs where the respective performances, before and after applying the proposed approach, were evaluated. Then, the robust control approach proposed was compared with three different control strategies for autonomous vehicles. The first comparison study  was performed with the robust Linear Quadratic Regulator (RLQR)  proposed by \cite{barbosa2019robust}. The second comparison was performed with the the standard $\mathcal{H}_\infty$ controller \citep{hu2016robust}, widely used in automotive applications \citep{ li2018vehicle, zhao2018displacement}. The third was performed with the Linear Quadratic Regulator presented in \citep{bertsekas2000dynamic}. These comparisons were made by  simulating the path-following task of  heavy vehicle and varying the mass in four different cases. That is, increasing the nominal payload of the vehicle in 0, 100, 200 and 300\%. 

This paper is organized as follows: \autoref{sec:2} presents the motivations of the problem, where the robust control problem is presented and mathematically formulated as a MOP. \autoref{sec:3} explains the proposed MO-LSP algorithm in detail. \autoref{sec:4} describes the experimental setup, where the generic and the applied robust control models are defined. Finally, \autoref{sec:5} shows the conclusions. 

% SECTION 2
\section{Motivation and problem description} 
\label{sec:2}

The main purpose of this work is to enhance the performance of a path-following control of a vehicle via multiobjective optimization. With this aim, a state-space description of the vehicle's lateral motion and the equations of a robust controller will be provided. The former demands information of a series of vehicle parameters, while the latter demands the solution of a control engineering problem.

The continuous time state-space equation was obtained by arranging the vehicle's parameters in a mathematical description of the laws of motion that govern the vehicle's movement. This motion equation is then discretized, as it is associated to discrete-time MOPs. In parallel, a series of MOPs was solved using the state-space equations to pursue uncertainty matrices, which optimize control performance, i.e., minimize path-following error for realistic values of steering. 
In order to estimate uncertainty, a MOP novel formulation is applied. Lack of information is precisely what defines an uncertain value and, consequently, hinders any estimation problems. Using MOP, this work presents an alternative to numerical estimation, based solely on the performance of the results, regardless of any physical meaning they must convey related to the control plant. The innovation in the optimization problem (i.e., the adaptation as a MO-LSP) further improves its performance.

The next subsections present the state-space model of a non-articulated truck, as well as a free body diagram describing the physical meaning of its parameters. Then, the RLQR is presented, which results in the controller to be used for path-following. 

\subsection{Vehicle model}

\begin{figure}[ht]
    \centering
    \includegraphics[scale=0.45]{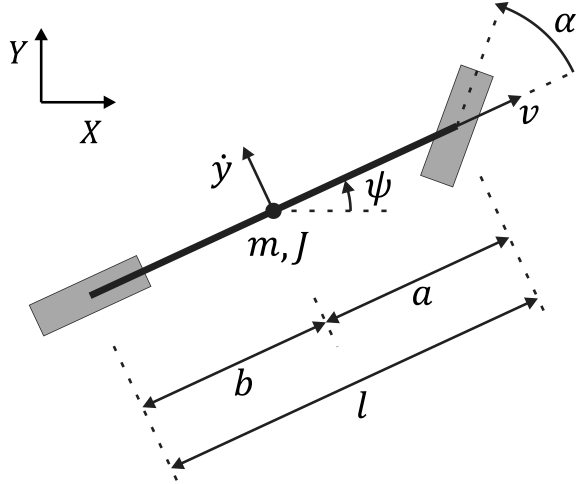}
	\caption{Single-track model for a non-articulated vehicle.}
	\label{single}
\end{figure}

A mathematical modeling of the vehicle is necessary to describe its lateral behavior. To this end, \autoref{single} shows a general free-body diagram of the vehicle and \autoref{parameters} details its parameters. This behavior is furthermore represented in $x=[\dot{y},\dot{\psi}, \rho,\theta]^T$ and described in the state-space form as:

\begin{equation}
\label{eq:ssdynamic}
\begin{bmatrix}
m & 0 & 0 & 0\\
0 & J & 0 & 0\\
0 & 0 & 1 & 0\\
0 & 0 & 0 & 1
\end{bmatrix}
\dot{x} = 
\begin{bmatrix}
\frac{-c_{1}-c_{2}}{v} & \frac{bc_{1}-ac_{1}-mv^{2}}{v} & 0 & 0\\
\frac{bc_{2}-ac_{1}}{v} & \frac{-a^{2}c_{1}-b^{2}c_{1}}{v} & 0 & 0\\
1 & 0 & 0 & v\\
0 & 1 & 0 & 0
\end{bmatrix}x+
\begin{bmatrix}
c_{1}\\
ac_{1}\\
0\\
0
\end{bmatrix}\alpha,
\end{equation}
%--
%--
where $c_{1}$, $c_{2}$ are the cornering stiffness coefficients of the vehicle's front and rear axles, respectively.

Equation (\ref{eq:ssdynamic}) was obtained by combining the linear single-track model and the path-following equations presented by \cite{van2012analysis} and  \cite{skjetne2001nonlinear}, respectively. A similar procedure was used by \cite{barbosa2019robust} to obtain the lateral model of an articulated vehicle.

\begin{table}[H]
\caption{Description of vehicle parameters}
\label{parameters}
\begin{center}
\begin{tabular}{c l c}
\hline
Parameter & Meaning & Unit\\
\hline
$a$ & Distance from the front axle to the vehicle center of gravity & $m$\\
$b$ & Distance from the vehicle rear axle to the tractor center of gravity & $m$\\
$l$ & Vehicle wheelbase & $m$\\
$v$ & Forward velocity & $m/s$\\
$\dot{y}$ & Lateral velocity & $m/s$\\
$m$ & Vehicle mass & $kg$\\
$J$ & Vehicle moment of inertia & $kg \hspace{1mm} m^{2}$\\
$\psi$ & Vehicle yaw & $rad$\\
$\alpha$ & Steering angle & $rad$\\
\hline
\end{tabular}
\end{center}
\end{table}
%--

In order to perform the vehicle path-following, a robust recursive regulator was used and is now briefly presented.

\subsection{Robust Linear Quadratic Control}\label{sec:2.2}

The aim of the RLQR is to minimize a given cost function, subject to the maximum influence of parametric uncertainties. Thus, it gives an optimal feedback gain $K_i$, generating an optimal feedback law $u_i=K_ix_i$. This section briefly describes the robust recursive regulator presented by \cite{Cerri2014} and in \cite{Cerri2009}, which is used to evaluate the novel method. 

Now, consider the following discrete-time linear system subject to parametric uncertainties
\begin{equation}
\label{eq:statespaceDT}
x_{i+1} = (F_{i} + \delta F_{i})x_{i} + (G_{i}  + \delta G_{i})u_{i},
\end{equation}
where $i = 0,\hdots, N$, $x_{i} \in \mathbb{R}^{n}$ is the state vector, $u_{i} \in \mathbb{R}^{m}$ is the control input, and $F_{i} \in \mathbb{R}^{n \times n}$ and $G_{i} \in \mathbb{R}^{n \times m}$ are known nominal model matrices. Moreover, the parametric uncertainties are represented by the uncertainty matrices $\delta F_{i}$ and $\delta G_{i}$ and modeled as
%--
\begin{equation}
\label{eq:RLQRuncertainties}
\begin{bmatrix}
\delta {F}_{i} & \delta G_{i}
\end{bmatrix} = H_{i} \Delta_{i} \begin{bmatrix}
E_{F_{i}} & E_{G_{i}}
\end{bmatrix},
\end{equation}
for all $i = 0,\hdots, N$, where $H_{i} \in \mathbb{R}^{n \times p}$; $E_{F_{i}} \in \mathbb{R} ^{l \times n}$ and $E_{G_{i}} \in \mathbb{R} ^{l \times m}$ are known matrices; and $\Delta_{i} \in \mathbb{R}^{p \times l}$ is a contraction matrix such that $||\Delta|| \leq 1$.

Then, the RLQR results from the solution of the optimization problem \citep{Cerri2014}:

\begin{equation}
\label{eq:rlqrminmax}
\underset{x_{i+1},u_{i}}{min} \ \underset{{\delta} F_{i},{\delta} G_{i}}{max} {\bar{J}^{\mu}_{i}(x_{i+1},u_{i},{\delta} F_{i},{\delta} G_{i})},
\end{equation}
where $\bar{J}^{\mu}_{i}$ is the cost function, defined as
\begin{align}
\label{eq:rqlqrcostfun}
\nonumber
& \bar{J}^{\mu}_i(x_{i+1},u_{i},{\delta} F_{i},{\delta} G_{i}) = \\
& \begin{bmatrix}
x_{i+1}\\
u_i
\end{bmatrix}^T
\begin{bmatrix}
P^{r}_{i+1} & 0 \\
0 & R_i
\end{bmatrix} 
\begin{bmatrix}
x_{i+1}\\
u_i
\end{bmatrix} + \Phi^T 
\begin{bmatrix}
Q_i & 0\\
0 & \mu I
\end{bmatrix}\Phi,
\end{align}
with fixed penalty parameter $\mu > 0$, weighing matrices $Q_{i} \succ 0$, $R_{i} \succ 0$, $P_{i+1} \succ 0$ and
\begin{equation*}
\Phi = \left\{
\begin{bmatrix}
0 & 0 \\
I & -G_i-\delta G_i
\end{bmatrix}
\begin{bmatrix}
x_{i+1}\\
u_i
\end{bmatrix}
- \begin{bmatrix}
-I\\
F_i+{\delta} F_i
\end{bmatrix}
x_i\right\}.
\end{equation*}

The solution of the optimization problem (\ref{eq:rlqrminmax})-(\ref{eq:rqlqrcostfun}), as solved by \cite{Cerri2014} and applied by \cite{barbosa2019robust}, is presented below.

Thus, with the purpose of calculating the optimal cost function, control input and state trajectory, the following theorem shows a framework given in terms of an array of matrices.
%--
\begin{theorem}
\label{theorem2}
For each $\mu > 0$ in the optimization problem (\ref{eq:rlqrminmax})-(\ref{eq:rqlqrcostfun}), the optimal solution is given by
%--
\begin{equation} 
\label{exp_a}
\begin{bmatrix}
x^{\ast}_{i+1}\\
u^{\ast}_{i}\\
\tilde{J}^{\mu}_{i}(x^{\ast}_{i+1},u^{\ast}_{i})
\end{bmatrix}=\begin{bmatrix}
I & 0 & 0 \\
0 & I & 0 \\
0 & 0 & x_{i}^{T}
\end{bmatrix}^{T}\begin{bmatrix}
L_{i,\mu}\\
K_{i,\mu}\\
P_{i,\mu}
\end{bmatrix}x_{i},
\end{equation}
%--
where the closed-loop system matrix $L_i$ and the feedback gain $K_i$ result from the recursion
\begin{equation}
\label{eq:rlqrsolution}
\begin{bmatrix}
L_{i,\mu} \\ K_{i,\mu} \\ P_{i,\mu}
\end{bmatrix} =
\begin{bmatrix}
0 & 0 & -I & \mathcal{F}_{i} & 0 & 0\\
0 & 0 & 0 & 0 & 0 & I\\ 
0 & 0 & 0 & 0 & I & 0
\end{bmatrix}
\Xi^{-1}
\begin{bmatrix}
0 \\ 0 \\ -I \\ \mathcal{F}_{i} \\ 0 \\ 0
\end{bmatrix},
\end{equation}
with
\begin{equation}
\Xi = \begin{bmatrix}
P^{-1}_{i+1} & 0 & 0 & 0 & I & 0 \\
0 & R^{-1}_{i} & 0 & 0 & 0 & I \\
0 & 0 & Q^{-1}_{i} & 0 & 0 & 0 \\
0 & 0 & 0 & \Sigma_{i,\mu,\hat \lambda} & \mathcal{I} & -\mathcal{G}_{i} \\
I & 0 & 0 & \mathcal{I}^{T} & 0 & 0 \\
0 & I & 0 & -\mathcal{G}^{T} & 0 & 0
\end{bmatrix}, \nonumber
\end{equation}
\begin{align*}
& \Sigma_{i,\mu,\hat \lambda} = \begin{bmatrix}
\mu^{-1}I - \hat{\lambda}^{-1}H_{i}H_{i}^{T}  & 0 \\
0 & \hat{\lambda}^{-1}I
\end{bmatrix}, \nonumber \\
& \mathcal{I} = \begin{bmatrix}
I \\ 0
\end{bmatrix}, \
\mathcal{G}_{i} = \begin{bmatrix}
G_{i} \\ E_{G_{i}}
\end{bmatrix}, \
\mathcal{F}_{i} = \begin{bmatrix}
F_{i} \\ E_{F_{i}}
\end{bmatrix},
\end{align*}
%--
\noindent where $P_{i+1}$ is the solution of the associated Riccati Equation and $\hat{\lambda}> \|\mu H_i^TH_i\|$ \citep{Sayed2001}. Furthermore, alternatively one has
%--

\begin{equation}
\label{alternative}
\begin{aligned}
P_{i,\mu} = &L^{T}_{i,\mu}P_{i+1}L_{i,\mu} + K_{i,\mu}R_{i}K_{i,\mu} + Q_{i}+\\
&(\mathcal{I}L_{i,\mu} - \mathcal{G}_{i}K_{i,\mu} - \mathcal{F}_{i})^{T}\Sigma_{i,\mu}^{-1}(\mathcal{I}L_{i,\mu}-\mathcal{G}_{i}K_{i,\mu}-\mathcal{F}_{i}) \succ 0.
\end{aligned}
\end{equation}
%--
\end{theorem}
%--

\vspace{2mm}
\begin{center}
\begin{algorithm2e}[H]
\SetAlgoLined
\textbf{Uncertain model:} Consider the model (\ref{eq:statespaceDT})-(\ref{eq:RLQRuncertainties}) and criterion (\ref{eq:rlqrminmax})-(\ref{eq:rqlqrcostfun}) with known\\ $F_{i}$, $G_{i}$, $E_{F_{i}}$, $E_{G_{i}}$, $Q_{i} \succ 0$, and $R_{i} \succ 0$ for all $i$.\\
\textbf{Initial conditions:} Define $x_{0}$ and $P_{i,N}\succeq{0}$.\\
\textbf{Step 1:} \textit{(Backward)} For all $i=N-1,\ldots,0$, compute\\
$\hfill \begin{bmatrix}
L_{i}\\
K_{i}\\
P_{i}
\end{bmatrix}
=
\begin{bmatrix}
0 & 0 & 0 \\
0 & 0 & 0 \\
0 & 0 & -I \\
0 & 0 & F_{i} \\
0 & 0 & E_{F_{i}} \\
I & 0 & 0 \\
0 & I & 0
\end{bmatrix}^{T}
\begin{bmatrix}
P_{i+1}^{-1}& 0 & 0 & 0 & 0 & I & 0\\
0 & R_{i}^{-1} & 0 & 0 & 0 & 0 & I\\
0 & 0 & Q_{i}^{-1} & 0 & 0 & 0 & 0\\
0 & 0 & 0 & 0 & 0 & I & -G_{i}\\
0 & 0 & 0 & 0 & 0 & 0 & -E_{G_{i}}\\
I & 0 & 0 & I & 0 & 0 & 0\\
0 & I & 0 & -G_{i}^{T} & -E_{G_{i}}^{T} & 0 & 0
\end{bmatrix}^{-1}
\begin{bmatrix}
0\\
0\\
-I\\
F_{i}\\
E_{F_{i}}\\
0\\
0
\end{bmatrix}.\hfill $\\
\textbf{Step 2:} \textit{(Forward)} For each $i=0,...,N-1$, obtain\\
$\hfill \begin{bmatrix}
x^{*}_{i+1}\\
u^{*}_{i}
\end{bmatrix}
=
\begin{bmatrix}
L_{i}\\
K_{i}
\end{bmatrix}
x^{*}_{i},\hfill $ \\
with the total cost given by $J_{r}^{*}=x_{0}^{T}P_{0}x_{0}$.
\\
\caption{The Robust Linear Quadratic Regulator}
\label{rlqr_algor}
\end{algorithm2e}
\end{center}
\vspace{2mm}

In every iteration of (\ref{alternative}), the matrix $P_{i,\mu}$ is finite and $\mathcal{I}L_{i,\mu}-\mathcal{G}_{i}K_{i,\mu}-\mathcal{F}_{i} \rightarrow 0$ when $\mu \rightarrow \infty$ \citep{Cerri2014}. Thus,
%--
\begin{equation}
\label{feedback}
\begin{aligned}
&L_{i,\infty} = F_{i}+G_{i}K_{i,\infty}\\
&E_{F_{i}} + E_{G_{i}}K_{i,\infty} = 0,
\end{aligned}
\end{equation}
%--
and a sufficient condition that satisfies (\ref{feedback}) is 
%--
\begin{equation}
\label{eq:rank_cond}
rank\,\big(\begin{bmatrix}
E_{F_{i}} & E_{G_{i}}
\end{bmatrix}\big) = rank\,\big(E_{G_{i}}\big).
\end{equation}
%--
Convergence and stability analysis of the RLQR are performed by direct identification with the standard optimal regulator problem for systems not subject to uncertainties. As in the standard LQR, the stability is directly related to the positiveness of the solution of the Riccati equation \citep{Cerri2014}. The reader is referred to \cite{Cerri2014} for further details on convergence and stability.

Notice that the formulation of the RLQR in \autoref{rlqr_algor} demands information on the matrix uncertainties $E_{F_i}$ and $E_{G_i}$. This information is not readily available. Thus, this shortcoming is circumvented in \cite{barbosa2019robust}  by calculating $E_{F_i}$ and $E_{G_i}$ via an algebraic equation. For the path-following control problem in this paper, these matrices aim to model the maximum weight variation, relating it to the rows of the state-space matrices which are most affected by maximum variations in a parameter (in this case, the maximum variation in the truck payload). Despite being effective, the method presented by \cite{barbosa2019robust} lacks any optimality criteria, which may lead to suboptimal performance of the controller. In order to overcome this, this paper proposes an uncertainty estimation method based on multiobjective problems, which is presented in the following section.

% SECTION 3
\section{The proposed method} 
\label{sec:3}

This section discusses the proposed MO-LSP algorithm. For the sake of clarity, a brief formulation of the multiobjective problem is described. Next, the MO-LSP formulation is presented in detail. Lastly, the application of the MO-LSP in robust control optimization domain is analyzed.   

\subsection{Multiobjective problem formulation } 
\label{sec:3.1}

The continuous-time MOP is mathematically represented as: 
\begin{equation} 
\begin{split} \label{eq:minF}
 & \min \textbf{F} \left( \textbf{z} \right) = \left( f_{1} \left( \textbf{z} \right), f_{2} \left( \textbf{z} \right),  \ldots , f_{\textbf{m}} \left( \textbf{z} \right) \right)^{T} \\
 & \textnormal{Subject to } \textbf{z} \in \Omega, \\
\end{split} 
\end{equation} 
\noindent where $\textbf{F} : \Omega \to \mathbb{R}^{\textbf{m}}$ constitutes a set of $\textbf{m}$ objective functions, $\mathbb{R}^{\textbf{m}}$ represents the feasible objective space, and $\textbf{z} = \left( z_{1}, z_{2},  \ldots, z_{\textbf{n}} \right)^{T}$ is the \textbf{n}-dimensional decision vector from the search region $\Omega \in \mathbb{R}^{\textbf{n}}$. For the MOP formulation applied to the RLQR, the decision vector represents the uncertainty matrices and the penalty parameter, as $\textbf{z}_{i} = [E_{F_i}, E_{G_i}, (\log_{10}(\mu_i))^2]^T$, while $\textbf{F} \left( \textbf{z}_{i}  \right)$ denotes the regulation errors associated to the control design.

The concept of dominance is applied to evaluate the solutions of the optimization problem, where $\textbf{z}^{*}$ dominates any decision vector $\textbf{z}$, denoted as $\textbf{z}^{*} \prec \textbf{z}$, iff $\forall i_{1} \in \{ 1, \ldots, \textbf{m} \}$, $f_{i_{1}} \left( \textbf{z}^{*} \right) \le f_{i_{1}} \left( \textbf{z} \right)$, and at least for one index $i_{2} \in \{ 1, \ldots, \textbf{m} \}$, $f_{i_{2}} \left(\textbf{z}^{*} \right) < f_{i_{2}} \left( \textbf{z} \right)$. A $\textbf{z}^{*}$ value  is defined as a Pareto optimal, and the set of all Pareto optimal solutions compose the Pareto set (PS), as: 
\begin{align}\label{eq:ps}
   PS =  \{ \textbf{z} \in \Omega \textnormal{ } | \textnormal{ } \textbf{z} \textnormal{ is Pareto optimal} \}.
\end{align} 
\noindent Finally, the composition of the objective vectors of the PS define the Pareto front (PF): 
\begin{align}\label{eq:pf}
   PF = \{ \textbf{F} \left( \textbf{z} \right) \in \mathbb{R}^{\textbf{m}} \textnormal{ } | \textnormal{ } \textbf{z} \in PS \}.
\end{align} 

A MOEA aims to define the optimal PF in optimization problems and several algorithms can be applied to search the decision vectors in nondominated search problems. In order to enhance the MOEAs convergence, this work proposes the MO-LSP algorithm.

\subsection{Multiobjective Local Search Procedure}
\label{sec:3.2}

The MO-LSP algorithm is proposed as an adaptation of the LSP, applied to enhance performance in MOEAs that are already established in the literature. For this method, a MOEA main structure is preserved and after each iteration of mutation and selected procedures, the MO-LSP performs a local search around the $\widehat{K}$ individuals that compose the two best evaluated PF. Thus, the $\widehat{K}$ fittest values are stored in $\widehat{\textbf{y}}_{j} \in \mathbb{R}^{\textbf{m}}$, $j = 1, \ldots, \widehat{K}$, where $\widehat{\textbf{y}}_{j} = \left(f_{j_{1}} , \ldots, f_{j_{\textbf{m}}} \right)$, and their respective decision vectors are stored in $\widehat{\textbf{z}}_{j} \in \Omega$, where $\widehat{\textbf{z}}_{j} = \left(z_{j_{1}} , \ldots, z_{j_{\textbf{n}}} \right)$.

Since a PF is usually composed of more than one solution, the proposed algorithm selects a single decision vector from the $\widehat{K}$ individuals to lead the local search. To this end, the MO-LSP evaluates the \textbf{m}-dimensional Lebesgue measure, denoted as $\Gamma_{\textbf{m}}$. It calculates the area between each $j$ fitness value from $\widehat{\textbf{y}}_{j}$ and an established and unchanged reference vector $\textbf{r} \in \mathbb{R}^{\textbf{m}}$, defined by the user, such that $\widehat{\textbf{y}}_{j} \prec \textbf{r}$. The individual with greater $\Gamma_{\textbf{m}}$ value is removed from $\widehat{\textbf{z}}_{j}$ and is stored in the variable $\widehat{\textbf{d}}$ that will lead the local search operation.

\begin{figure}[ht]
    \centering
    \includegraphics[scale=0.1]{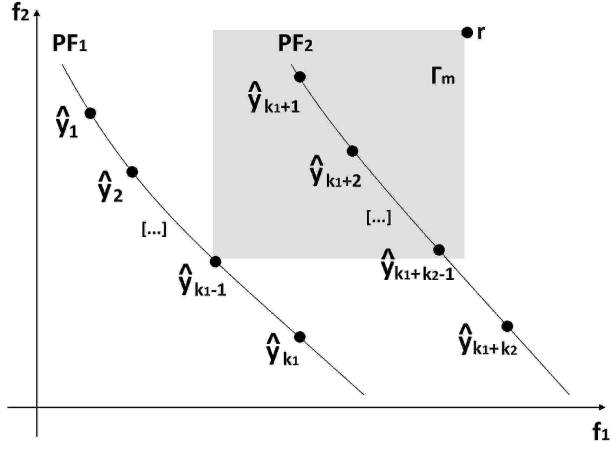}
	\caption{Selection of the leader variable $\widehat{\textbf{d}}$ in a two-dimensional minimization problem.}
	\label{fig:mo-lsp}
\end{figure}

\autoref{fig:mo-lsp} shows the selection of the leader variable $\widehat{\textbf{d}}$ in a minimization MOP with two objective functions. In this problem, matrix $\widehat{\textbf{y}}$ is composed of $\widehat{K} = k_{1} + k_{2}$ individuals from the first two PF, represented by PF$_{1}$ and PF$_{2}$, respectively. The search reference vector $\textbf{r}$ delimits the $\Gamma_{\textbf{m}}$ value from each $\widehat{\textbf{y}}$ member, where the $\widehat{\textbf{y}}_{k_{1} - 1}$ individual  has the highest $\Gamma_{\textbf{m}}$ value and, therefore, $\widehat{\textbf{d}} = \widehat{\textbf{z}}_{k_{1} - 1}$.

%--
%--

Afterwards, the algorithm defines the $\Lambda_{h}$ operator to delimit the local search space as:
\begin{align}\label{eq:delta}
    \Lambda_{h} = \widehat{\textbf{d}} - \widehat{\textbf{z}}_{h},
\end{align} 
\noindent with $h =  1, \ldots, \widehat{K} -1$. If any of the independent variables $w =  1, \ldots, \textbf{n}$ from $\widehat{\textbf{z}}_{h}$ and $\widehat{\textbf{d}}$ have the same value, the algorithm applies the following restriction constraint condition to preserve search initiative:  
\begin{equation} 
\begin{split} \label{eq:delta_cond}
\textnormal{\textbf{if }} \Lambda_{h_{w}} = 0, \textnormal{\textbf{then}} \\
\Lambda_{h_{w}} = 1. 
\end{split}
\end{equation}

For the last steps, the MO-LSP uses the same procedure proposed by \cite{de2019soft} to the LSP, wherein the mutated values $\widehat{\textbf{m}}_{j} \in \Omega$, with $\widehat{\textbf{m}}_{j} = \left(\widehat{m}_{j_{1}} , \ldots, \widehat{m}_{j_{\textbf{n}}} \right)$, are calculated in two independent stages. In the first one, the algorithm calculates the $\widehat{\textbf{m}}_{1}$ value by performing  a local search around the region defined by $\widehat{\textbf{d}}$. Following this, the second stage defines the $\widehat{\textbf{m}}_{j+1}$ values around the region delimited by $\widehat{\textbf{z}}$ and $\Lambda$, as described by: 
\begin{align}\label{eq:bestM1}
	\widehat{\textbf{m}}_{1_{w}} = \widehat{\textbf{d}}_{1_{w}} + \widehat{\textbf{d}}_{1_{w}} \times \kappa \times (-1)^{\phi}, 	 
\end{align} 
\noindent and
\begin{align}\label{eq:bestM}
	\widehat{\textbf{m}}_{(h+1)_{w}} = \widehat{\textbf{z}}_{h_{w}} + \Lambda_{h_{w}}\times \kappa \times (-1)^{\phi},  
\end{align}
\noindent where $\kappa$ is a random number between 0 and 1, and $\phi$ is a random integer number that can be $1$ or $2$.

\begin{figure}[ht]
    \centering
    \includegraphics[scale=0.2]{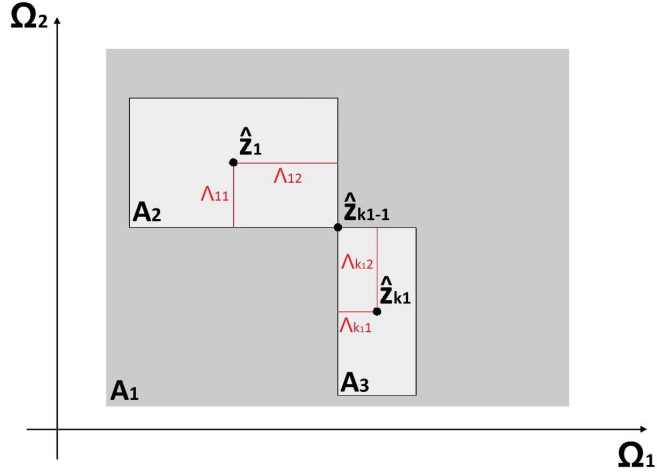}
	\caption{The MO-LSP in a two-dimensional optimization problem. Source: adapted from \cite{de2019soft}. }
	\label{fig:search}
\end{figure}

The MO-LSP in a two-dimensional optimization problem is presented in \autoref{fig:search}, with three decision vectors $\widehat{\textbf{z}}$ from the $\widehat{\textbf{y}}_{j}$ matrix. In this situation, the individual $\widehat{\textbf{y}}_{k_{1} - 1}$ presents the highest $\Gamma_{\textbf{m}}$ value, thus its respective decision vector, $\widehat{\textbf{z}}_{k_{1} - 1}$, leads the local mutation procedure. First, the dimensions of the $\widehat{\textbf{z}}_{k_{1} - 1}$ vector defines the search space $A_1$ using (\ref{eq:bestM1}). Next, the $\Lambda_{h}$ values are calculated using the difference of the reference decision vector and vectors $\widehat{\textbf{z}}_{k_{1}}$ and $\widehat{\textbf{z}}_{1}$ (\ref{eq:delta}), and search spaces $A_2$ and $A_3$ are defined by (\ref{eq:bestM}).  

%--
%--

Subsequently to the mutation step, the algorithm performs a constraint handling procedure to ensure that each independent variable $w$ belongs to the search limits proposed in the problem. In case the mutated value is higher than the upper limit $\textbf{U}_{w}$, this value is replaced by a random number chosen between 75 to 100\% of the search value. Similarly, if the mutated value is smaller than the lower limit $\textbf{L}_{w}$, this value will be replaced by a random number chosen among the 25\% minor numbers from the feasible search space. This procedure can be summarized as:
\begin{equation} 
\begin{split} \label{eq:ch}
 & \textnormal{\textbf{if }} \widehat{\textbf{m}}_{j_{w}} < \textbf{L}_{w}, \textnormal{\textbf{then}}\\
 & \ \ \ \ \widehat{\textbf{m}}_{j_{w}} = f_r \left( 0.25 \textbf{ss}_{w} \right) \\
 & \textnormal{\textbf{if }} \widehat{\textbf{m}}_{j_{w}} > \textbf{U}_{w}, \textnormal{\textbf{then}} \\
 & \ \ \ \ \widehat{\textbf{m}}_{j_{w}} = f_r \left( 0.25 \textbf{ss}_{w} \right) +  0.75 \textbf{ss}_{w},
\end{split} 
\end{equation} 
\noindent where $\textbf{ss}_{w}$ represents the range of the feasible search space and $f_r$ is a function that yields a random number between $0$ and the input value. In conclusion, the algorithm uses the original MOEA function-based selection operator where all decision vectors from $\widehat{\textbf{d}}$, $\widehat{\textbf{z}}$ and $\widehat{\textbf{m}}$ are ranked together, and the final selected individuals are returned to the main population. The multiobjective optimization procedure with MO-LSP is summarized in \autoref{alg:mo-lsp}, where the variable $c_{max}$ is the total number of iterations to fulfill the termination criterion defined by the user.

\vspace{2mm}
\begin{algorithm2e}[H]
\SetAlgoLined
\textbf{Input:} $N$ (population size), $M$ objectives \\
\textbf{Output:} $P$ (final population)  \\
Initialize a random population; \\
\textbf{while} $c < c_{max}$ \\
\scalebox{0.8}{\textbf{ 1}} \ \ \ Execute MOEA; \\
\scalebox{0.8}{\textbf{ 2}} \ \ \ Define $\widehat{\textbf{z}}$ and $\widehat{\textbf{y}}$ from PF$_{1}$ and PF$_{2}$; \\
\scalebox{0.8}{\textbf{ 3}} \ \ \ Select $\widehat{\textbf{d}}$ using $\Gamma_{t}$; \\
\scalebox{0.8}{\textbf{ 4}} \ \ \ Remove $\widehat{\textbf{d}}$ from $\widehat{\textbf{z}}$ population; \\
\scalebox{0.8}{\textbf{ 5}} \ \ \ Calculate $\Lambda$ using (\ref{eq:delta}); \\
\scalebox{0.8}{\textbf{ 6}} \ \ \ Verify condition defined by (\ref{eq:delta_cond}); \\
\scalebox{0.8}{\textbf{ 7}} \ \ \ Calculate $\widehat{\textbf{m}}_{1}$ using (\ref{eq:bestM1}); \\
\scalebox{0.8}{\textbf{ 9}} \ \ \ Calculate $\widehat{\textbf{m}}_{(h+1)}$ using (\ref{eq:bestM}); \\
\scalebox{0.8}{\textbf{10}} \ \ \ Verify boundary constraints proposed in (\ref{eq:ch}); \\
\scalebox{0.8}{\textbf{11}} \ \ \ Perform selection technique; \\
\scalebox{0.8}{\textbf{12}} \ \ \ Return the best individuals to the main population; \\
\scalebox{0.8}{\textbf{13}} \ \ \ $c = c + 1$; \\
\caption{Multiobjective Local Search Procedure}
\label{alg:mo-lsp}
\end{algorithm2e}
\vspace{2mm}

\subsection{MO-LSP for robust control optimization }
\label{sec:3.3}

\begin{figure}[ht]
    \centering
    \includegraphics[scale=0.45]{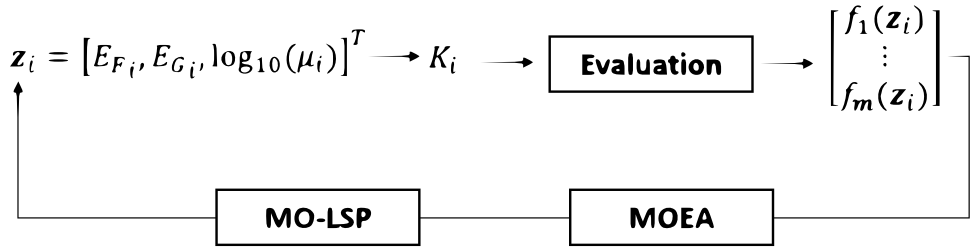}
	\caption{Robust control optimization using the MO-LSP algorithm.}
	\label{fig:flux_mop_control}
\end{figure}

The MO-LSP algorithm limits the search space around highly qualified solutions. This way, the method enhances performance and is suitable for large size problems, such as the design of robust control methods based on input and output measurements. To illustrate this procedure, an optimization loop is shown in \autoref{fig:flux_mop_control}.

%--
%--

First, an individual $\textbf{z}_{i}$ is randomly chosen in the feasible search space, as presented in \autoref{sec:3.1}. The gain $K_i$ for each $\textbf{z}_{i}$ value is calculated with the matrices $Q$, $R$, $F$ and $G$ of the RLQR, as presented in \autoref{sec:2.2}. Then, the vehicle's performance is evaluated in a simulation environment with the calculated gain. This evaluation returns $m$ errors associated with the objective functions $f_{1} \left( \textbf{z}_{i} \right), f_{2} \left( \textbf{z}_{i} \right),  \ldots , f_{\textbf{m}} \left( \textbf{z}_{i} \right)$, as in (\ref{eq:minF}). Finally, a MOEA performs an optimization procedure to decrease the associated errors, and the MO-LSP algorithm adjusts the values returned by the MOEA to define the $\textbf{z}_{i}$ configuration (i.e., uncertainties matrices $E_{F_i}$ and $E_{G_i}$, and $(\log_{10}(\mu_i))^2$ value) for the best set of solutions in the formulation of control design until the stop criterion is met.

% SECTION 4
\section{Experiments and results}
\label{sec:4} 

This section presents the numerical results and discusses the efficiency of the MO-LSP in optimizing the RLQR performance. Thus, the MO-LSP was evaluated in two different situations: using a generic robust control model and an applied control model. Furthermore, the final system's response and performance are discussed in this section. To validate the performance of the proposed MO-LSP, 10 well established and competitive MOEAs are considered for comparison. This implementation considered $50$ independent simulations in the statistical analysis. Furthermore, all experimental designs were carried using the MATLAB platform for evolutionary multiobjective optimization \citep{tian2017platemo}, on Ubuntu $16.04$ OS with an Intel core $i7-7700$ processor $4.2$ GHz. The MOEAs used to validate the effectiveness of the proposed approach are first introduced, then the parameter settings are given and finally the metrics used to evaluate their performance are described.

\subsection{Experimental design}
\label{sec:4.ed}          

\subsubsection{Comparative algorithms}
\label{sec:4.ca}

\begin{enumerate} 

\item NSGA-II \citep{deb2002fast} is an improved version of NSGA \citep{srinivas1994muiltiobjective}, with a fast nondominated sorting procedure, an elitist strategy, a parameter-less approach and a simple yet efficient constraint-handling method. The algorithm uses a crowded-comparison approach to replace the well-known sharing function approach used by the NSGA, which decreases overall complexity and does not require parameter definition by the user. 

\item NSGA-III \citep{deb2013evolutionary} is a version of NSGA-II with significant changes in its selection operator. For NSGA-III, diversity among population members is maintained by supplying and adaptively updating a number of direction vectors, where the algorithm proposes a systematic analysis of new population members based on the supplied reference points.

\item $\theta$-DEA \citep{yuan2015new} is an algorithm that combines the advantages of two models, NSGA-III and MOEA/D \citep{zhang2007moea}. A new $\theta$-dominance operation is proposed to enhance the convergence, by keeping the strength of NSGA-III in diversity preservation while it explores the fitness evaluation scheme in MOEA/D.  

\item MOMBI-II \citep{hernandez2015improved} is an improved version of MOMBI \citep{gomez2013mombi} that overcomes the loss of diversity for high dimensionality. MOMBI-II is an R$2$ indicator-based optimizer, that uses achievement scalar function values and population statistical information for ranking procedure. 

\item MOEA/IGD-NS \citep{tian2016multi} is an algorithm that considers Inverted Generational Distance (IGD) metric and noncontributing solutions (NS) for selection procedures. NS are non-dominated individuals that do not represent the nearest neighbor of any reference point from Pareto optimal front, and they are ignored on IGD calculation. Thus, MOEA/IGD-NS considers IGD to keep diversity and convergence, and the sum of the minimum distance from each NS, to enhance IGD values with few NS individuals. 

\item EFR-RR \citep{yuan2015balancing} is an enhanced version of the EFR \citep{yuan2014evolutionary} with a ranking restriction scheme to promote convergence and maintain the diversity and the distribution of solutions during the evolutionary process. This algorithm evaluates the perpendicular distance from a solution to a weight vector and considers to rank only the fitness functions with close corresponding weight vectors in the objective space. 

\item MaOEA-DDFC \citep{cheng2015many} is an algorithm based on measures on directional diversity (DD) and favorable convergence (FC). The convergence performance is measured based on the Chebyshev function and favorable weight, and the environmental selection considers diversity and convergence in a tournament-like manner to select the most promising convergence performance individual. 

\item SPEA/R \citep{jiang2017strength} is a SPEA \citep{zitzler1999comparison} version that adopts a diversity-first-and-convergence-second selection strategy to keep a balance between diversity and convergence. The algorithm evaluates a set of reference directions to delimit independent subregions in the objective space and perform the search procedure. After the genetic operators, SPEA/R uses an objective normalization strategy to merge the parent and offspring population, and each individual is associated to a respective subregion.  

\item SPEA2+SDE \citep{li2013shift} is a SPEA2 \citep{zitzler2001spea2} variant that uses a shift-based density estimation (SDE) strategy that covers the distribution and convergence information of individuals, and to make Pareto-based algorithms suitable for many-objective optimization. Considering density estimators, SDE allocates poor convergence individuals into crowded sparse regions, assigning them with a high-density value and easily eliminating them during the evolutionary process. 

\item BiGE \citep{li2015bi} is a model that considers the individuals’ proximity and crowding degree to convert a many-objective optimization problem into a bi-goal (objective) optimization problem, and perform the genetic operators using Pareto dominance in a bi-goal domain. BiGE performs individual comparison strategies in the mating and environmental selection, where the proximity and crowding degree are considered to estimate individuals’ performance.

\end{enumerate}

% ------------------------------- %

\subsubsection{Parameter settings}
\label{sec:4.ps}

The same population size was addressed for all MOEAs. This value is controlled by division parameters and cannot be arbitrarily specified \citep{yuan2015new}. Then the divisions $(H_{1}, H_{2})$ were set to $(12,0)$, and each MOEA was initialized with a random population with $92$ individuals. The reproduction parameters were set as crossover probability $p_{c} = 1.0$, mutation probability $p_{m} = 0.5$, distribution index of polynomial mutation $\eta_{m} = 20$, and all the other parameters were set as suggested by their corresponding references. To define the termination criterion, an experimental analysis was performed to find a relation between performance and computational cost, and a good balance was obtained by $10000$ executed evaluations functions per simulation. 

\subsubsection{Measures of  performance}
\label{sec:4.mofp}

The performance metrics used to evaluate the quality of non-dominated values obtained by the MOEAs, known as PF approximations \citep{garcia2019comparison}, can be classified into indicators of cardinality, convergence, distribution and spread \citep{audet2018performance}. These metrics consider the PF approximations obtained by the search algorithms, defined as $\mathcal{P}$ values, the optimal PF, represented by $\mathcal{P}^{*}$, and $| \mathcal{P} |$ and $| \mathcal{P}^{*} |$ as the number of elements from $\mathcal{P}$ and $\mathcal{P}^{*}$, respectively. Also, the MOP defined in \autoref{sec:2} has an unknown optimal PF solution. Consequently, it is necessary to specify a set of points in the objective space and use it as $\mathcal{P}^{*}$ for all MOEAs \citep{audet2018performance}. In this sense, this paper considers five popular performance indicators, based on the previous classification. They are described as follows.  

\begin{enumerate}

    \item Inverted Generational Distance ($IGD$) \citep{coello2005solving}:  is a popular convergence indicator used to evaluate the distance between $\mathcal{P}$ and $\mathcal{P}^{*}$: 
    \begin{align}\label{eq:igd}
        IGD(\mathcal{P},\mathcal{P}^{*}) = \frac{1}{|\mathcal{P}^{*}|} \sum_{ \textbf{k} \in \mathcal{P}^{*}} d_{\textbf{k}},  
    \end{align} 
    \noindent where $d_{\textbf{k}}$ is the minimal Euclidian distance between $\textbf{k} \in \mathcal{P}^{*}$ and all the $\textbf{i} \in \mathcal{P}$ non-dominated points from $\mathcal{P}$:
    \begin{align}\label{eq:dk}
        d_{\textbf{k}} = \min \limits_{\textbf{i} \in P} \left( \sum_{\textbf{j}} \left( \mathcal{P}^{*}_{\textbf{k}_{\textbf{j}}} - \mathcal{P}_{\textbf{i}_{\textbf{j}}} \right)^{2} \right)^{\frac{1}{2}},
    \end{align} 
    \noindent and $\textbf{j} = 1, \ldots, \textbf{m} $, with $\textbf{m}$ objectives functions. More proximity between $\mathcal{P}$ and the optimal PF is desired. Consequently, lower $IGD$ values represent better evaluated populations. 
    
    \item Spacing (SP) \citep{schott1995fault}: is a distribution and spread indicator that aims to evaluate the arrangement of the members in $\mathcal{P}$. IT is calculated as:
    \begin{align}\label{eq:sp}
        SP(\mathcal{P}) = \sqrt{ \frac{1}{|\mathcal{P}| - 1} \sum_{\textbf{i} \in \mathcal{P}} \left( \overline{d} - d_{\textbf{i}}  \right) },
    \end{align} 
    \noindent where $d_{\textbf{i}}$ is the Euclidian distance between a point $\textbf{i} \in \mathcal{P}$ and its closest neighbor $\textbf{h} \in \mathcal{P}$ from the same obtained non-dominated set $\mathcal{P}$ as: 
    \begin{align}\label{eq:di}
        d_{i} = \min \limits_{(\textbf{i},\textbf{h}) \in \mathcal{P}, \textbf{i} \ne \textbf{h}} \left( \sum_{\textbf{j}} \left( \mathcal{P}_{\textbf{i}_{\textbf{j}}} - \mathcal{P}_{\textbf{h}_{\textbf{j}}} \right)^{2} \right)^{\frac{1}{2}},
    \end{align} 
    \noindent with $\overline{d}$ as the mean value of $d_{\textbf{i}}$. SP metric evaluates according to the spacing between  the values. 
 
    \item Hypervolume (HV) \citep{while2006faster}: is a measure of convergence and distribution properties, that calculates the volume of the region between a reference point $\widehat{\textbf{r}} \in \mathbb{R}^{m}$ and $\mathcal{P}$, such that $\mathcal{P} \prec \widehat{\textbf{r}}$. The HV metric is defined as: 
    \begin{align}\label{eq:hv}
        HV(\mathcal{P},\widehat{\textbf{r}}) = \Gamma_{\textbf{m}} \left( \bigcup_{\textbf{i} \in \mathcal{P}} \left[ \mathcal{P}_{\textbf{i}}; \widehat{\textbf{r}} \right] \right), 
    \end{align} 
    \noindent with $\Gamma_{\textbf{m}}$ as the \textbf{m}-dimensional Lebesgue measure. Large values of HV represent better convergence of the algorithms. The reference point is fixed for all the MOEAs, and it is set to $[25, 25, 25]$ and to $[5000, 10000, 5000, 5000]$ for the generic and the applied control models, respectively.
    
    \item Pure Divesity (PD) \citep{wang2016diversity}: is a metric inspired by a measure of biodiversity that calculates an accumulation of the dissimilarity in the population, represented by: 
    \begin{align}\label{eq:pd}
        PD(\mathcal{P}) = \max \limits_{\textbf{i} \in \mathcal{P}} \left( PD(\mathcal{P} - \textbf{i}) + d_{s}(\textbf{i}, \mathcal{P} - \textbf{i}) \right),
    \end{align} 
    \noindent where $d_{s}$ represents the dissimilarity of an individual ($\sigma_{\textbf{i}}$) to a population ($\Upsilon$) as: 
    \begin{align}\label{eq:ds}
        d_{s}(\sigma,\Upsilon) = \max \limits_{\sigma_{i} \in \Upsilon} \left( dissimilarity(\sigma,\sigma_{\textbf{i}}) \right),
    \end{align} 
    \noindent where the dissimilarity measures the different degrees of $\sigma$ to other individuals from the population $\Upsilon$.
    Non-dominated solutions of $\mathcal{P}$ with good diversity should provide the maximal degree of dissimilarity and, consequently, maximum amount of information \citep{wang2016diversity}. Hence, large values of PD are desired. 
    \item Spread \citep{wang2010multi}: this metric reflects the variance of the distances between neighboring Pareto solutions \citep{zhou2019decomposition}, and it is defined as: 
    \begin{align}\label{eq:sprd}
        Spread(\mathcal{P}) = \frac{\sum_{\textbf{i} = 1}^{\textbf{m}} d(e_{\textbf{i}}, \mathcal{P}) + \sum_{\chi \in \mathcal{P}} |d(x,\mathcal{P}) - \overline{d}|}{ \sum_{\textbf{i} = 1}^{\textbf{m}} d(e_{\textbf{i}}, \mathcal{P}) + |\mathcal{P}| * \overline{d} },  
    \end{align} 
    \noindent where $e_{\textbf{i}}$ represents the $\textbf{i}$-th extreme solution in $\mathcal{P}^{*}$, $d(e_{\textbf{i}}, \mathcal{P})$ is the minimal Euclidian distance between $e_{\textbf{i}}$ and all the non-dominated points from $\mathcal{P}$, as defined in (\ref{eq:dk}), and $d(\chi, \mathcal{P})$ is the Euclidian distance between a point $\chi \in \mathcal{P}$ and its closest neighbor, as represented in (\ref{eq:di}). Finally, $\overline{d}$ is defined as: 
    \begin{align}\label{eq:dm}
        \overline{d} = \frac{1}{|\mathcal{P}^{*}|} \sum_{\chi \in \mathcal{P}^{*}} d(\chi,\mathcal{P}).  
    \end{align} 
    \noindent Lower Spread values are desired, which better represent a non-dominated set uniformly distributed.   

\end{enumerate}

\subsection{Generic control model} 
\label{sec:4.4}

In this section, an application  involving a generic control defined by a continuous test model with known nominal matrix is presented as a MOP. The evolutionary search aims to propose the best controller parameters. For this procedure, decision vectors with $\textbf{n} = 5$ independent variables define the uncertainty matrices in a problem with $\textbf{m} = 3$ objective functions, represented by the regulation errors of the 3 state variables. The optimization problem is summarized as 
\begin{equation} 
\begin{split} \label{eq:minF2}
 & \min \textbf{F} \left( \textbf{z} \right) = \left( f_{1} \left( \textbf{z} \right), f_{2} \left( \textbf{z} \right),  f_{3} \left( \textbf{z} \right) \right)^{T} \\
 & \textbf{z} = \left( z_{1}, z_{2},  \ldots, z_{5} \right)^{T}, \\
\end{split} 
\end{equation} 
\noindent where $f_{1}$, $f_{2}$ and $f_{3}$ are the regulation errors of variables $x_1$, $x_2$ and $x_3$, respectively. The $\textbf{z}$ independent variables were initialized by random real numbers between $0$ and $200$, as $\textbf{z} \in \mathbb{R}$ and $\textbf{z} \in [ 0, \textnormal{ } 200]$. The system matrices of the generic model used in this evaluation are
%--
\begin{align*}F=
\begin{bmatrix}
0.9& 0.8 &0.7\\ 0.01 & 0.1 & 0.3\\ 0.0 & 0.25 & 0.1
\end{bmatrix},
\ \nonumber
G=\begin{bmatrix}0.6\\ 0.1\\ 0.25\end{bmatrix}, \ \nonumber
M=\begin{bmatrix}
1\\1\\1
\end{bmatrix}, \ \nonumber
E_F=\begin{bmatrix}
0.1\\0.2\\0.2
\end{bmatrix}^T, \ \nonumber
E_G = \begin{bmatrix}
0.1
\end{bmatrix}.
\end{align*}
%--

The mean and standard deviation performances of the $PF$ obtained by the selected MOEAs in their original version, and with the addition of the MO-LSP algorithm are shown in \autoref{tab:result_1}. To analyze the proposed algorithm results, a one-way analysis of variance (ANOVA) is performed for the multiple comparison of the results of each MOEA in its original version and also when applying the MO-LSP. This comparison was implemented using Tukey's honestly significant difference criterion, based on the Studentized range distribution (confidence level of 95\%). 

\begin{table*}[!htbp]
\scriptsize
\caption{The average and ($\pm$) the standard deviation over the measures of performance of the MOEAs’ in their original versions and with addition of the proposed MO-LSP, on the MOP for the generic control design.} 
\label{tab:result_1}
\begin{center}
\scriptsize
\scalebox{0.7}{% 
\begin{tabular}{ c  c  c  c  c  c  c  c  c  c  c}
\hline \\
\textbf{MOEA} & 
\textbf{IGD}  & \textbf{ } & 
\textbf{SP}   & \textbf{ } &
\textbf{HV}   & \textbf{ } &
\textbf{PD}   & \textbf{ } &
\textbf{Spread} 
\\ 
\scriptsize
& 
\\ \hline
% 1. NSGA-II \textbf{ }$^{a}$
%----------------------------------------------------------%
\linebreak \\ 
NSGA-II/MO-LSP
&  \textbf{2.6514E+1 $\pm$ 2.67E-1}$^{a,*}$ % IGD
& \textbf{}
&  \textbf{1.1815E-1 $\pm$ 3.73E-2}$^{a}$ % SP
& \textbf{}
&  \textbf{1.3644E+4 $\pm$ 1.23E+1}$^{a,*}$ % HV
& \textbf{}
&  \textbf{2.2943E+5 $\pm$ 2.23E+4}$^{a,*}$ % PD
& \textbf{}
&  \textbf{9.6881E-1 $\pm$ 4.57E-3} % Spread
\\
%-----------------%
NSGA-II
&  2.6867E+1 $\pm$ 2.18E-1 % IGD
& \textbf{}
&  8.2553E-2 $\pm$ 1.52E-2 % SP
& \textbf{}
&  1.3628E+4 $\pm$ 9.22E+0 % HV
& \textbf{}
&  2.0761E+5 $\pm$ 2.45E+4 % PD
& \textbf{}
&  9.6998E-1 $\pm$ 3.08E-3 % Spread
\\
% 2. NSGA-III 
%----------------------------------------------------------%
\linebreak \\ 
NSGA-III/MO-LSP
&  \textbf{2.6670E+1 $\pm$ 3.54E-1}$^{a}$ % IGD
& \textbf{}
&  \textbf{1.3804E-1 $\pm$ 5.73E-2}$^{a}$ % SP
& \textbf{}
&  \textbf{1.3625E+4 $\pm$ 1.59E+1}$^{a}$ % HV
& \textbf{}
&  \textbf{1.8573E+5 $\pm$ 2.39E+4}$^{a}$ % PD
& \textbf{}
&  \textbf{9.7854E-1 $\pm$ 6.11E-3}$^{a}$ % Spread
\\
%-----------------%
NSGA-III
&  2.7007E+1 $\pm$ 4.26E-1 % IGD
& \textbf{}
&  1.0641E-1 $\pm$ 4.41E-2 % SP
& \textbf{}
&  1.3605E+4 $\pm$ 2.15E+1 % HV
& \textbf{}
&  1.5608E+5 $\pm$ 2.14E+4 % PD
& \textbf{}
&  9.8263E-1 $\pm$ 5.05E-3 % Spread
\\
% 3. $\theta$-DEA
%----------------------------------------------------------%
\linebreak \\ 
$\theta$-DEA/MO-LSP
&  \textbf{2.6690E+1 $\pm$ 3.10E-1}$^{a}$ % IGD
& \textbf{}
&  \textbf{1.3586E-1 $\pm$ 5.72E-2}$^{a}$ % SP
& \textbf{}
&  1.3572E+4 $\pm$ 2.73E+1 % HV
& \textbf{}
&  \textbf{1.6647E+5 $\pm$ 1.61E+4}$^{a}$ % PD
& \textbf{}
&  \textbf{9.7989E-1 $\pm$ 4.91E-3} % Spread
\\
%-----------------%
$\theta$-DEA
&  2.7022E+1 $\pm$ 4.63E-1 % IGD
& \textbf{}
&  9.6822E-2 $\pm$ 2.91E-2 % SP
& \textbf{}
&  \textbf{1.3576E+4 $\pm$ 2.35E+1} % HV
& \textbf{}
&  1.5562E+5 $\pm$ 2.45E+4 % PD
& \textbf{}
&  9.7995E-1 $\pm$ 4.57E-3 % Spread
\\
% 4. MOMBI-II
%----------------------------------------------------------%
\linebreak \\ 
MOMBI-II/MO-LSP
&  \textbf{2.6986E+1 $\pm$ 4.39E-1}$^{a}$ % IGD
& \textbf{}
&  \textbf{1.5640E-1 $\pm$ 6.57E-2}$^{a}$ % SP
& \textbf{}
&  1.3522E+4 $\pm$ 4.99E+2 % HV
& \textbf{}
&  \textbf{1.3280E+5 $\pm$ 2.83E+4}$^{a}$ % PD
& \textbf{}
&  9.9271E-1 $\pm$ 7.03E-3 % Spread
\\
%-----------------%
MOMBI-II
&  2.7469E+1 $\pm$ 7.39E-1 % IGD
& \textbf{}
&  1.0330E-1 $\pm$ 5.52E-2 % SP
& \textbf{}
&  \textbf{1.3595E+4 $\pm$ 3.77E+1} % HV
& \textbf{}
&  1.1381E+5 $\pm$ 3.36E+4 % PD
& \textbf{}
&  \textbf{9.9222E-1 $\pm$ 3.91E-3} % Spread
\\
% 5. MOEA/IGD-NS
%----------------------------------------------------------%
\linebreak \\ 
MOEA/IGD-NS/MO-LSP
&  \textbf{2.6583E+1 $\pm$ 2.50E-1}$^{a}$ % IGD
& \textbf{}
&  \textbf{1.2679E-1 $\pm$ 4.78E-2}$^{a}$ % SP
& \textbf{}
&  \textbf{1.3625E+4 $\pm$ 1.71E+1}$^{a}$ % HV
& \textbf{}
&  \textbf{2.0069E+5 $\pm$ 1.56E+4}$^{a}$ % PD
& \textbf{}
&  9.6113E-1 $\pm$ 3.72E-3 % Spread
\\
%-----------------%
MOEA/IGD-NS
&  2.6847E+1 $\pm$ 2.77E-1 % IGD
& \textbf{}
&  9.4555E-2 $\pm$ 3.72E-2 % SP
& \textbf{}
&  1.3612E+4 $\pm$ 1.25E+1 % HV
& \textbf{}
&  1.8549E+5 $\pm$ 1.53E+4 % PD
& \textbf{}
&  \textbf{9.6076E-1 $\pm$ 3.87E-3}$^{*}$ % Spread
\\
% 6. EFR-RR
%----------------------------------------------------------%
\linebreak \\ 
EFR-RR/MO-LSP
&  \textbf{2.6598E+1 $\pm$ 2.88E-1} % IGD
& \textbf{}
&  \textbf{1.9211E-1 $\pm$ 6.80E-2}$^{a,*}$ % SP
& \textbf{}
&  \textbf{1.3638E+4 $\pm$ 1.33E+1}$^{a}$ % HV
& \textbf{}
&  \textbf{1.7061E+5 $\pm$ 1.64E+4}$^{a}$ % PD
& \textbf{}
&  9.8057E-1 $\pm$ 7.49E-3 % Spread
\\
%-----------------%
{EFR-RR}
&  2.6626E+1 $\pm$ 3.09E-1 % IGD
& \textbf{}
&  1.6012E-1 $\pm$ 5.94E-2 % SP
& \textbf{}
&  1.3630E+4 $\pm$ 1.05E+1 % HV
& \textbf{}
&  1.6122E+5 $\pm$ 1.35E+4 % PD
& \textbf{}
&  \textbf{9.8021E-1 $\pm$ 4.26E-3} % Spread
\\
% 7. MaOEA-DDFC
%----------------------------------------------------------%
\linebreak \\ 
MaOEA-DDFC/MO-LSP
&  \textbf{2.6580E+1 $\pm$ 3.01E-1}$^{a}$ % IGD
& \textbf{}
&  \textbf{1.5213E-1 $\pm$ 4.69E-2}$^{a}$ % SP
& \textbf{}
&  \textbf{1.3635E+4 $\pm$ 1.41E+1}$^{a}$ % HV
& \textbf{}
&  \textbf{1.8333E+5 $\pm$ 2.28E+4}$^{a}$ % PD
& \textbf{}
&  \textbf{9.8473E-1 $\pm$ 6.76E-3} % Spread
\\
%-----------------%
MaOEA-DDFC
&  2.6955E+1 $\pm$ 3.70E-1 % IGD
& \textbf{}
&  1.1452E-1 $\pm$ 3.14E-2 % SP
& \textbf{}
&  1.3620E+4 $\pm$ 1.51E+1 % HV
& \textbf{}
&  1.6074E+5 $\pm$ 1.89E+4 % PD
& \textbf{}
&  9.8554E-1 $\pm$ 4.16E-3 % Spread
\\
% 8. SPEA/R
%----------------------------------------------------------%
\linebreak \\ 
SPEA/R/MO-LSP
&  2.6616E+1 $\pm$ 3.54E-1 % IGD
& \textbf{}
&  \textbf{1.4497E-1 $\pm$ 4.85E-2} % SP
& \textbf{}
&  1.3621E+4 $\pm$ 1.66E+1 % HV
& \textbf{}
&  \textbf{1.8292E+5 $\pm$ 1.64E+4} % PD
& \textbf{}
&  \textbf{9.8104E-1 $\pm$ 4.92E-3} % Spread
\\
%-----------------%
SPEA/R
&  \textbf{2.6596E+1 $\pm$ 3.37E-1} % IGD
& \textbf{}
&  1.3316E-1 $\pm$ 4.00E-2 % SP
& \textbf{}
&  \textbf{1.3625E+4 $\pm$ 1.38E+1} % HV
& \textbf{}
&  1.8284E+5 $\pm$ 2.09E+4 % PD
& \textbf{}
&  9.8226E-1 $\pm$ 4.36E-3 % Spread
\\
% 9. SPEA2+SDE
%----------------------------------------------------------%
\linebreak \\ 
SPEA2+SDE/MO-LSP
&  \textbf{2.6819E+1 $\pm$ 2.78E-1}$^{a}$ % IGD
& \textbf{}
&  \textbf{1.3680E-1 $\pm$ 3.92E-2}$^{a}$ % SP
& \textbf{}
&  \textbf{1.3625E+4 $\pm$ 1.17E+1}$^{a}$ % HV
& \textbf{}
&  \textbf{1.7099E+5 $\pm$ 1.48E+4}$^{a}$ % PD
& \textbf{}
&  \textbf{9.8566E-1 $\pm$ 3.34E-3} % Spread
\\
%-----------------%
SPEA2+SDE
&  2.7149E+1 $\pm$ 4.74E-1 % IGD
& \textbf{}
&  1.0764E-1 $\pm$ 4.89E-2 % SP
& \textbf{}
&  1.3615E+4 $\pm$ 2.46E+1 % HV
& \textbf{}
&  1.5231E+5 $\pm$ 2.21E+4 % PD
& \textbf{}
&  9.8583E-1 $\pm$ 3.63E-3 % Spread
\\
% 10. BiGE
%----------------------------------------------------------%
\linebreak \\ 
BiGE/MO-LSP
&  \textbf{2.6530E+1 $\pm$ 3.14E-1}$^{a}$ % IGD
& \textbf{}
&  \textbf{1.4666E-1 $\pm$ 5.16E-2}$^{a}$ % SP
& \textbf{}
&  \textbf{1.3623E+4 $\pm$ 1.64E+1}$^{a}$ % HV
& \textbf{}
&  \textbf{1.8345E+5 $\pm$ 1.73E+4}$^{a}$ % PD
& \textbf{}
&  \textbf{9.8084E-1 $\pm$ 4.55E-3} % Spread
\\
%-----------------%
BiGE
&  2.6966E+1 $\pm$ 3.96E-1 % IGD
& \textbf{}
&  1.0567E-1 $\pm$ 2.80E-2 % SP
& \textbf{}
&  1.3611E+4 $\pm$ 2.04E+1 % HV
& \textbf{}
&  1.6268E+5 $\pm$ 2.23E+4 % PD
& \textbf{}
&  9.8207E-1 $\pm$ 3.83E-3 % Spread
\\
\hline
\end{tabular}
}
\caption*{a: the model is significantly better (95\%). \\
          *: best overall result. }
\end{center}
\end{table*}

From \autoref{tab:result_1}, one can notice a significant improvement in the algorithm's performance by applying the proposed MOI-LSP algorithm. For IGD metric, the proposed MOEA/MO-LSP had better performance in terms of average and standard deviation in $9$ out of $10$ performance evaluations, $8$ results being significantly better with respect to Tukey's test. Only the SPEA/R algorithm had no improvement with MO-LSP on IGD metric, but the results had no difference using Tukey’s criterion. Following this, the SP metric presented better results for all MOEA/MO-LSP models when compared to their original versions, and the proposed method showed honestly statistical difference by $9$ algorithms' results.   

The proposed method provided better results with Tukey's honestly statistical difference in seven algorithms for the HV metric, where for the $\theta$-DEA, MOMBI-II and SPEA/R algorithms the MO-LSP did not improve performance, but no difference using Tukey’s criterion was obtained. Subsequently, the MO-LSP algorithm showed better performance for all the MOEAs in PD metric, and only for the SPEA/R algorithm the proposed model did not present honestly statistical difference over the results. 

Using the Spread metric, the MO-LSP had better performance in terms of average for $7$ algorithms, and only in NSGA-III/MO-LSP the model presented statistical difference. These results are consistent, since a local search procedure is limited by the front individuals and this may not change or decrease the uniformity of the non-dominated distribution set.

\begin{figure}[ht]
    \centering
    \includegraphics[scale=0.6]{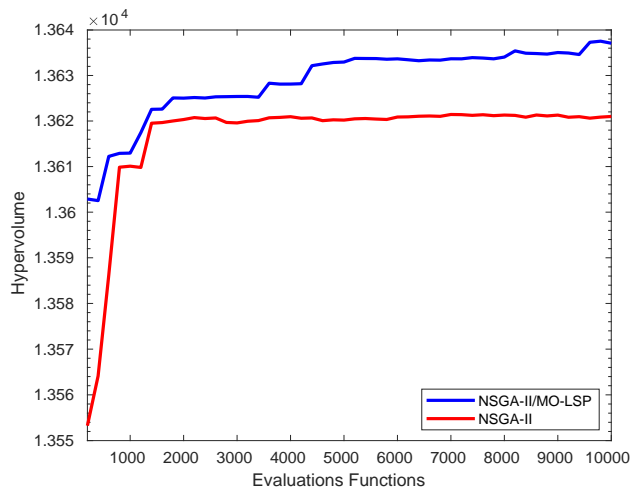}
	\caption{HV values returned by the NSGAII/MO-LSP and NSGAII algorithms for the generic control model over the optmization procedure.}
	\label{fig:hv-ge}
\end{figure}

Moreover, as indicated in \autoref{tab:result_1}, the NSGA-II/MO-LSP algorithm had the best overall results for IGD, HV and PD metrics, while EFR-RR/MO-LSP presented the best SP values. For the the Spread metric, the best overall results were achieved by the MOEA/IGD-NS algorithm. To investigate the influence of the MO-LSP, \autoref{fig:hv-ge} shows the evolution of the HV metric over the evaluation functions for the NSGAII/MO-LSP and NSGAII. As presented, the MO-LSP enhanced the MOEA's performance and convergence, by obtaining competitive PF values for the generic control optimization method.

%--
%--

The NSGA-II/MO-LSP was used to generate a PF to the generic control application due to its improved results. The parameters selection procedure is presented next.  

\subsubsection{Uncertain matrix definition}

\begin{figure}[ht]
    \centering
    \includegraphics[scale=0.65]{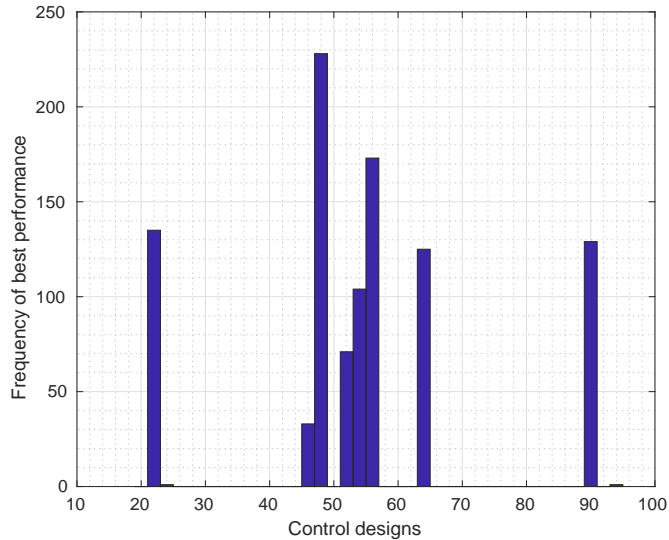}
	\caption{Individual performance of the $94$ control designs to the generic control model. The first $92$ values are composed by the PF decision vectors values, the $93^{rd}$ is a nominal control model and the $94^{th}$ considers the uncertainty matrix formed by random values.}
	\label{fig:fl-generic}
\end{figure}

The search performed by the NSGA-II/MO-LSP algorithm resulted in a PF composed of $92$ decision vectors. In order to select the global optimal value, represented by $\textbf{z}_{*}$, a comparison was performed considering $94$ control designs ($92$ decision vectors, one nominal control model and a design with random uncertainty variables). To select the fitness value, an evaluation with $1000$ independent simulations was carried out, each one with $\Delta$ randomly varying between $-1$ and $1$. The control design with less cumulative error is assigned as the winner for each simulation, and the individual performances are shown in \autoref{fig:fl-generic}. 

%-- 
%--

The performance of the individuals represented in \autoref{fig:fl-generic} was measured with $\Delta$ randomly varying for each test case. Then, one control loop is performed for each individual in the optimal set delivered by the NSGA-II/MO-LSP. The performance of each individual is evaluated in terms of the total mean square regulation error (i.e., the accumulated state-space variable error until the control system regulates them to zero). The histogram shows how many times each individual was the best performer. This demonstration serves two purposes: first, it shows that the individuals in the optimal set outperform both the randomly generated uncertainties and the zero uncertainties (nominal case); second, as the control algorithm demands one single individual to attribute value to the matrix uncertainties, the selection criterium is to choose the individual that performs the best in most cases (i.e., the most frequent individual in the performance histogram). 

As presented in \autoref{fig:fl-generic}, the $49^{th}$ PF decision vector represented the most competitive values, finding the best results in $222$ simulations. The random model achieved the best performance in just a single simulation, whilst the nominal did not yield competitive results in any of the simulations. The best uncertainty vector (individual $49^{th}$) is represented as

%--
\begin{align*} 
\textbf{z}_{*} =
\begin{bmatrix}
E_{F}^{T}\\ 
E_{G}^{T}\\
(\log_{10}(\mu))^2
\end{bmatrix}
\ \nonumber
= \begin{bmatrix}
19.933888  \\
19.999998  \\
20.000000  \\
16.339275  \\
20.000000  \\
\end{bmatrix}. \ \nonumber
\end{align*}
%--

\subsection{Applied control model}
\label{sec:4.5}

This section explores the applied control model presented in \autoref{sec:2}, discretized at a sampling time of $0.1 s$. The uncertainty matrices $E_{f}$ and $E_{g}$ are represented by decision vectors with $\textbf{n} = 7$ independent variables, and the optimization problem aims to minimize $\textbf{m} = 4$ objective functions, as
\begin{equation} 
\begin{split} \label{eq:minF3}
 & \min \textbf{F} \left( \textbf{z} \right) = \left( f_{1} \left( \textbf{z} \right), \ldots,  f_{4} \left( \textbf{z} \right) \right)^{T} \\
 & \textbf{z} = \left( z_{1}, z_{2},  \ldots, z_{7} \right)^{T}, \\
\end{split} 
\end{equation} 
\noindent where $f_{1}$ is the error of lateral displacement, $f_{2}$ is the regulation error of yaw rate, $f_{3}$ is the regulation error of lateral velocity,  and $f_{4}$ is the regulation error of vehicle orientation. 

The parameters of this RLQR design cannot be equal to zero, otherwise the controller would already be initialized in the solution. Therefore, $\textbf{z}$ independent variables were initialized by random real numbers, such as $\textbf{z} \in \mathbb{R}$ and $\textbf{z} \in (0, \textnormal{ } 500]$. \autoref{parameters-values} shows the vehicle's parameters and the method used to calculate the system matrices is similar to the one used in \cite{barbosa2019robust}, as shown in \ref{sec:uncertainties}. The resulting uncertainty system matrices are:
%--
\begin{equation*}
H =
\begin{bmatrix}
1\\
1\\
1\\
1
\end{bmatrix},\hspace{1mm}E_{F} = 
\begin{bmatrix}
6.85718E-4\\
-8.62010E-4\\
0\\
-6.6666E-3
\end{bmatrix}^{T}, \hspace{1mm}E_{G} =
\begin{bmatrix}
-6.66666E-3\\
-6.66666E-3
\end{bmatrix}^{T} \textnormal{and} \hspace{1mm} (\log_{10}(\mu))^2 = 81.
\end{equation*}
%--
\begin{table}[H]
\caption{Vehicle parameters values}
\label{parameters-values}
\begin{center}
\begin{tabular}{c l}
\hline
Parameter & Value\\
\hline
$a$  & $3.1870\hspace{1mm}m$\\
$b$  & $1.6180\hspace{1mm}m$\\
$l$ & $4.8050\hspace{1mm}m$\\
$v$ & $16.6667\hspace{1mm}m/s$\\
$m$ & $16030\hspace{1mm}kg$\\
Payload & $12550\hspace{1mm}kg$\\
$J$ & $2.1572\times10^5\hspace{1mm}kg \hspace{1mm} m^{2}$\\
$c_{1}$ & $1.0645\times10^5\hspace{1mm}N/rad$ \\
$c_{2}$ & $5.4042\times10^5\hspace{1mm}N/rad$\\
\hline
\end{tabular}
\end{center}
\end{table}

\autoref{tab:result_2} presents the statistical results of the PF, obtained by MOEAs over the presented performance measures, which considers their original versions and the proposed approach (MOEA/MO-LSP). To ensure the significant difference of the proposed application, the table presents an ANOVA test to multiple comparison procedures using Tukey's honestly significant difference criterion (confidence level of 95\%).

\begin{table*}[!thbp]
\scriptsize
\caption{The average and ($\pm$) the standard deviation over the measures of performance of the MOEAs’  on their original versions and with addition of the proposed MO-LSP, on the MOP for the applied control design.} 
\label{tab:result_2}
\begin{center}
\scriptsize 
\scalebox{0.7}{
\begin{tabular}{ c  c  c  c  c  c  c  c  c  c  c}
\hline \\
\textbf{MOEA} & 
\textbf{IGD}  & \textbf{ } & 
\textbf{SP}   & \textbf{ } &
\textbf{HV}   & \textbf{ } &
\textbf{PD}   & \textbf{ } &
\textbf{Spread} 
\\ 
\scriptsize  
& 
\\ \hline
\scriptsize
% 1. NSGA-II \textbf{ }$^{a}$
%----------------------------------------------------------%
\linebreak \\ 
NSGA-II/MO-LSP
&  \textbf{9.9567E+3 $\pm$ 3.75E+2}$^{a}$ % IGD
& \textbf{}
&  \textbf{9.1979E+1 $\pm$ 1.38E+2} % SP
& \textbf{}
&  \textbf{1.2347E+15 $\pm$ 5.49E+12}$^{a,*}$ % HV
& \textbf{}
&  \textbf{5.6215E+8 $\pm$ 2.28E+8}$^{a}$ % PD
& \textbf{}
&  \textbf{9.9212E-1 $\pm$ 1.72E-2} % Spread
\\
%-----------------%
NSGA-II
&  1.0153E+4 $\pm$ 3.65E+2 % IGD
& \textbf{}
&  7.7183E+1 $\pm$ 1.37E+2 % SP
& \textbf{}
&  1.2281E+15 $\pm$ 3.18E+12 % HV
& \textbf{}
&  4.0812E+8 $\pm$ 1.54E+8 % PD
& \textbf{}
&  9.9859E-1 $\pm$ 1.82E-2 % Spread
\\
% 2. NSGA-III 
%----------------------------------------------------------%
\linebreak \\ 
NSGA-III/MO-LSP
&  \textbf{1.0278E+4 $\pm$ 2.25E+2}$^{a}$ % IGD
& \textbf{}
&  \textbf{4.5133E+1 $\pm$ 3.72E+1} % SP
& \textbf{}
&  \textbf{1.2272E+15 $\pm$ 5.08E+12}$^{a}$ % HV
& \textbf{}
&  \textbf{2.9024E+8 $\pm$ 7.91E+7}$^{a}$ % PD
& \textbf{}
&  \textbf{9.9951E-1 $\pm$ 7.41E-3} % Spread
\\
%-----------------%
NSGA-III
&  1.0396E+4 $\pm$ 2.51E+2 % IGD
& \textbf{}
&  4.0309E+1 $\pm$ 6.73E+1 % SP
& \textbf{}
&  1.2168E+15 $\pm$ 1.10E+13 % HV
& \textbf{}
&  1.9930E+8 $\pm$ 7.99E+7 % PD
& \textbf{}
&  1.0012E+0 $\pm$ 8.90E-3 % Spread
\\
% 3. $\theta$-DEA
%----------------------------------------------------------%
\linebreak \\ 
$\theta$-DEA/MO-LSP
&  \textbf{1.0335E+4 $\pm$ 2.89E+2} % IGD
& \textbf{}
&  \textbf{5.9223E+1 $\pm$ 1.53E+2} % SP
& \textbf{}
&  \textbf{1.2172E+15 $\pm$ 1.66E+13} % HV
& \textbf{}
&  \textbf{2.0107E+8 $\pm$ 7.09E+7}$^{a}$ % PD
& \textbf{}
&  1.0050E+0 $\pm$ 2.18E-2 % Spread
\\
%-----------------%
$\theta$-DEA
&  1.0430E+4 $\pm$ 2.23E+2 % IGD
& \textbf{}
&  3.6280E+1 $\pm$ 8.01E+1 % SP
& \textbf{}
&  1.2129E+15 $\pm$ 1.37E+13 % HV
& \textbf{}
&  1.4122E+8 $\pm$ 4.14E+7 % PD
& \textbf{}
&  \textbf{1.0027E+0 $\pm$ 1.08E-2} % Spread
\\
% 4. MOMBI-II
%----------------------------------------------------------%
\linebreak \\ 
MOMBI-II/MO-LSP
&   \textbf{1.0301E+4 $\pm$ 1.92E+2}$^{a}$ % IGD
& \textbf{}
&   \textbf{3.0226E+1 $\pm$ 5.07E+1} % SP
& \textbf{}
&   \textbf{1.2325E+15 $\pm$ 3.49E+12}$^{a}$ % HV
& \textbf{}
&   \textbf{2.5432E+8 $\pm$ 7.47E+7}$^{a}$ % PD
& \textbf{}
&  1.0036E+0 $\pm$ 6.75E-3 % Spread
\\
%-----------------%
MOMBI-II
&  1.0414E+4 $\pm$ 1.53E+2 % IGD
& \textbf{}
&  2.5857E+1 $\pm$ 4.34E+1 % SP
& \textbf{}
&  1.2266E+15 $\pm$ 3.44E+12 % HV
& \textbf{}
&  1.8467E+8 $\pm$ 5.30E+7 % PD
& \textbf{}
&   \textbf{1.0025E+0 $\pm$ 4.25E-3} % Spread
\\
% 5. MOEA/IGD-NS
%----------------------------------------------------------%
\linebreak \\ 
MOEA/IGD-NS/MO-LSP
&  \textbf{1.0312E+4 $\pm$ 2.26E+2}$^{a}$ % IGD
& \textbf{}
&  \textbf{8.9270E+1 $\pm$ 1.38E+2}$^{a}$ % SP
& \textbf{}
&  \textbf{1.2307E+15 $\pm$ 5.98E+12}$^{a}$ % HV
& \textbf{}
&  \textbf{3.2759E+8 $\pm$ 7.18E+7}$^{a}$ % PD
& \textbf{}
&  9.9845E-1 $\pm$ 1.78E-2 % Spread
\\
%-----------------%
MOEA/IGD-NS
&  1.0407E+4 $\pm$ 2.28E+2 % IGD
& \textbf{}
&  3.8264E+1 $\pm$ 6.99E+1 % SP
& \textbf{}
&  1.2194E+15 $\pm$ 1.02E+13 % HV
& \textbf{}
&  2.3176E+8 $\pm$ 7.23E+7 % PD
& \textbf{}
&  \textbf{9.9352E-1 $\pm$ 1.02E-2} % Spread
\\
% 6. EFR-RR
%----------------------------------------------------------%
\linebreak \\ 
EFR-RR/MO-LSP
&  \textbf{9.4054E+3 $\pm$ 2.47E+2}$^{a}$ % IGD
& \textbf{}
&  \textbf{4.1226E+2 $\pm$ 3.53E+2}$^{a}$ % SP
& \textbf{}
&  \textbf{1.2326E+15 $\pm$ 4.15E+12}$^{a}$ % HV
& \textbf{}
&  \textbf{7.6329E+8 $\pm$ 2.26E+8}$^{a,*}$ % PD
& \textbf{}
&  1.0576E+0 $\pm$ 4.14E-2 % Spread
\\
%-----------------%
EFR-RR
&  9.9348E+3 $\pm$ 4.71E+2 % IGD
& \textbf{}
&  2.5012E+2 $\pm$ 3.51E+2 % SP
& \textbf{}
&  1.2247E+15 $\pm$ 7.60E+12 % HV
& \textbf{}
&  4.0443E+8 $\pm$ 2.41E+8 % PD
& \textbf{}
&  \textbf{1.0312E+0 $\pm$ 4.73E-2}$^{a,*}$ % Spread
\\
% 7. MaOEA-DDFC
%----------------------------------------------------------%
\linebreak \\ 
MaOEA-DDFC/MO-LSP
&  \textbf{1.0158E+4 $\pm$ 2.71E+2}$^{a}$ % IGD
& \textbf{}
&  \textbf{9.5584E+1 $\pm$ 9.46E+1}$^{a,*}$ % SP
& \textbf{}
&  \textbf{1.2325E+15 $\pm$ 3.90E+12}$^{a}$ % HV
& \textbf{}
&  \textbf{3.9193E+8 $\pm$ 6.45E+7}$^{a}$ % PD
& \textbf{}
&  1.0037E+0 $\pm$ 1.40E-2 % Spread
\\
%-----------------%
MaOEA-DDFC
&  1.0372E+4 $\pm$ 2.33E+2 % IGD
& \textbf{}
&  3.9554E+1 $\pm$ 5.29E+1 % SP
& \textbf{}
&  1.2195E+15 $\pm$ 7.20E+12 % HV
& \textbf{}
&  2.5277E+8 $\pm$ 7.80E+7 % PD
& \textbf{}
&  \textbf{9.9677E-1 $\pm$ 6.44E-3}$^{a}$ % Spread
\\
% 8. SPEA/R
%----------------------------------------------------------%
\linebreak \\ 
SPEA/R/MO-LSP
&  \textbf{9.8804E+4 $\pm$ 4.70E+2}$^{a}$ % IGD
& \textbf{}
&  \textbf{1.6476E+2 $\pm$ 2.59E+2}$^{a}$ % SP
& \textbf{}
&  \textbf{1.2285E+15 $\pm$ 4.94E+12}$^{a}$ % HV
& \textbf{}
&  \textbf{5.8282E+8 $\pm$ 4.06E+8}$^{a}$ % PD
& \textbf{}
&  1.0176E+0 $\pm$ 2.87E-2 % Spread
\\
%-----------------%
SPEA/R
&  1.0382E+4 $\pm$ 1.95E+2 % IGD
& \textbf{}
&  4.5938E+1 $\pm$ 8.07E+1 % SP
& \textbf{}
&  1.2173E+15 $\pm$ 1.06E+13 % HV
& \textbf{}
&  2.1226E+8 $\pm$ 7.46E+7 % PD
& \textbf{}
& \textbf{1.0049E+0 $\pm$ 1.31E-2}$^{a}$ % Spread
\\
% 9. SPEA2+SDE
%----------------------------------------------------------%
\linebreak \\ 
SPEA2+SDE/MO-LSP
&  \textbf{1.0445E+4 $\pm$ 6.60E+1}$^{a}$ % IGD
& \textbf{}
&  \textbf{2.5946E+1 $\pm$ 1.38E+1} % SP
& \textbf{}
&  \textbf{1.2303E+15 $\pm$ 4.62E+12}$^{a}$ % HV
& \textbf{}
&  \textbf{2.0872E+8 $\pm$ 5.14E+7}$^{a}$ % PD
& \textbf{}
&  9.9855E-1 $\pm$ 3.46E-3 % Spread
\\
%-----------------%
SPEA2+SDE
&  1.0528E+4 $\pm$ 5.12E+1 % IGD
& \textbf{}
&  1.8283E+1 $\pm$ 2.76E+1 % SP
& \textbf{}
&  1.2231E+15 $\pm$ 5.03E+12 % HV
& \textbf{}
&  1.5596E+8 $\pm$ 2.15E+7 % PD
& \textbf{}
&  \textbf{9.9670E-1 $\pm$ 4.78E-3} % Spread
\\
% 10. BiGE
%----------------------------------------------------------%
\linebreak \\ 
BiGE/MO-LSP
&  \textbf{1.0149E+4 $\pm$ 2.83E+2}$^{a,*}$ % IGD
& \textbf{}
&  \textbf{7.3337E+1 $\pm$ 9.46E+1}$^{a}$ % SP
& \textbf{}
&  \textbf{1.2327E+15 $\pm$ 4.39E+12}$^{a}$ % HV
& \textbf{}
&  \textbf{3.4968E+8 $\pm$ 1.05E+8}$^{a}$ % PD
& \textbf{}
&  1.0066E+0 $\pm$ 1.34E-2 % Spread
\\
%-----------------%
BiGE
&  1.0405E+4 $\pm$ 1.66E+2 % IGD
& \textbf{}
&  3.3735E+1 $\pm$ 5.53E+1 % SP
& \textbf{}
&  1.22385E+15 $\pm$ 5.65E+12 % HV
& \textbf{}
&  2.2544E+8 $\pm$ 6.24E+7 % PD
& \textbf{}
&  \textbf{1.0014E+0 $\pm$ 6.98E-3}$^{a}$ % Spread
\\
\hline
\end{tabular}
}
\caption*{a: the model is significantly better (95\%). \\
          *: best overall result. }
\end{center}
\end{table*}

From \autoref{tab:result_2} it is possible to conclude that the proposed application enhanced the MOEAs convergence (rate) to the applied control model. The MOEA/MO-LSP improved the results in terms of average and standard deviation in all applications for IGD, SP, HV and PD metrics. 

For the HV metric, the proposed application presented Tukey's honestly statistical difference in $9$ out of $10$ results, only the $\theta$-DEA results had no statistical significance level. Furthermore, the proposed application had no difference using Tukey’s criterion in SP metric in $5$ applications. It shows that a local search optimization may not significantly affect the spread of individuals in the applied control application for some MOEAs. However, decreases on the IGD metric proved the non-dominated values were improved. Also, for the PD metric, the proposed MOEA/MO-LSP obtained honestly statistical difference for all control applications. 

Considering the $Spread$ metric, the MO-LSP could not improve the algorithms performance, presenting only for the better results NSGA-II and NSGA-III algorithms, with no statistical significance level. In general, the comparisons showed no difference when using Tukey’s criterion for $5$ algorithms. As previously presented, the front individuals limit a local search procedure, which may decrease the uniformity of the non-dominated distribution set. It is interesting to observe that MO-LSP led the uncertain matrices elements to increase the HV value, by adjusting the values to the better prepared individual and, in this case, this procedure decreased the $Spread$ performance. 

From \autoref{tab:result_2}, one can conclude that the best overall result for the IGD metric was presented by the BiGE/MO-LSP algorithm, the MaOEA-DDFC/MO-LSP algorithm showed the best result for the SP metric, the best HV result was defined by the NSGA-II/MO-LSP algorithm, and the best overall PD and Spread metrics were proposed by the EFR-RR/MO-LSP and EFR-RR algorithms, respectively. 

\begin{figure}[ht]
    \centering
    \includegraphics[scale=0.6]{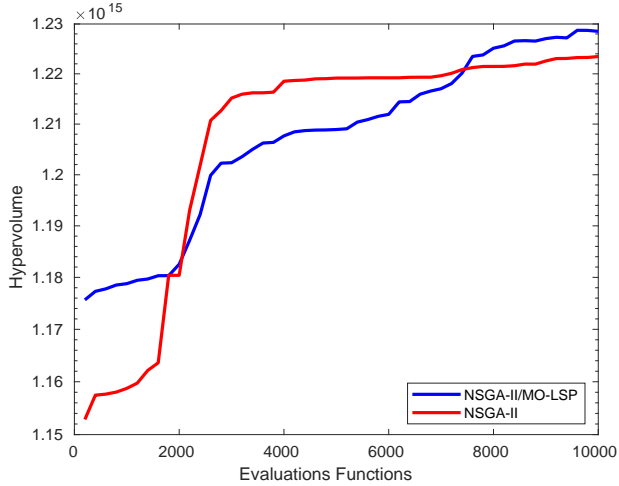}
	\caption{$HV$ values returned by the NSGA-II/MO-LSP and NSGA-II algorithms for the applied control model over the optmization procedure.}
	\label{fig:hv-ap}
\end{figure}

In addition, \autoref{fig:hv-ap} presents the evolution of the HV metric over the evaluation functions for the NSGA-II/MO-LSP and NSGA-II algorithms for the applied control optimization method, where the MO-LSP algortihm provides improved PF values. To perform an experimental analysis, the NSGA-II/MO-LSP algorithm was also used to generate a PF to the applied control application, and the parameters selection procedure is presented as follows.  

%--
%--

\subsubsection{Uncertain matrix definition} \label{umd}

In order to verify the global optimal value $\textbf{z}_{*}$ to the applied control model, an optimization search was performed by the NSGA-II/MO-LSP algorithm, resulting in a PF with $92$ decision vectors. Additionally, a comparison procedure was carried out considering $95$ control designs, being the first $92$ defined by the PF decision vectors, the $93^{rd}$ is a nominal control design, the $94^{th}$ is the algebraic manipulation proposed by \cite{barbosa2019robust} and, finally, the $95^{th}$ design represented an uncertainty matrix composed of random values. 

\begin{figure}[ht]
    \centering
    \includegraphics[scale=0.65]{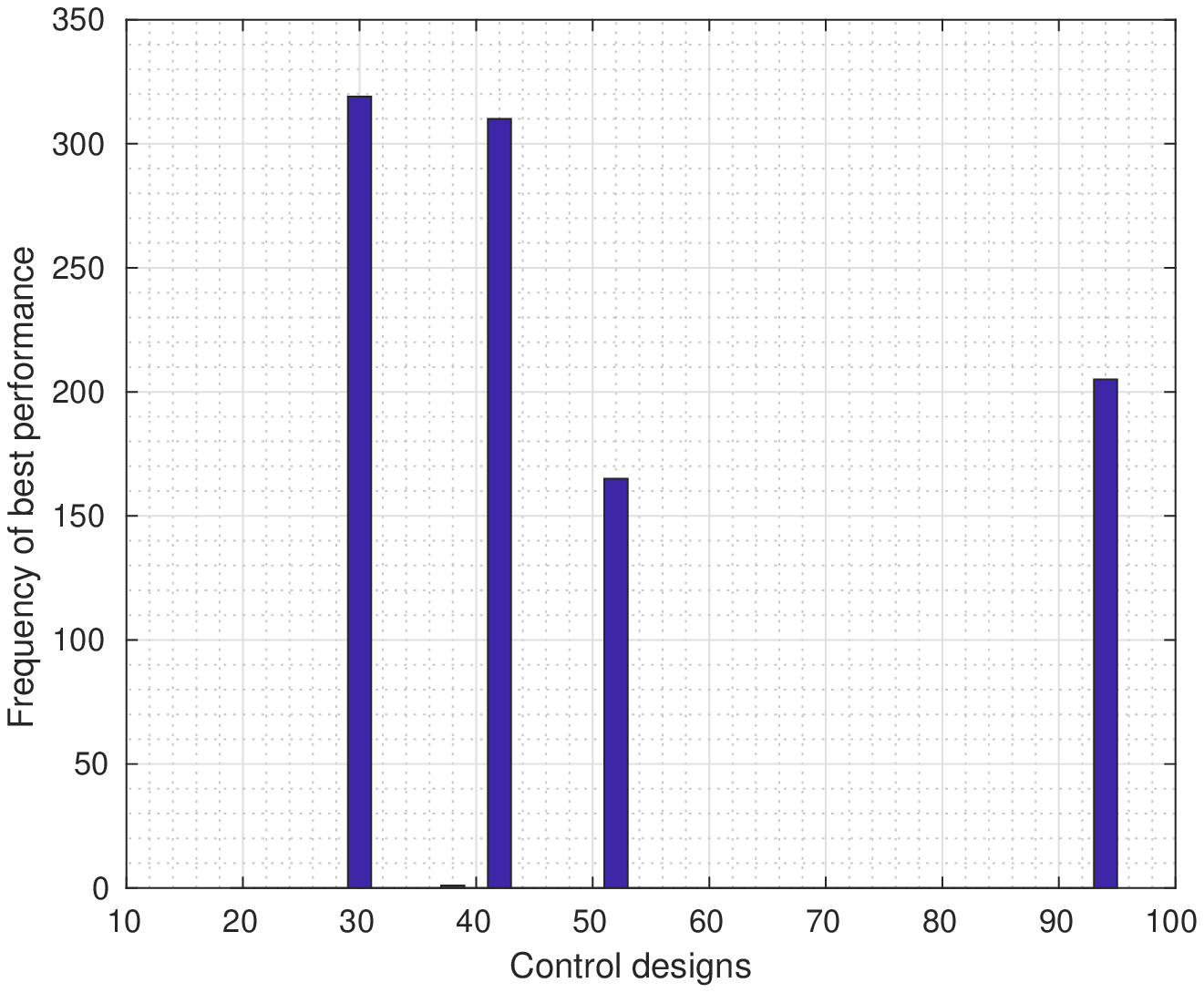}
	\caption{Individual performance of the $95$ control designs to the applied control model, being the first $92$ values composed of $PF$ decision vectors values, the $93^{rd}$ is a nominal control model, the $94^{th}$ the configuration proposed by \cite{barbosa2019robust}, and the $95^{th}$ considers the uncertainty matrix formed by random values.}
	\label{fig:fl-applied}
\end{figure}

Analogically to what was previously done, the performance of the individuals represented in \autoref{fig:fl-applied} is considered in the same way as in the generic example, where an evaluation with $1000$ independent simulations with $\Delta$ randomly varying between $-1$ and $1$ was performed to select the control design with less cumulative error in all states as the fitness value. 

%--
%--

The control design with the best results is proposed by the $31^{st}$  decision vector, which presented improved performance in $319$ simulations. Also, the $41^{st}$ decision vector had a similar performance, with enhanced performance in $310$ simulations. Furthermore, the algebraic manipulation control design outperformed the others in $205$ simulations, while the nominal and random models did not achieve any competitive results in the simulations. The optimal $\textbf{z}_{*}$ value to the uncertainty matrix (individual $31^{th}$) is defined as:

%--
\begin{align*} 
\textbf{z}_{*} =
\begin{bmatrix}
E_{F}^{T}\\ 
E_{G}^{T}\\ 
(\log_{10}(\mu))^2
\end{bmatrix}
\ \nonumber
= \begin{bmatrix}
4.45032  \\
57.84300  \\
21.90069  \\
455.67736  \\
48.05012  \\
44.98551  \\
280.28421  \\
\end{bmatrix}. \ \nonumber
\end{align*}
%--

\subsubsection{System response and performance of the applied control model}
\label{sec:sr_ap}

The performance of the robust recursive controller for the vehicle model, shown in \autoref{single} and  represented  in the state-space Equation (\ref{eq:ssdynamic}), was evaluated to minimize the path-following error, given a reference path and smoother commands for realistic steering values. This comparison took into account the performance of  four different control approaches for path following. First, the multiobjective approach optimized by the NSGA-II/MO-LSP algorithm (RLQR$^{MOP}$) (see \autoref{umd}). Second, the method used by \cite{barbosa2019robust}(RLQR$^{A}$). Third, the standard $\mathcal{H}_\infty$ controller subject to uncertainties on the payload of the vehicle \citep{hu2016robust}. And lastly, the standard LQR controller.

\begin{figure}[t]
    \centering
    \includegraphics[scale=0.35]{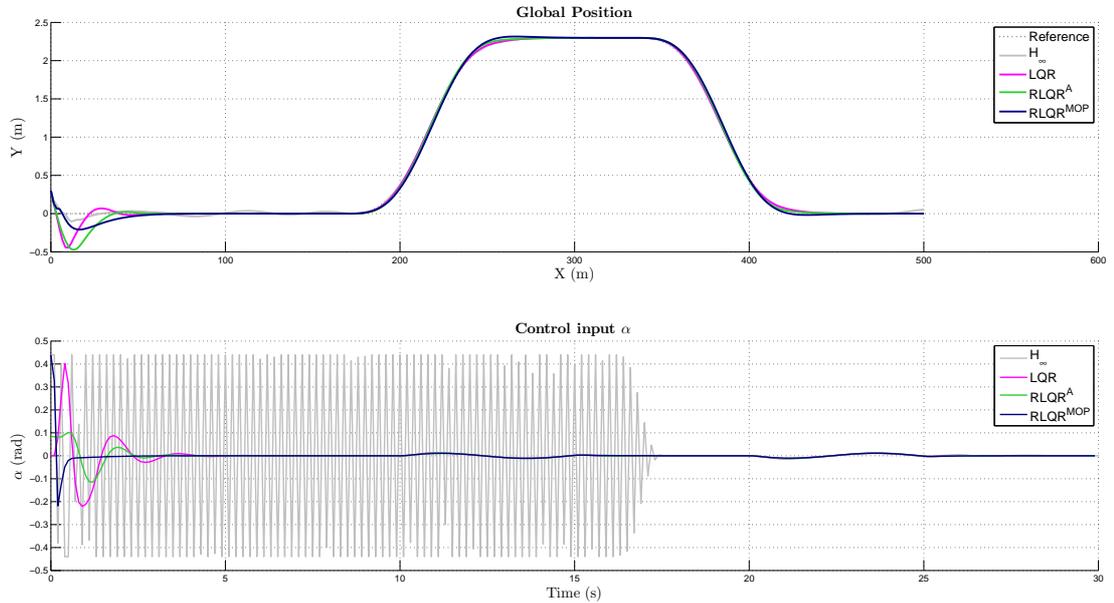}
    \caption{Global position and steering angle for nominal payload.}
    \label{fig:sys_resp_0}
\end{figure}

\begin{figure}[ht]
    \centering
    \includegraphics[scale=0.35]{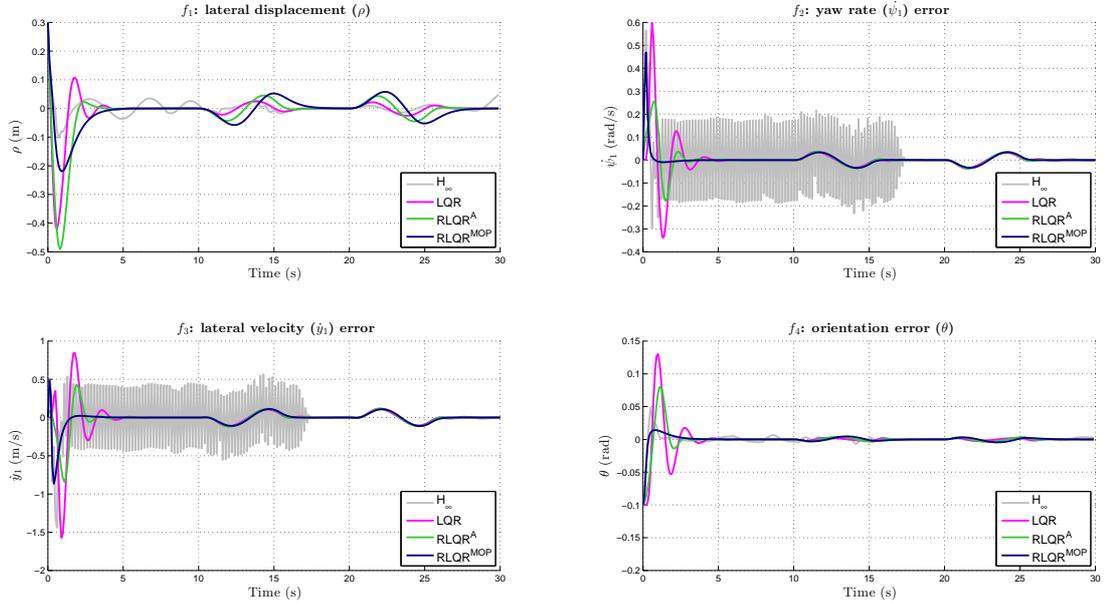}
    \caption{Controller performances for nominal payload.}
    \label{fig:sys_erro_0}
\end{figure}

\begin{figure}[t]
    \centering
    \includegraphics[scale=0.35]{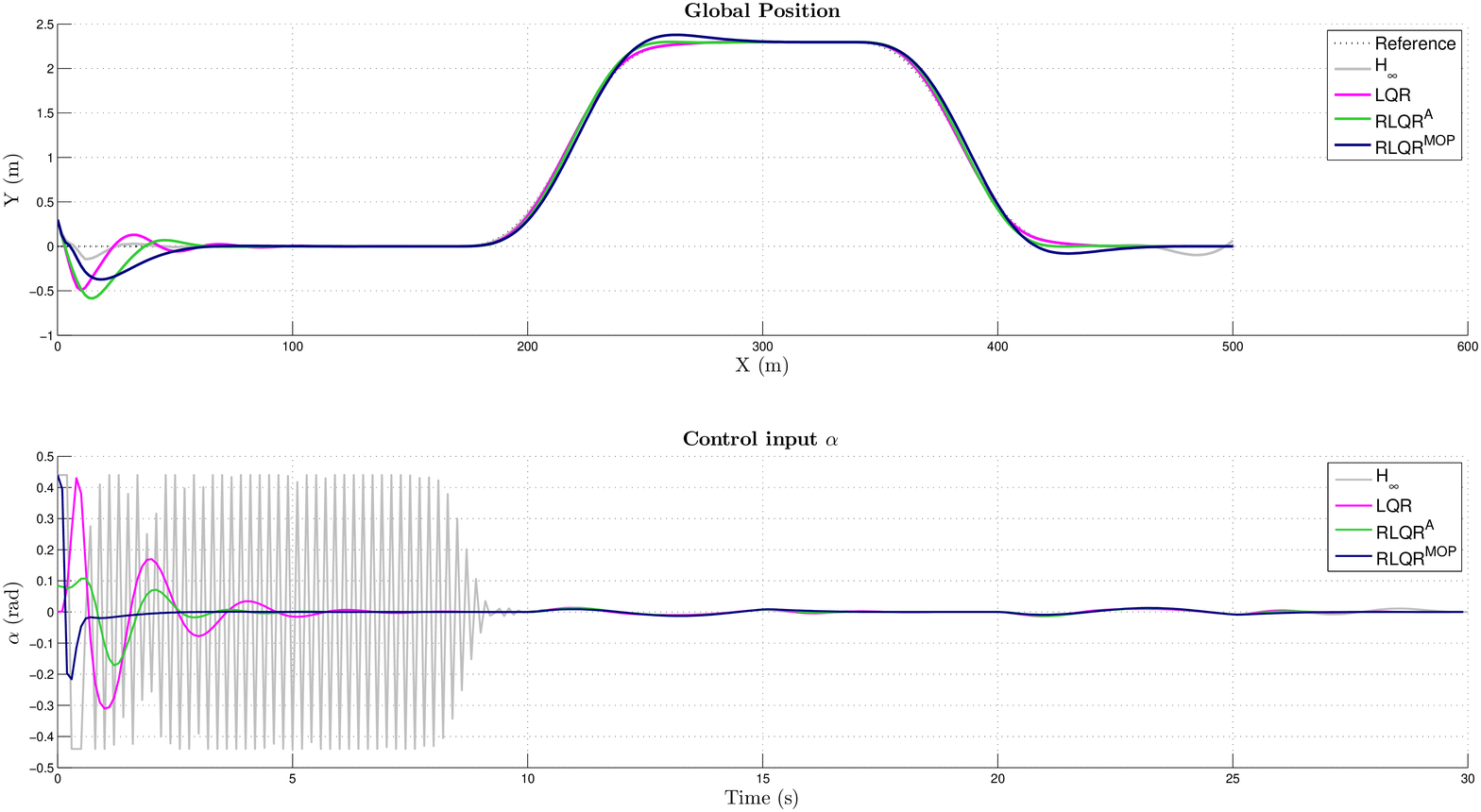}
    \caption{Global position and steering angle for $100\%$ of overload.}
    \label{fig:sys_resp_100}
\end{figure}

\begin{figure}[ht]
    \centering
    \includegraphics[scale=0.35]{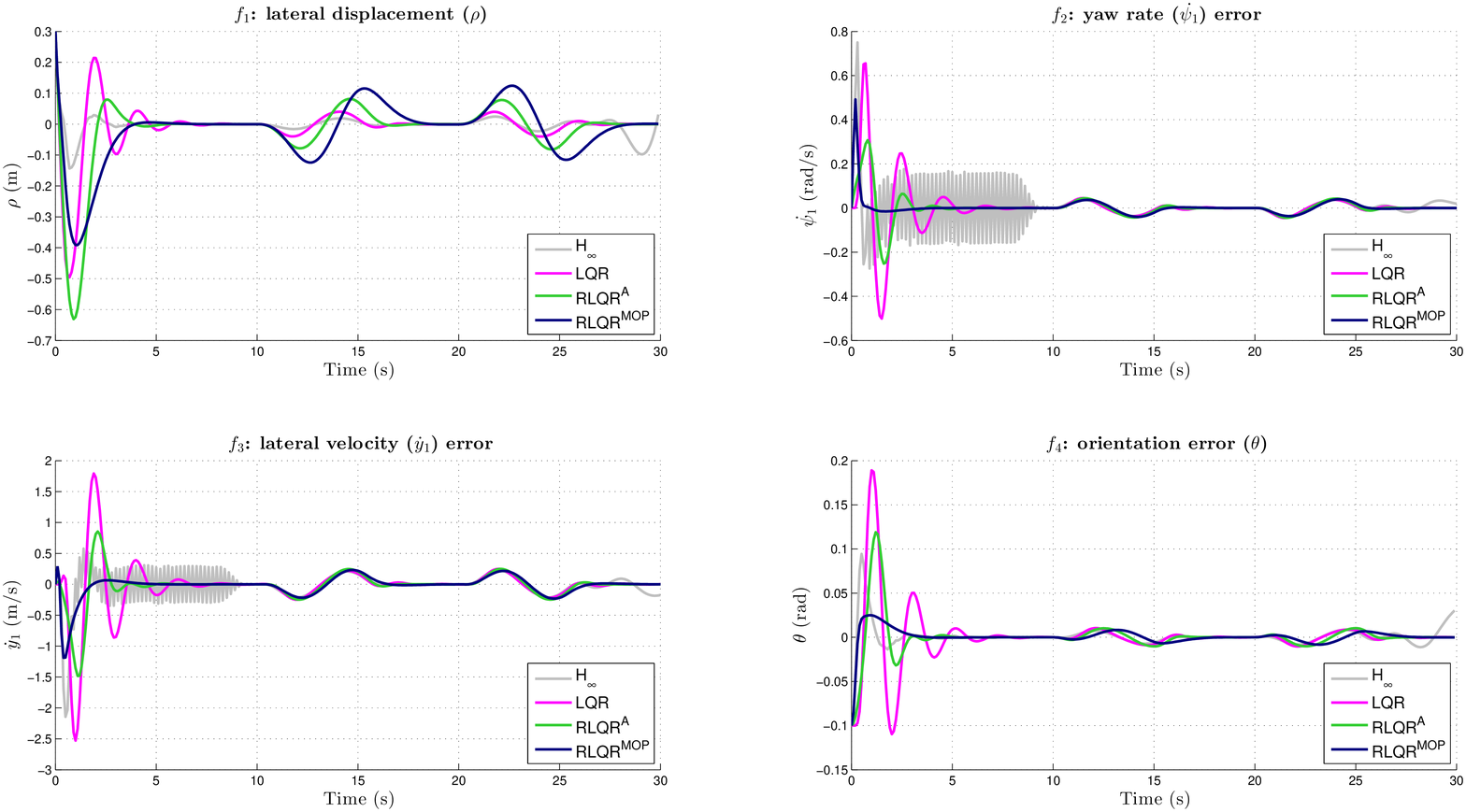}
    \caption{Controller performances for $100\%$ of overload.}
    \label{fig:sys_erro_100}
\end{figure}

\begin{figure}[t]
    \centering
    \includegraphics[scale=0.35]{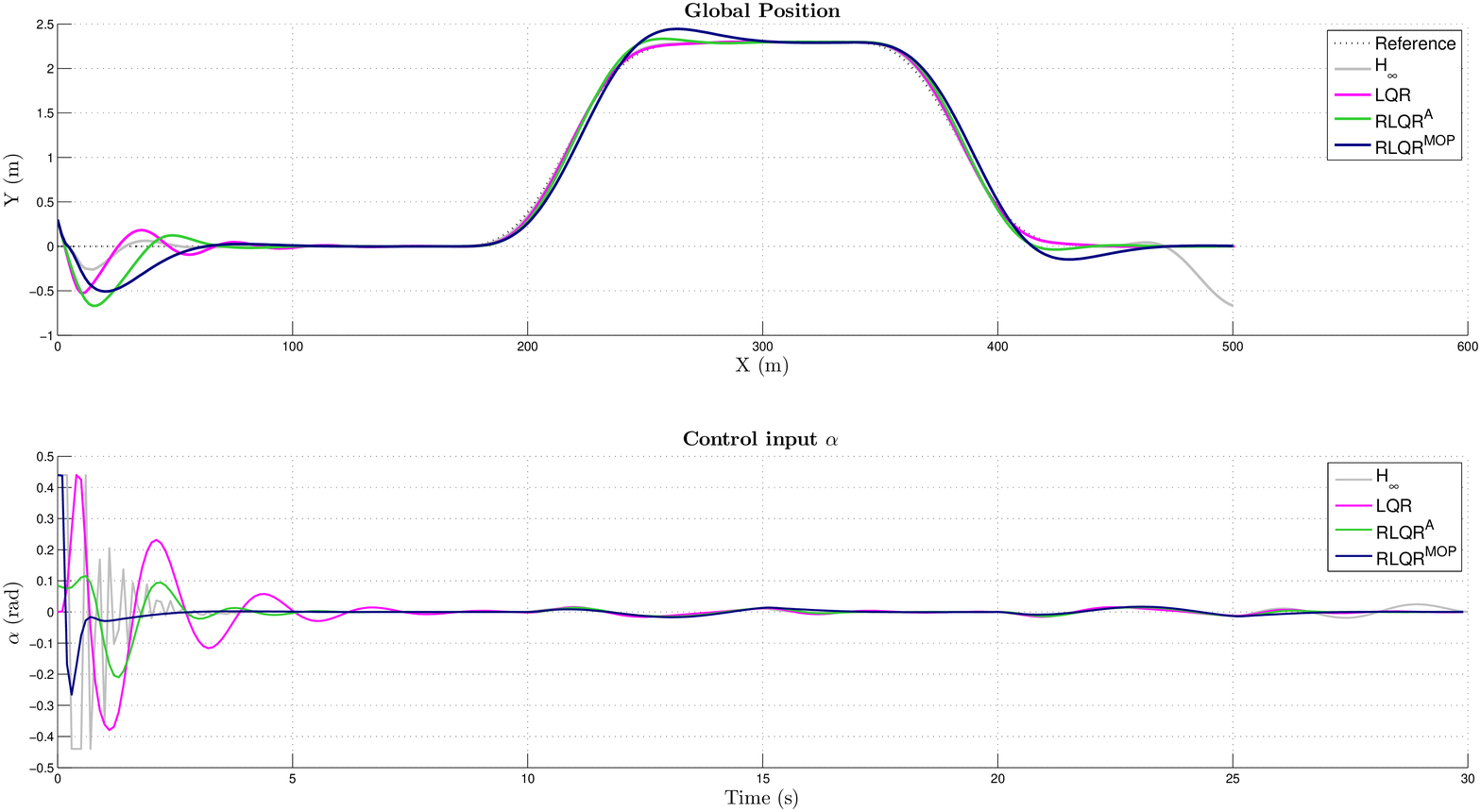}
    \caption{Global position and steering angle for $200\%$ of overload.}
    \label{fig:sys_resp_200}
\end{figure}

\begin{figure}[ht]
    \centering
    \includegraphics[scale=0.35]{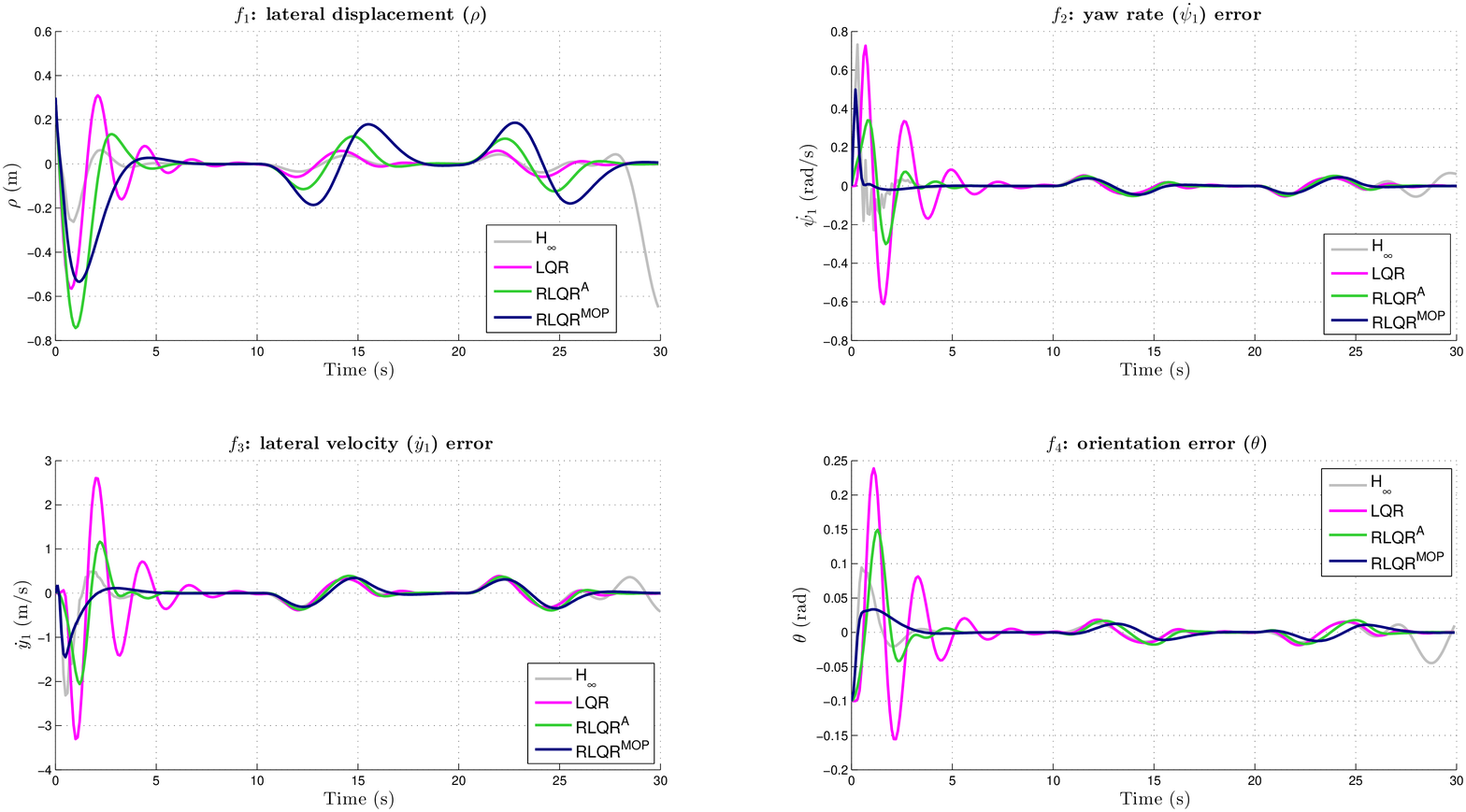}
    \caption{Controller performances for $200\%$ of overload.}
    \label{fig:sys_erro_200}
\end{figure}

\begin{figure}[t]
    \centering
    \includegraphics[scale=0.35]{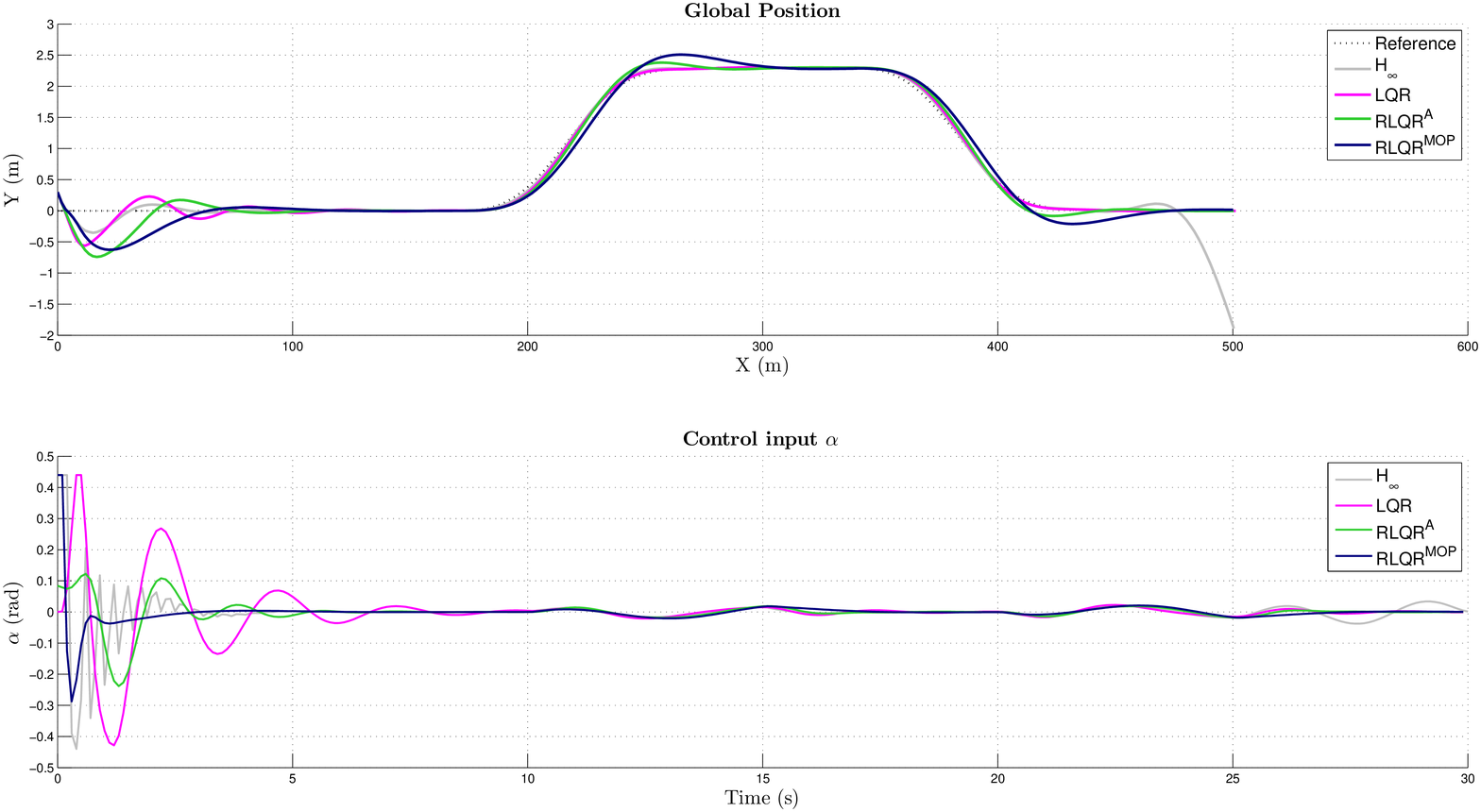}
    \caption{Global position and steering angle for $300\%$ of overload.}
    \label{fig:sys_resp_300}
\end{figure}

\begin{figure}[ht]
    \centering
    \includegraphics[scale=0.35]{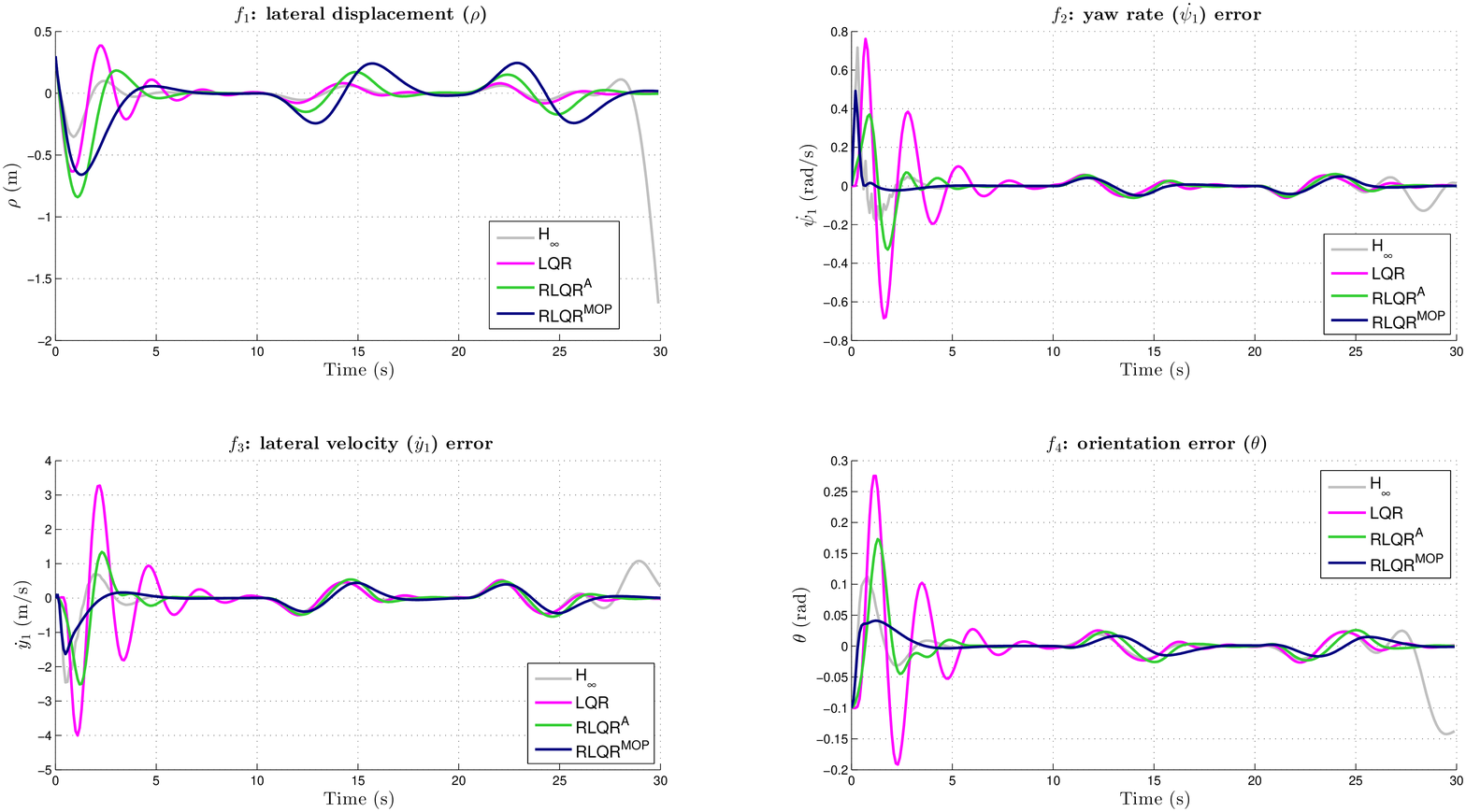}
    \caption{Controller performances for $300\%$ of overload.}
    \label{fig:sys_erro_300}
\end{figure}

The lateral displacement ($f_{1}$), the regulation error of yaw rate ($f_{2}$), the regulation error of lateral velocity ($f_{3}$) and the regulation error of vehicle orientation ($f_{4}$) were evaluated in four different vehicle payload conditions:
\begin{itemize}
    \item The first case considers the nominal payload of the vehicle.   The performances of the controllers related to  global positions and steering angles are shown  in \autoref{fig:sys_resp_0}, and those related to the objective functions are shown in \autoref{fig:sys_erro_0}. 
    \item The second case takes into account twice ($100\%$ overload) the nominal payload of the vehicle. The performances of the controllers can be seen in \autoref{fig:sys_resp_100}  and \autoref{fig:sys_erro_100}. 
    \item The third case evaluates three times ($200\%$ overload) the nominal payload of the vehicle. The results  provided by the controllers are shown in \autoref{fig:sys_resp_200}  and \autoref{fig:sys_erro_200}. 
    \item Finally, the last case considers four times ($300\%$ overload)  the nominal payload of the vehicle. The performances of the controllers are detailed in \autoref{fig:sys_resp_300} and \autoref{fig:sys_erro_300}. 
\end{itemize}
\noindent Furthermore, the Mean Square Error (MSE) values of the $4$ performance metrics ($f_1$, $f_2$, $f_3$ and $f_4$), obtained by the controllers in the path-following simulations, are displayed in \autoref{tab:mse}. 

%--
%--
%--
%--
%--
%--
%--
%--
%--
%--
%--
%--
%--
%--
%--
%--

% -------------------------------------------------------
\begin{table}[!thb]
\centering
\caption{MSE values for RLQR$^{MOP}$, RLQR$^{A}$, $\mathcal{H}_\infty$ and LQR controllers in the four objective problems: $f_{1}$ lateral displacement ($\rho$); $f_{2}$ yaw rate ($\dot{\psi_{1}}$) error; $f_{3}$ lateral velocity ($\dot{y}_{1}$); and $f_{4}$ orientation error ($\theta$).} 
\label{tab:mse}
\scalebox{0.82}{%
\begin{tabular}{ c  c  c  c  c  c }
\hline
Objective function & Controller & \multicolumn{4}{ c }{Overload}  
\\ 
\cline{3-6} & &
0\% & 100\% & 200\% & 300\% 
\\ \hline
%----------------------------------------------------------%
\linebreak \\ 
\multirow{4}{*}{$f_{1}$} & 
{RLQR$^{MOP}$}
& 2.72161E-3
& 9.52776E-3
& 2.02697E-2
& 3.46455E-2
\\
%-----------------%
&
{RLQR$^{A}$}
& 6.72252E-3
& 1.29370E-2
& 2.06950E-2
& 3.00461E-2
\\
%-----------------%
&
{$\mathcal{H}_\infty$}
& \textbf{0.83135E-3}
& \textbf{1.08787E-3}
& \textbf{1.09022E-2}
& 4.19582E-2
\\
%-----------------%
&
{LQR}
& 3.92518E-3
& 6.95036E-3
& 1.10944E-2
& \textbf{1.61736E-2}
\\
\hline
\linebreak \\ 
\multirow{4}{*}{$f_{2}$} & 
{RLQR$^{MOP}$}
& \textbf{1.44405E-3}
& \textbf{1.67467E-3}
& \textbf{1.85903E-3}
& \textbf{1.99537E-3}
\\
%-----------------%
&
{RLQR$^{A}$}
& 1.84585E-3
& 3.01635E-3
& 4.09464E-3
& 4.98721E-3
\\
%-----------------%
&
{$\mathcal{H}_\infty$}
& 1.85683E-2
& 1.03793E-2
& 4.64690E-3
& 5.18353E-3
\\
%-----------------%
&
{LQR}
& 5.54885E-3
& 1.04151E-2
& 1.56832E-2
& 2.02004E-2
\\
\hline
\linebreak \\ 
\multirow{4}{*}{$f_{3}$} & 
{RLQR$^{MOP}$}
& \textbf{1.17762E-2}
& \textbf{3.07187E-2}
& \textbf{5.65872E-2}
& \textbf{8.72447E-2}
\\
%-----------------%
&
{RLQR$^{A}$}
& 1.67254E-2
& 6.45166E-2
& 1.34078E-1
& 2.15567E-1
\\
%-----------------%
&
{$\mathcal{H}_\infty$}
& 1.09009E-1
& 7.74970E-2
& 9.98572E-2
& 1.95922E-1
\\
%-----------------%
&
{LQR}
& 4.54013E-2
& 1.71043E-1
& 3.76471E-1
& 6.23971E-1
\\
\hline
\linebreak \\ 
\multirow{4}{*}{$f_{4}$} & 
{RLQR$^{MOP}$}
& \textbf{7.35717E-5} 
& \textbf{1.00885E-4}
& \textbf{1.44767E-4}
& \textbf{2.02602E-4}
\\
%-----------------%
&
{RLQR$^{A}$}
& 2.42387E-4
& 4.22745E-4
& 6.48997E-4
& 8.95945E-4
\\
%-----------------%
&
{$\mathcal{H}_\infty$}
& 8.74106E-5
& 1.94354E-4
& 3.50360E-4
& 1.29590E-3
\\
%-----------------%
&
{LQR}
& 4.59179E-4
& 1.02236E-3
& 1.82643E-3
& 2.72205E-3
\\
\hline
\end{tabular}
}
\end{table}
% -------------------------------------------------------

The first case (nominal payload of the vehicle) is presented in \autoref{fig:sys_resp_0}, where the $\mathcal{H}_\infty$ controller yielded the  best performance in terms of maximum deviation from the reference path. Nevertheless, it presented the worst performance for the control action, with an oscillatory behavior and higher energy expenditure. Also, the control signal of the $\mathcal{H}_\infty$ controller saturated at $-0.45$ radians, which is harmful for actual applications. The RLQR$^{A}$ and LQR controllers also yielded some level of  oscillation, mainly until the $5$ seconds of operation. The RLQR$^{MOP}$ outperformed the other controllers in terms of control signals, it expended less energy to efficiently stabilize the system. Details of the controller performances can be seen  in \autoref{fig:sys_erro_0} and \autoref{tab:mse}, respectively, for four objective functions. The $\mathcal{H}_\infty$ controller yielded the smallest error with respect to $f_1$, followed by RLQR$^{MOP}$, LQR and RLQR$^{A}$ controllers. Considering $f_2$ and $f_3$, the RLQR$^{MOP}$ presented the best performance, followed by the RLQR$^{A}$, LQR and $\mathcal{H}_\infty$ controllers. The RLQR$^{MOP}$ outperformed the other controllers regarding $f_4$, followed by the $\mathcal{H}_\infty$, RLQR$^{A}$ and LQR controllers. 

The second case ($100\%$ overload) is presented in \autoref{fig:sys_resp_100}. In this case the $\mathcal{H}_\infty$ yielded less deviation from the reference path. However, it presented an oscillatory control input with the highest energy expenditure, which saturated at $-0.45$ radians. Also, the RLQR$^{A}$ and LQR presented oscillatory behaviors around the reference, while the RLQR$^{MOP}$ yielded a control signal with less energy expenditure. The objective functions were detailed in \autoref{fig:sys_erro_100} and \autoref{tab:mse}, where the $\mathcal{H}_\infty$ yielded the best performance in $f_1$, followed by LQR, RLQR$^{MOP}$ and RLQR$^{A}$ controllers. Regarding $f_2$ and $f_3$, the best results are presented by the RLQR$^{MOP}$, followed by the RLQR$^{A}$, $\mathcal{H}_\infty$ and LQR controllers. Regarding $f_4$, the RLQR$^{MOP}$ yielded the best performance followed by the $\mathcal{H}_\infty$, RLQR$^{A}$ and LQR controllers.  

The third case ($200\%$ overload) is presented in \autoref{fig:sys_resp_200}. It shows  that the $\mathcal{H}_\infty$ controller presented the best performance in terms of deviation from the reference path in the first $100$ seconds of the simulation. However, it lost the ability to follow the reference  after $400$ seconds, approximately. In addition, the $\mathcal{H}_\infty$ controller yielded the worst oscillatory control actions, with saturation at $-0.45$ radians until $1$ second of the simulation. The RLQR$^{A}$ and LQR also yielded equivalent levels of oscillatory behaviors for path-following and control signals.The RLQR$^{MOP}$ outperformed the other controllers regarding reference following, and presented control action with less energy expenditure, considering $f_2$, $f_3$ and $f_4$ objective functions. A detailed explanation is displayed in \autoref{fig:sys_erro_200} and \autoref{tab:mse}. Even though $\mathcal{H}_\infty$ controller  presented an unstable behavior at the end of the simulation, it yielded the best performance in $f_1$, followed by LQR, RLQR$^{MOP}$ and RLQR$^{A}$ controllers. 

The fourth case ($300\%$ overload) is presented in \autoref{fig:sys_resp_300}. The $\mathcal{H}_\infty$ lost stability at the end of the simulation and yielded the worst control action in terms of oscillation. The objective functions for this case are presented in \autoref{fig:sys_erro_300} and \autoref{tab:mse}. The LQR yielded the best performance in $f_1$, followed by the RLQR$^{A}$, RLQR$^{MOP}$ and $\mathcal{H}_\infty$. The RLQR$^{MOP}$ outperformed the other controllers for the metrics $f_2$, $f_3$ and $f_4$.

% SECTION 5
\section{Conclusions}
\label{sec:5}

This work has proposed a self-adaptative robust recursive procedure via multiobjective optimization for lateral control applied to autonomous heavy vehicles. For this method, the system's dynamic is defined as discrete-time MOP and the uncertainty matrices are adjusted in order to enhance the control performance, based on the measured input variables and path-following errors, regardless of any system’s physical meaning about the uncertainty matrices. In addition, this paper has proposed the MO-LSP algorithm as a new local search procedure to enhance the performance of MOEAs. 

The experiments were carried out in a generic and an applied robust control models, the former being with five uncertainty parameters variables and three objective functions, and the latter with seven uncertainty parameters variables and four objective functions. The optimization method aimed to decrease the errors associated with lateral displacement, yaw rate regulation, lateral velocity and the regulation of vehicle orientation. For both the generic and the applied models discussed in \autoref{sec:4.4} and \autoref{sec:4.5}, respectively,  the combination of the MO-LSP algorithm with a MOEA is compared with $10$ state-of-art MOEAs for the control designs in $5$ performance metrics.

As shown in \autoref{tab:result_1} and discussed in \autoref {sec:4.4}, the generic model optimized using the MO-LSP obtained: (\textit{i.}) $9$ better results, $8$ being statistically different, in IGD metric; (\textit{ii.})  $10$ better results, $9$ being statistically different, in $SP$ metric; (\textit{iii.}) $7$ statistically better results in HV metric; (\textit{iv.}) better solutions in all the cases in PD metric, $9$ being statistically different; (\textit{v.}) and, finally, $7$ better results, with $1$ statistically different solution, in Spread metric. Furthermore, \autoref{tab:result_2} details the applied control model performance discussed in \autoref{sec:4.5}, where the MO-LSP provided: (\textit{i.}) better results for all the cases in IGD metric, $9$ being statistically different; (\textit{ii.})  better results for all the cases in $SP$ metric, $5$ being statistically different; (\textit{iii.}) $9$ better results, with $1$ result statistically different, in HV metric; (\textit{iv.}) all the cases with statistically better results in PD metric; (\textit{v.}) and $2$ better cases not statistically different in Spread metric. Though the proposed local search algorithm affects the spread of the individuals for the applied control, it significantly improves the performance in all the other metrics for the generic and applied control model.

The RLQR controller provides better performance for the vehicle's lateral dynamic regulator, improving robustness, stability, smoothness and safety \citep{barbosa2019robust}. By applying robust control optimization as a MOP in RLQR design, a performance improvement was noticed when compared to the standard algebraic manipulations of the model presented in (\ref{eq:ssdynamic}). Comparing to the original breakthrough RLQR, the standard LQR and $\mathcal{H}_\infty$ controllers, the proposed approach had better performance in $3$ out of $4$ independent measures of performance.

The main advantages of the proposed approach include the fact that (\textit{i.}) the MO-LSP  can be added for any MOEA already established in literature with few changes in the main code; (\textit{ii.}) the proposed algorithm is useful for large size problems and applications; (\textit{iii.}) a control method defined by MOP is less dependent on mathematical formulation for the parameters design; (\textit{iv.}) with a multiobjective formulation, different objective functions can be easily added to the problem; and (\textit{v.}) a control designer can choose a parameter configuration that best optimizes a specific objective. 

On the other hand, the presented method has some limitations: (\textit{i.}) The MO-LSP uses the individuals presented in the PF to define a local search area and perform the mutation procedure presented in (\ref{eq:bestM1}) and (\ref{eq:bestM}). Since the size of the PF can be equal or similar to the size of the total population, the number of executed mutation operations can be equal on both processes (MOEA main search and MO-LSP local search). Thus, the MO-LSP can significantly increase the processing time in cases with large PF. This can result in poor performance, as a local search method is expected to have minimal influence on code processing time and maximum influence on the resulting performance. (\textit{ii.})  The classical mathematical formulation of a robust control design needs a defined number of encoding operations to return a valid result. However, since MO-LSP is a search method operation, the training time to return the best control setting is usually longer than the previous mathematical formulation. (\textit{iii.})  As with other machine learning methods, the MO-LSP cannot guarantee an optimal configuration in just one training process. Therefore, several training procedures must be performed to define the best set of variables.

To overcome these limitations, an optimization method including parallel computing will be explored to decrease training time. For future work, the MO-LSP will be applied to a real heavy vehicle, where the local search method will also define the restriction conditions presented in the control plant together with parameter optimizations. In addition, we intend to apply  the MO-LSP in planning optimization, motion coordination and real-time state and parameter estimation of autonomous driving systems.

\section*{Acknowledgements}
This research was financed in part by the Brazilian National Council for Scientific and Technological Development - CNPq (grants 465755/2014-3 and 304201/2018-9), by the Coordination of Improvement of Higher Education Personnel - Brazil - CAPES (Finance Code 001 and 88887.136349/2017-00), the S\~{a}o Paulo Research Foundation - FAPESP (grant 2014/50851-0) and the Minas Gerais Research Foundation - FAPEMIG (grant PPM 00337/17). 

\section*{Conflict of interest}

None declared.

\appendix
 
\section{Uncertainties matrices}
\label{sec:uncertainties}

To estimate the uncertainties matrices $E_{F}$, $E_{G}$ and $H$ in (\ref{eq:RLQRuncertainties}), consider the inertia uncertainties related to maximum and minimum values of payload, which result, respectively, in  $m_{p_{max}}$ and $m_{p_{min}}$. Once these values of mass are determined, the maximum variations of the matrices $F$ and $G$ are calculated according to:
\begin{equation}
\Gamma_{F} = F_{m_{p_{min}}}-F_{m_{p_{max}}},
\end{equation}
\begin{equation}
\Gamma_{G} = G_{m_{p_{min}}}-G_{m_{p_{max}}},
\end{equation}
where $F_{m_{p_{min}}}$, $G_{m_{p_{min}}}$, $F_{m_{p_{max}}}$ and $G_{m_{p_{max}}}$ are the discretized state-space matrices in (\ref{eq:ssdynamic}) corresponding to $m_{p_{max}}$ and $m_{p_{min}}$. Next, select the row in $\Gamma_{F}$ that is most affected by mass variations (in this example, it is the first row).

Notice that $m_{p_{min}}$ and $m_{p_{min}}$ values correspond to the unloaded and $100\%$ of overload vehicle operation. Therefore, this range of uncertainty may cause large $E_F$ and $E_G$ values, denoted as $E_{F_{100\%}}$ and $E_{G_{100\%}}$. By choosing these values during the control design, robustness and stability are ensured within such range of mass variability. On the other hand, this decreases the performance of the nominal case. Therefore, lower $E_{F}$ and $E_{G}$ values were taken into account to overcome this problem. This method is capable of improving the robustness of the proposed scheme without jeopardizing the system performance during nominal operation.

Thus, the matrices $E_{F_i}$ and $E_{G_i}$ are calculated as:

%--
\begin{equation}
E_{F_{i}} = 
\begin{bmatrix}
1\\
1\\
1\\
0.1\\
\end{bmatrix}^T
\begin{bmatrix}
\underset{\Gamma_{F_{1,1}}}{arg}(|\Gamma_{F_{1,1}}|) & 0 & 0 & 0 \\ 0 & \underset{\Gamma_{F_{1,2}}}{arg}(|\Gamma_{F_{1,2}}|) & 0 & 0\\ 0 & 0 & \underset{\Gamma_{F_{1,3}}}{arg}(|\Gamma_{F_{1,3}}|) & 0 \\ 0 & 0 & 0 & \underset{\Gamma_{F_{1,4}}}{arg}(|\Gamma_{F_{1,4}}|)
\end{bmatrix}
\end{equation}
%--
\begin{equation}
\label{Eg_calc}
E_{G_{i}} = 
\begin{bmatrix}
0.1 \\ 
0.1
\end{bmatrix}^T
\begin{bmatrix}
\underset{\Gamma_{F_{1,j}}}{arg} \{|\Gamma_{G_{1,j}}|\} & 0\\
0 & \underset{\Gamma_{F_{1,j}}}{arg} \{|\Gamma_{G_{1,j}}|\}
\end{bmatrix}
\end{equation}
%--
and $H_{i} = [1,1,1,1]^T$. Then, the uncertainties matrices are obtained through (\ref{eq:RLQRuncertainties}) as
%--
\begin{equation*}
\begin{bmatrix}
\delta {F}_{i} & \delta G_{i}
\end{bmatrix} = H_{i} \Delta_{i} \begin{bmatrix}
E_{F_{i}} & E_{G_{i}}
\end{bmatrix},
\end{equation*}
%--
where $\Delta_{i}$ is a scalar represented by the mass variation.

% BIBLIOGRAPHY
\bibliographystyle{model5-names}\biboptions{authoryear}
\bibliography{main.bib}

\begin{thebibliography}{67}
\expandafter\ifx\csname natexlab\endcsname\relax\def\natexlab#1{#1}\fi
\providecommand{\url}[1]{\texttt{#1}}
\providecommand{\href}[2]{#2}
\providecommand{\path}[1]{#1}
\providecommand{\DOIprefix}{doi:}
\providecommand{\ArXivprefix}{arXiv:}
\providecommand{\URLprefix}{URL: }
\providecommand{\Pubmedprefix}{pmid:}
\providecommand{\doi}[1]{\href{http://dx.doi.org/#1}{\path{#1}}}
\providecommand{\Pubmed}[1]{\href{pmid:#1}{\path{#1}}}
\providecommand{\bibinfo}[2]{#2}
\ifx\xfnm\relax \def\xfnm[#1]{\unskip,\space#1}\fi
%Type = Article
\bibitem[{Aguirre et~al.(2017)Aguirre, Teixeira, Barbosa, Teixeira, Campos \&
  Mendes}]{aguirre2017cep}
\bibinfo{author}{Aguirre, L.~A.}, \bibinfo{author}{Teixeira, B. O.~S.},
  \bibinfo{author}{Barbosa, B. H.~G.}, \bibinfo{author}{Teixeira, A.~F.},
  \bibinfo{author}{Campos, M. C. M.~M.}, \& \bibinfo{author}{Mendes, E. M.
  A.~M.} (\bibinfo{year}{2017}).
\newblock \bibinfo{title}{Development of soft sensors for permanent downhole
  gauges in deepwater oil wells}.
\newblock {\it \bibinfo{journal}{Control Engineering Practice}\/},  {\it
  \bibinfo{volume}{65}\/}, \bibinfo{pages}{83 -- 99}. \URLprefix
  \url{https://www.sciencedirect.com/science/article/abs/pii/S0967066117301284}.
  \DOIprefix\doi{https://doi.org/10.1016/j.conengprac.2017.06.002}.
%Type = Article
\bibitem[{Alam et~al.(2015)Alam, Besselink, Turri, Martensson \&
  Johansson}]{alam2015heavy}
\bibinfo{author}{Alam, A.}, \bibinfo{author}{Besselink, B.},
  \bibinfo{author}{Turri, V.}, \bibinfo{author}{Martensson, J.}, \&
  \bibinfo{author}{Johansson, K.~H.} (\bibinfo{year}{2015}).
\newblock \bibinfo{title}{Heavy-duty vehicle platooning for sustainable freight
  transportation: A cooperative method to enhance safety and efficiency}.
\newblock {\it \bibinfo{journal}{IEEE Control Systems Magazine}\/},  {\it
  \bibinfo{volume}{35}\/}, \bibinfo{pages}{34--56}. \URLprefix
  \url{https://ieeexplore.ieee.org/abstract/document/7286902}.
  \DOIprefix\doi{10.1109/MCS.2015.2471046}.
%Type = Article
\bibitem[{Alatas et~al.(2008)Alatas, Akin \& Karci}]{alatas2008modenar}
\bibinfo{author}{Alatas, B.}, \bibinfo{author}{Akin, E.}, \&
  \bibinfo{author}{Karci, A.} (\bibinfo{year}{2008}).
\newblock \bibinfo{title}{Modenar: Multi-objective differential evolution
  algorithm for mining numeric association rules}.
\newblock {\it \bibinfo{journal}{Applied Soft Computing}\/},  {\it
  \bibinfo{volume}{8}\/}, \bibinfo{pages}{646--656}. \URLprefix
  \url{https://www.sciencedirect.com/science/article/abs/pii/S156849460700049X}.
  \DOIprefix\doi{https://doi.org/10.1016/j.asoc.2007.05.003}.
%Type = Article
\bibitem[{Arabmaldar et~al.(2021)Arabmaldar, Mensah \&
  Toloo}]{arabmaldar2021robust}
\bibinfo{author}{Arabmaldar, A.}, \bibinfo{author}{Mensah, E.~K.}, \&
  \bibinfo{author}{Toloo, M.} (\bibinfo{year}{2021}).
\newblock \bibinfo{title}{Robust worst-practice interval dea with
  non-discretionary factors}.
\newblock {\it \bibinfo{journal}{Expert Systems with Applications}\/},  (p.
  \bibinfo{pages}{115256}). \URLprefix
  \url{https://www.sciencedirect.com/science/article/abs/pii/S0957417421006886}.
  \DOIprefix\doi{https://doi.org/10.1016/j.eswa.2021.115256}.
%Type = Article
\bibitem[{Audet et~al.(2018)Audet, Bigeon, Cartier, Le~Digabel \&
  Salomon}]{audet2018performance}
\bibinfo{author}{Audet, C.}, \bibinfo{author}{Bigeon, J.},
  \bibinfo{author}{Cartier, D.}, \bibinfo{author}{Le~Digabel, S.}, \&
  \bibinfo{author}{Salomon, L.} (\bibinfo{year}{2018}).
\newblock \bibinfo{title}{Performance indicators in multiobjective
  optimization}.
\newblock {\it \bibinfo{journal}{Optimization Online}\/}, . \URLprefix
  \url{http://www.optimization-online.org/DB_HTML/2018/10/6887.html}.
%Type = Article
\bibitem[{{Barbosa} et~al.(2019){Barbosa}, {Aguirre} \&
  {Braga}}]{barbosaiet2019}
\bibinfo{author}{{Barbosa}, B. H.~G.}, \bibinfo{author}{{Aguirre}, L.~A.}, \&
  \bibinfo{author}{{Braga}, A.~P.} (\bibinfo{year}{2019}).
\newblock \bibinfo{title}{Piecewise affine identification of a hydraulic
  pumping system using evolutionary computation}.
\newblock {\it \bibinfo{journal}{IET Control Theory Applications}\/},  {\it
  \bibinfo{volume}{13}\/}, \bibinfo{pages}{1394--1403}.
%Type = Article
\bibitem[{Barbosa et~al.(2011)Barbosa, Aguirre, Martinez \&
  Braga}]{barbosa2011}
\bibinfo{author}{Barbosa, B. H.~G.}, \bibinfo{author}{Aguirre, L.~A.},
  \bibinfo{author}{Martinez, C.~B.}, \& \bibinfo{author}{Braga, A.~P.}
  (\bibinfo{year}{2011}).
\newblock \bibinfo{title}{Black and gray-box identification of a hydraulic
  pumping system}.
\newblock {\it \bibinfo{journal}{Control Systems Technology, IEEE Transactions
  on}\/},  {\it \bibinfo{volume}{19}\/}, \bibinfo{pages}{398 --406}. \URLprefix
  \url{https://ieeexplore.ieee.org/document/5422811}.
  \DOIprefix\doi{10.1109/TCST.2010.2042600}.
%Type = Article
\bibitem[{Barbosa et~al.(2019)Barbosa, Marcos, da~Silva, Terra \&
  Junior}]{barbosa2019robust}
\bibinfo{author}{Barbosa, F.~M.}, \bibinfo{author}{Marcos, L.~B.},
  \bibinfo{author}{da~Silva, M.~M.}, \bibinfo{author}{Terra, M.~H.}, \&
  \bibinfo{author}{Junior, V.~G.} (\bibinfo{year}{2019}).
\newblock \bibinfo{title}{Robust path-following control for articulated
  heavy-duty vehicles}.
\newblock {\it \bibinfo{journal}{Control Engineering Practice}\/},  {\it
  \bibinfo{volume}{85}\/}, \bibinfo{pages}{246--256}. \URLprefix
  \url{https://www.sciencedirect.com/science/article/abs/pii/S0967066118303216}.
  \DOIprefix\doi{https://doi.org/10.1016/j.conengprac.2019.01.017}.
%Type = Book
\bibitem[{Bertsekas et~al.(2000)}]{bertsekas2000dynamic}
\bibinfo{author}{Bertsekas, D.~P.} et~al. (\bibinfo{year}{2000}).
\newblock {\it \bibinfo{title}{Dynamic programming and optimal control: Vol.
  1}\/}.
\newblock \bibinfo{publisher}{Athena scientific Belmont}.
%Type = Inproceedings
\bibitem[{Cerri et~al.(2009)Cerri, Terra \& Ishihara}]{Cerri2009}
\bibinfo{author}{Cerri, J.~P.}, \bibinfo{author}{Terra, M.~H.}, \&
  \bibinfo{author}{Ishihara, J.~Y.} (\bibinfo{year}{2009}).
\newblock \bibinfo{title}{Recursive robust regulator for discrete-time
  state-space systems}.
\newblock In {\it \bibinfo{booktitle}{2009 American Control Conference}\/}.
\newblock \bibinfo{publisher}{{IEEE}}.
\newblock \URLprefix \url{https://doi.org/10.1109/acc.2009.5160553}.
  \DOIprefix\doi{10.1109/acc.2009.5160553}.
%Type = Article
\bibitem[{Cheng et~al.(2015)Cheng, Yen \& Zhang}]{cheng2015many}
\bibinfo{author}{Cheng, J.}, \bibinfo{author}{Yen, G.~G.}, \&
  \bibinfo{author}{Zhang, G.} (\bibinfo{year}{2015}).
\newblock \bibinfo{title}{A many-objective evolutionary algorithm with enhanced
  mating and environmental selections}.
\newblock {\it \bibinfo{journal}{IEEE Transactions on Evolutionary
  Computation}\/},  {\it \bibinfo{volume}{19}\/}, \bibinfo{pages}{592--605}.
  \URLprefix \url{https://ieeexplore.ieee.org/document/7090975}.
  \DOIprefix\doi{10.1109/TEVC.2015.2424921}.
%Type = Article
\bibitem[{Coello \& Cort{\'e}s(2005)}]{coello2005solving}
\bibinfo{author}{Coello, C. A.~C.}, \& \bibinfo{author}{Cort{\'e}s, N.~C.}
  (\bibinfo{year}{2005}).
\newblock \bibinfo{title}{Solving multiobjective optimization problems using an
  artificial immune system}.
\newblock {\it \bibinfo{journal}{Genetic Programming and Evolvable
  Machines}\/},  {\it \bibinfo{volume}{6}\/}, \bibinfo{pages}{163--190}.
  \URLprefix \url{https://link.springer.com/article/10.1007/s10710-005-6164-x}.
  \DOIprefix\doi{https://doi.org/10.1007/s10710-005-6164-x}.
%Type = Article
\bibitem[{Deb \& Jain(2013)}]{deb2013evolutionary}
\bibinfo{author}{Deb, K.}, \& \bibinfo{author}{Jain, H.}
  (\bibinfo{year}{2013}).
\newblock \bibinfo{title}{An evolutionary many-objective optimization algorithm
  using reference-point-based nondominated sorting approach, part i: solving
  problems with box constraints}.
\newblock {\it \bibinfo{journal}{IEEE transactions on evolutionary
  computation}\/},  {\it \bibinfo{volume}{18}\/}, \bibinfo{pages}{577--601}.
  \URLprefix \url{https://ieeexplore.ieee.org/document/6600851}.
  \DOIprefix\doi{10.1109/TEVC.2013.2281535}.
%Type = Article
\bibitem[{Deb et~al.(2002)Deb, Pratap, Agarwal \& Meyarivan}]{deb2002fast}
\bibinfo{author}{Deb, K.}, \bibinfo{author}{Pratap, A.},
  \bibinfo{author}{Agarwal, S.}, \& \bibinfo{author}{Meyarivan, T.}
  (\bibinfo{year}{2002}).
\newblock \bibinfo{title}{A fast and elitist multiobjective genetic algorithm:
  Nsga-ii}.
\newblock {\it \bibinfo{journal}{IEEE transactions on evolutionary
  computation}\/},  {\it \bibinfo{volume}{6}\/}, \bibinfo{pages}{182--197}.
  \URLprefix \url{https://ieeexplore.ieee.org/document/996017}.
  \DOIprefix\doi{10.1109/4235.996017}.
%Type = Article
\bibitem[{Decerle et~al.(2019)Decerle, Grunder, El~Hassani \&
  Barakat}]{decerle2019memetic}
\bibinfo{author}{Decerle, J.}, \bibinfo{author}{Grunder, O.},
  \bibinfo{author}{El~Hassani, A.~H.}, \& \bibinfo{author}{Barakat, O.}
  (\bibinfo{year}{2019}).
\newblock \bibinfo{title}{A memetic algorithm for multi-objective optimization
  of the home health care problem}.
\newblock {\it \bibinfo{journal}{Swarm and evolutionary computation}\/},  {\it
  \bibinfo{volume}{44}\/}, \bibinfo{pages}{712--727}. \URLprefix
  \url{https://www.sciencedirect.com/science/article/abs/pii/S2210650218300518}.
  \DOIprefix\doi{https://doi.org/10.1016/j.swevo.2018.08.014}.
%Type = Article
\bibitem[{Delgado et~al.(2008)Delgado, Cuellar \&
  Pegalajar}]{delgado2008multiobjective}
\bibinfo{author}{Delgado, M.}, \bibinfo{author}{Cuellar, M.~P.}, \&
  \bibinfo{author}{Pegalajar, M.~C.} (\bibinfo{year}{2008}).
\newblock \bibinfo{title}{Multiobjective hybrid optimization and training of
  recurrent neural networks}.
\newblock {\it \bibinfo{journal}{IEEE Transactions on Systems, Man, and
  Cybernetics, Part B (Cybernetics)}\/},  {\it \bibinfo{volume}{38}\/},
  \bibinfo{pages}{381--403}. \URLprefix
  \url{https://ieeexplore.ieee.org/abstract/document/4415532}.
  \DOIprefix\doi{10.1109/TSMCB.2007.912937}.
%Type = Article
\bibitem[{Garcia \& Trinh(2019)}]{garcia2019comparison}
\bibinfo{author}{Garcia, S.}, \& \bibinfo{author}{Trinh, C.~T.}
  (\bibinfo{year}{2019}).
\newblock \bibinfo{title}{Comparison of multi-objective evolutionary algorithms
  to solve the modular cell design problem for novel biocatalysis}.
\newblock {\it \bibinfo{journal}{Processes}\/},  {\it \bibinfo{volume}{7}\/},
  \bibinfo{pages}{361}. \URLprefix
  \url{https://www.biorxiv.org/content/10.1101/616078v1}.
  \DOIprefix\doi{https://doi.org/10.1101/616078}.
%Type = Inproceedings
\bibitem[{G{\'o}mez \& Coello(2013)}]{gomez2013mombi}
\bibinfo{author}{G{\'o}mez, R.~H.}, \& \bibinfo{author}{Coello, C. A.~C.}
  (\bibinfo{year}{2013}).
\newblock \bibinfo{title}{Mombi: A new metaheuristic for many-objective
  optimization based on the r2 indicator}.
\newblock In {\it \bibinfo{booktitle}{2013 IEEE Congress on Evolutionary
  Computation}\/} (pp. \bibinfo{pages}{2488--2495}).
\newblock \bibinfo{organization}{IEEE}.
\newblock \URLprefix
  \url{https://ieeexplore.ieee.org/abstract/document/6557868}.
  \DOIprefix\doi{10.1109/CEC.2013.6557868}.
%Type = Article
\bibitem[{Guedes et~al.(2016)Guedes, Ferreira \& Barbosa}]{GUEDES201665}
\bibinfo{author}{Guedes, J.~D.}, \bibinfo{author}{Ferreira, D.~D.}, \&
  \bibinfo{author}{Barbosa, B.~H.} (\bibinfo{year}{2016}).
\newblock \bibinfo{title}{A non-intrusive approach to classify electrical
  appliances based on higher-order statistics and genetic algorithm: a smart
  grid perspective}.
\newblock {\it \bibinfo{journal}{Electric Power Systems Research}\/},  {\it
  \bibinfo{volume}{140}\/}, \bibinfo{pages}{65 -- 69}. \URLprefix
  \url{http://www.sciencedirect.com/science/article/pii/S0378779616302516}.
  \DOIprefix\doi{https://doi.org/10.1016/j.epsr.2016.06.042}.
%Type = Article
\bibitem[{He et~al.(2017)He, Tian, Jin, Zhang \& Pan}]{he2017radial}
\bibinfo{author}{He, C.}, \bibinfo{author}{Tian, Y.}, \bibinfo{author}{Jin,
  Y.}, \bibinfo{author}{Zhang, X.}, \& \bibinfo{author}{Pan, L.}
  (\bibinfo{year}{2017}).
\newblock \bibinfo{title}{A radial space division based evolutionary algorithm
  for many-objective optimization}.
\newblock {\it \bibinfo{journal}{Applied Soft Computing}\/},  {\it
  \bibinfo{volume}{61}\/}, \bibinfo{pages}{603--621}. \URLprefix
  \url{https://www.sciencedirect.com/science/article/abs/pii/S1568494617305069}.
  \DOIprefix\doi{https://doi.org/10.1016/j.asoc.2017.08.024}.
%Type = Article
\bibitem[{Held et~al.(2018)Held, Fl{\"a}rdh \&
  M{\aa}rtensson}]{held2018optimal}
\bibinfo{author}{Held, M.}, \bibinfo{author}{Fl{\"a}rdh, O.}, \&
  \bibinfo{author}{M{\aa}rtensson, J.} (\bibinfo{year}{2018}).
\newblock \bibinfo{title}{Optimal speed control of a heavy-duty vehicle in
  urban driving}.
\newblock {\it \bibinfo{journal}{IEEE Transactions on Intelligent
  Transportation Systems}\/},  {\it \bibinfo{volume}{20}\/},
  \bibinfo{pages}{1562--1573}. \URLprefix
  \url{https://ieeexplore.ieee.org/abstract/document/8437156}.
  \DOIprefix\doi{10.1109/TITS.2018.2853264}.
%Type = Inproceedings
\bibitem[{Hern{\'a}ndez~G{\'o}mez \&
  Coello~Coello(2015)}]{hernandez2015improved}
\bibinfo{author}{Hern{\'a}ndez~G{\'o}mez, R.}, \&
  \bibinfo{author}{Coello~Coello, C.~A.} (\bibinfo{year}{2015}).
\newblock \bibinfo{title}{Improved metaheuristic based on the r2 indicator for
  many-objective optimization}.
\newblock In {\it \bibinfo{booktitle}{Proceedings of the 2015 Annual Conference
  on Genetic and Evolutionary Computation}\/} (pp. \bibinfo{pages}{679--686}).
\newblock \URLprefix \url{https://dl.acm.org/doi/10.1145/2739480.2754776}.
  \DOIprefix\doi{https://doi.org/10.1145/2739480.2754776}.
%Type = Article
\bibitem[{Hu et~al.(2016)Hu, Jing, Wang, Yan \& Chadli}]{hu2016robust}
\bibinfo{author}{Hu, C.}, \bibinfo{author}{Jing, H.}, \bibinfo{author}{Wang,
  R.}, \bibinfo{author}{Yan, F.}, \& \bibinfo{author}{Chadli, M.}
  (\bibinfo{year}{2016}).
\newblock \bibinfo{title}{Robust $\mathcal{H}_{\infty}$ output-feedback control
  for path following of autonomous ground vehicles}.
\newblock {\it \bibinfo{journal}{Mechanical Systems and Signal Processing}\/},
  {\it \bibinfo{volume}{70}\/}, \bibinfo{pages}{414--427}. \URLprefix
  \url{https://www.sciencedirect.com/science/article/abs/pii/S0888327015004124}.
  \DOIprefix\doi{https://doi.org/10.1016/j.ymssp.2015.09.017}.
%Type = Article
\bibitem[{Jiang \& Yang(2017)}]{jiang2017strength}
\bibinfo{author}{Jiang, S.}, \& \bibinfo{author}{Yang, S.}
  (\bibinfo{year}{2017}).
\newblock \bibinfo{title}{A strength pareto evolutionary algorithm based on
  reference direction for multiobjective and many-objective optimization}.
\newblock {\it \bibinfo{journal}{IEEE Transactions on Evolutionary
  Computation}\/},  {\it \bibinfo{volume}{21}\/}, \bibinfo{pages}{329--346}.
  \URLprefix \url{https://ieeexplore.ieee.org/document/7886269}.
  \DOIprefix\doi{10.1109/TEVC.2016.2592479}.
%Type = Article
\bibitem[{Kati et~al.(2016)Kati, K{\"o}ro{\u{g}}lu \&
  Fredriksson}]{kati2016robust}
\bibinfo{author}{Kati, M.~S.}, \bibinfo{author}{K{\"o}ro{\u{g}}lu, H.}, \&
  \bibinfo{author}{Fredriksson, J.} (\bibinfo{year}{2016}).
\newblock \bibinfo{title}{Robust lateral control of an a-double combination via
  $\mathcal{H}_{\infty}$ and generalized h2 static output feedback}.
\newblock {\it \bibinfo{journal}{IFAC-PapersOnLine}\/},  {\it
  \bibinfo{volume}{49}\/}, \bibinfo{pages}{305--311}.
%Type = Article
\bibitem[{Koduru et~al.(2008)Koduru, Dong, Das, Welch, Roe \&
  Charbit}]{koduru2008multiobjective}
\bibinfo{author}{Koduru, P.}, \bibinfo{author}{Dong, Z.}, \bibinfo{author}{Das,
  S.}, \bibinfo{author}{Welch, S.~M.}, \bibinfo{author}{Roe, J.~L.}, \&
  \bibinfo{author}{Charbit, E.} (\bibinfo{year}{2008}).
\newblock \bibinfo{title}{A multiobjective evolutionary-simplex hybrid approach
  for the optimization of differential equation models of gene networks}.
\newblock {\it \bibinfo{journal}{IEEE Transactions on Evolutionary
  Computation}\/},  {\it \bibinfo{volume}{12}\/}, \bibinfo{pages}{572--590}.
  \URLprefix \url{https://ieeexplore.ieee.org/abstract/document/4469887}.
  \DOIprefix\doi{10.1109/TEVC.2008.917202}.
%Type = Article
\bibitem[{L{\'a}zaro et~al.(2018)L{\'a}zaro, Jim{\'e}nez \&
  Takeda}]{lazaro2018improving}
\bibinfo{author}{L{\'a}zaro, J.~L.}, \bibinfo{author}{Jim{\'e}nez, {\'A}.~B.},
  \& \bibinfo{author}{Takeda, A.} (\bibinfo{year}{2018}).
\newblock \bibinfo{title}{Improving cash logistics in bank branches by coupling
  machine learning and robust optimization}.
\newblock {\it \bibinfo{journal}{Expert Systems With Applications}\/},  {\it
  \bibinfo{volume}{92}\/}, \bibinfo{pages}{236--255}. \URLprefix
  \url{https://www.sciencedirect.com/science/article/abs/pii/S0957417417306474}.
  \DOIprefix\doi{https://doi.org/10.1016/j.eswa.2017.09.043}.
%Type = Article
\bibitem[{Li et~al.(2018)Li, Jing, Wang \& Chen}]{li2018vehicle}
\bibinfo{author}{Li, C.}, \bibinfo{author}{Jing, H.}, \bibinfo{author}{Wang,
  R.}, \& \bibinfo{author}{Chen, N.} (\bibinfo{year}{2018}).
\newblock \bibinfo{title}{Vehicle lateral motion regulation under unreliable
  communication links based on robust $\mathcal{H}_\infty$ output-feedback
  control schema}.
\newblock {\it \bibinfo{journal}{Mechanical Systems and Signal Processing}\/},
  {\it \bibinfo{volume}{104}\/}, \bibinfo{pages}{171--187}. \URLprefix
  \url{https://www.sciencedirect.com/science/article/abs/pii/S0888327017304892}.
  \DOIprefix\doi{https://doi.org/10.1016/j.ymssp.2017.09.012}.
%Type = Article
\bibitem[{Li et~al.(2013)Li, Yang \& Liu}]{li2013shift}
\bibinfo{author}{Li, M.}, \bibinfo{author}{Yang, S.}, \& \bibinfo{author}{Liu,
  X.} (\bibinfo{year}{2013}).
\newblock \bibinfo{title}{Shift-based density estimation for pareto-based
  algorithms in many-objective optimization}.
\newblock {\it \bibinfo{journal}{IEEE Transactions on Evolutionary
  Computation}\/},  {\it \bibinfo{volume}{18}\/}, \bibinfo{pages}{348--365}.
  \URLprefix \url{https://ieeexplore.ieee.org/document/6516892}.
  \DOIprefix\doi{10.1109/TEVC.2013.2262178}.
%Type = Article
\bibitem[{Li et~al.(2015)Li, Yang \& Liu}]{li2015bi}
\bibinfo{author}{Li, M.}, \bibinfo{author}{Yang, S.}, \& \bibinfo{author}{Liu,
  X.} (\bibinfo{year}{2015}).
\newblock \bibinfo{title}{Bi-goal evolution for many-objective optimization
  problems}.
\newblock {\it \bibinfo{journal}{Artificial Intelligence}\/},  {\it
  \bibinfo{volume}{228}\/}, \bibinfo{pages}{45--65}. \URLprefix
  \url{https://www.sciencedirect.com/science/article/pii/S0004370215000995}.
  \DOIprefix\doi{https://doi.org/10.1016/j.artint.2015.06.007}.
%Type = Article
\bibitem[{Malikopoulos(2015)}]{malikopoulos2015multiobjective}
\bibinfo{author}{Malikopoulos, A.~A.} (\bibinfo{year}{2015}).
\newblock \bibinfo{title}{A multiobjective optimization framework for online
  stochastic optimal control in hybrid electric vehicles}.
\newblock {\it \bibinfo{journal}{IEEE Transactions on Control Systems
  Technology}\/},  {\it \bibinfo{volume}{24}\/}, \bibinfo{pages}{440--450}.
  \URLprefix \url{https://ieeexplore.ieee.org/document/7174519}.
  \DOIprefix\doi{10.1109/TCST.2015.2454444}.
%Type = Inproceedings
\bibitem[{Moe et~al.(2018)Moe, Rustad \& Hanssen}]{moe2018machine}
\bibinfo{author}{Moe, S.}, \bibinfo{author}{Rustad, A.~M.}, \&
  \bibinfo{author}{Hanssen, K.~G.} (\bibinfo{year}{2018}).
\newblock \bibinfo{title}{Machine learning in control systems: An overview of
  the state of the art}.
\newblock In {\it \bibinfo{booktitle}{International Conference on Innovative
  Techniques and Applications of Artificial Intelligence}\/} (pp.
  \bibinfo{pages}{250--265}).
\newblock \bibinfo{organization}{Springer}.
\newblock \URLprefix
  \url{https://link.springer.com/chapter/10.1007/978-3-030-04191-5\_23}.
  \DOIprefix\doi{https://doi.org/10.1007/978-3-030-04191-5\_23}.
%Type = Article
\bibitem[{Mohammadzadeh \& Taghavifar(2020)}]{mohammadzadeh2020novel}
\bibinfo{author}{Mohammadzadeh, A.}, \& \bibinfo{author}{Taghavifar, H.}
  (\bibinfo{year}{2020}).
\newblock \bibinfo{title}{A novel adaptive control approach for path tracking
  control of autonomous vehicles subject to uncertain dynamics}.
\newblock {\it \bibinfo{journal}{Proceedings of the Institution of Mechanical
  Engineers, Part D: Journal of Automobile Engineering}\/},  {\it
  \bibinfo{volume}{234}\/}, \bibinfo{pages}{2115--2126}. \URLprefix
  \url{https://journals.sagepub.com/doi/abs/10.1177/0954407019901083}.
  \DOIprefix\doi{https://doi.org/10.1177/0954407019901083}.
%Type = Article
\bibitem[{Van~de Molengraft-Luijten et~al.(2012)Van~de Molengraft-Luijten,
  Besselink, Verschuren \& Nijmeijer}]{van2012analysis}
\bibinfo{author}{Van~de Molengraft-Luijten, M.}, \bibinfo{author}{Besselink,
  I.~J.}, \bibinfo{author}{Verschuren, R.}, \& \bibinfo{author}{Nijmeijer, H.}
  (\bibinfo{year}{2012}).
\newblock \bibinfo{title}{Analysis of the lateral dynamic behaviour of
  articulated commercial vehicles}.
\newblock {\it \bibinfo{journal}{Vehicle system dynamics}\/},  {\it
  \bibinfo{volume}{50}\/}, \bibinfo{pages}{169--189}. \URLprefix
  \url{https://www.tandfonline.com/doi/abs/10.1080/00423114.2012.676650}.
  \DOIprefix\doi{https://doi.org/10.1080/00423114.2012.676650}.
%Type = Article
\bibitem[{de~Morais et~al.(2020)de~Morais, Marcos, Bueno, de~Resende, Terra \&
  Grassi~Jr}]{demorais2020vision}
\bibinfo{author}{de~Morais, G.~A.}, \bibinfo{author}{Marcos, L.~B.},
  \bibinfo{author}{Bueno, J. N.~A.}, \bibinfo{author}{de~Resende, N.~F.},
  \bibinfo{author}{Terra, M.~H.}, \& \bibinfo{author}{Grassi~Jr, V.}
  (\bibinfo{year}{2020}).
\newblock \bibinfo{title}{Vision-based robust control framework based on deep
  reinforcement learning applied to autonomous ground vehicles}.
\newblock {\it \bibinfo{journal}{Control Engineering Practice}\/},  {\it
  \bibinfo{volume}{104}\/}, \bibinfo{pages}{104630}. \URLprefix
  \url{http://dx.doi.org/10.1016/j.conengprac.2020.104630}.
  \DOIprefix\doi{10.1016/j.conengprac.2020.104630}.
%Type = Article
\bibitem[{de~Morais et~al.(2019)de~Morais, Barbosa, Ferreira \&
  Paiva}]{de2019soft}
\bibinfo{author}{de~Morais, G. A.~P.}, \bibinfo{author}{Barbosa, B. H.~G.},
  \bibinfo{author}{Ferreira, D.~D.}, \& \bibinfo{author}{Paiva, L.~S.}
  (\bibinfo{year}{2019}).
\newblock \bibinfo{title}{Soft sensors design in a petrochemical process using
  an evolutionary algorithm}.
\newblock {\it \bibinfo{journal}{Measurement}\/},  {\it
  \bibinfo{volume}{148}\/}, \bibinfo{pages}{106920}. \URLprefix
  \url{https://www.sciencedirect.com/science/article/pii/S0263224119307778}.
  \DOIprefix\doi{https://doi.org/10.1016/j.measurement.2019.106920}.
%Type = Article
\bibitem[{Nguyen et~al.(2020)Nguyen, Rath, Guerra, Palhares \&
  Zhang}]{nguyen2020robust}
\bibinfo{author}{Nguyen, A.-T.}, \bibinfo{author}{Rath, J.},
  \bibinfo{author}{Guerra, T.-M.}, \bibinfo{author}{Palhares, R.}, \&
  \bibinfo{author}{Zhang, H.} (\bibinfo{year}{2020}).
\newblock \bibinfo{title}{Robust set-invariance based fuzzy output tracking
  control for vehicle autonomous driving under uncertain lateral forces and
  steering constraints}.
\newblock {\it \bibinfo{journal}{IEEE Transactions on Intelligent
  Transportation Systems}\/}, . \URLprefix
  \url{https://ieeexplore.ieee.org/abstract/document/9204811}.
  \DOIprefix\doi{10.1109/TITS.2020.3021292}.
%Type = Inproceedings
\bibitem[{Oliveira et~al.(2018)Oliveira, Cirillo, Wahlberg
  et~al.}]{oliveira2018combining}
\bibinfo{author}{Oliveira, R.}, \bibinfo{author}{Cirillo, M.},
  \bibinfo{author}{Wahlberg, B.} et~al. (\bibinfo{year}{2018}).
\newblock \bibinfo{title}{Combining lattice-based planning and path
  optimization in autonomous heavy duty vehicle applications}.
\newblock In {\it \bibinfo{booktitle}{2018 IEEE Intelligent Vehicles Symposium
  (IV)}\/} (pp. \bibinfo{pages}{2090--2097}).
\newblock \bibinfo{organization}{IEEE}.
%Type = Article
\bibitem[{Qasem \& Shamsuddin(2011)}]{qasem2011radial}
\bibinfo{author}{Qasem, S.~N.}, \& \bibinfo{author}{Shamsuddin, S.~M.}
  (\bibinfo{year}{2011}).
\newblock \bibinfo{title}{Radial basis function network based on time variant
  multi-objective particle swarm optimization for medical diseases diagnosis}.
\newblock {\it \bibinfo{journal}{Applied Soft Computing}\/},  {\it
  \bibinfo{volume}{11}\/}, \bibinfo{pages}{1427--1438}. \URLprefix
  \url{https://www.sciencedirect.com/science/article/abs/pii/S1568494610000906}.
  \DOIprefix\doi{https://doi.org/10.1016/j.asoc.2010.04.014}.
%Type = Inproceedings
\bibitem[{Rodriguez \& Fathy(2018)}]{rodriguez2018speed}
\bibinfo{author}{Rodriguez, M.}, \& \bibinfo{author}{Fathy, H.}
  (\bibinfo{year}{2018}).
\newblock \bibinfo{title}{Speed trajectory optimization for a heavy-duty truck
  traversing multiple signalized intersections: A dynamic programming study}.
\newblock In {\it \bibinfo{booktitle}{2018 IEEE Conference on Control
  Technology and Applications (CCTA)}\/} (pp. \bibinfo{pages}{1454--1459}).
\newblock \bibinfo{organization}{IEEE}.
%Type = Article
\bibitem[{Sayed(2001)}]{Sayed2001}
\bibinfo{author}{Sayed, A.~H.} (\bibinfo{year}{2001}).
\newblock \bibinfo{title}{A framework for state-space estimation with uncertain
  models}.
\newblock {\it \bibinfo{journal}{IEEE Transactions on Automatic Control}\/},
  {\it \bibinfo{volume}{46}\/}, \bibinfo{pages}{998--1013}. \URLprefix
  \url{https://ieeexplore.ieee.org/abstract/document/935054}.
  \DOIprefix\doi{10.1109/9.935054}.
%Type = Techreport
\bibitem[{Schott(1995)}]{schott1995fault}
\bibinfo{author}{Schott, J.~R.} (\bibinfo{year}{1995}).
\newblock {\it \bibinfo{title}{Fault tolerant design using single and
  multicriteria genetic algorithm optimization.}\/}.
\newblock \bibinfo{type}{Technical Report} AIR FORCE INST OF TECH
  WRIGHT-PATTERSON AFB OH.
%Type = Article
\bibitem[{Shahnejat-Bushehri et~al.(2021)Shahnejat-Bushehri,
  Tavakkoli-Moghaddam, Boronoos \& Ghasemkhani}]{shahnejat2021robust}
\bibinfo{author}{Shahnejat-Bushehri, S.}, \bibinfo{author}{Tavakkoli-Moghaddam,
  R.}, \bibinfo{author}{Boronoos, M.}, \& \bibinfo{author}{Ghasemkhani, A.}
  (\bibinfo{year}{2021}).
\newblock \bibinfo{title}{A robust home health care routing-scheduling problem
  with temporal dependencies under uncertainty}.
\newblock {\it \bibinfo{journal}{Expert Systems with Applications}\/},  (p.
  \bibinfo{pages}{115209}). \URLprefix
  \url{https://www.sciencedirect.com/science/article/abs/pii/S0957417421006424}.
  \DOIprefix\doi{https://doi.org/10.1016/j.eswa.2021.115209}.
%Type = Article
\bibitem[{Shin et~al.(2005)Shin, Lee, Kim \& Zhang}]{shin2005multiobjective}
\bibinfo{author}{Shin, S.-Y.}, \bibinfo{author}{Lee, I.-H.},
  \bibinfo{author}{Kim, D.}, \& \bibinfo{author}{Zhang, B.-T.}
  (\bibinfo{year}{2005}).
\newblock \bibinfo{title}{Multiobjective evolutionary optimization of dna
  sequences for reliable dna computing}.
\newblock {\it \bibinfo{journal}{IEEE transactions on evolutionary
  computation}\/},  {\it \bibinfo{volume}{9}\/}, \bibinfo{pages}{143--158}.
  \URLprefix \url{https://ieeexplore.ieee.org/abstract/document/1413256}.
  \DOIprefix\doi{10.1109/TEVC.2005.844166}.
%Type = Inproceedings
\bibitem[{Skjetne \& Fossen(2001)}]{skjetne2001nonlinear}
\bibinfo{author}{Skjetne, R.}, \& \bibinfo{author}{Fossen, T.~I.}
  (\bibinfo{year}{2001}).
\newblock \bibinfo{title}{Nonlinear maneuvering and control of ships}.
\newblock In {\it \bibinfo{booktitle}{MTS/IEEE Oceans 2001. An Ocean Odyssey.
  Conference Proceedings (IEEE Cat. No. 01CH37295)}\/} (pp.
  \bibinfo{pages}{1808--1815}).
\newblock \bibinfo{organization}{IEEE} volume~\bibinfo{volume}{3}.
\newblock \URLprefix
  \url{https://ieeexplore.ieee.org/abstract/document/968121}.
  \DOIprefix\doi{10.1109/OCEANS.2001.968121}.
%Type = Article
\bibitem[{Srinivas \& Deb(1994)}]{srinivas1994muiltiobjective}
\bibinfo{author}{Srinivas, N.}, \& \bibinfo{author}{Deb, K.}
  (\bibinfo{year}{1994}).
\newblock \bibinfo{title}{Muiltiobjective optimization using nondominated
  sorting in genetic algorithms}.
\newblock {\it \bibinfo{journal}{Evolutionary computation}\/},  {\it
  \bibinfo{volume}{2}\/}, \bibinfo{pages}{221--248}. \URLprefix
  \url{https://www.mitpressjournals.org/doi/abs/10.1162/evco.1994.2.3.221}.
  \DOIprefix\doi{https://doi.org/10.1162/evco.1994.2.3.221}.
%Type = Article
\bibitem[{Sun et~al.(2021)Sun, Wang, Zhang \& Cao}]{sun2021robust}
\bibinfo{author}{Sun, H.}, \bibinfo{author}{Wang, Y.}, \bibinfo{author}{Zhang,
  J.}, \& \bibinfo{author}{Cao, W.} (\bibinfo{year}{2021}).
\newblock \bibinfo{title}{A robust optimization model for
  location-transportation problem of disaster casualties with triage and
  uncertainty}.
\newblock {\it \bibinfo{journal}{Expert Systems with Applications}\/},  {\it
  \bibinfo{volume}{175}\/}, \bibinfo{pages}{114867}. \URLprefix
  \url{https://www.sciencedirect.com/science/article/abs/pii/S0957417421003080}.
  \DOIprefix\doi{https://doi.org/10.1016/j.eswa.2021.114867}.
%Type = Article
\bibitem[{Terra et~al.(2014)Terra, Cerri \& Ishihara}]{Cerri2014}
\bibinfo{author}{Terra, M.~H.}, \bibinfo{author}{Cerri, J.~P.}, \&
  \bibinfo{author}{Ishihara, J.~Y.} (\bibinfo{year}{2014}).
\newblock \bibinfo{title}{Optimal robust linear quadratic regulator for systems
  subject to uncertainties}.
\newblock {\it \bibinfo{journal}{{IEEE} Transactions on Automatic Control}\/},
  {\it \bibinfo{volume}{59}\/}, \bibinfo{pages}{2586--2591}. \URLprefix
  \url{https://doi.org/10.1109/tac.2014.2309282}.
  \DOIprefix\doi{10.1109/tac.2014.2309282}.
%Type = Article
\bibitem[{Tian et~al.(2017)Tian, Cheng, Zhang \& Jin}]{tian2017platemo}
\bibinfo{author}{Tian, Y.}, \bibinfo{author}{Cheng, R.},
  \bibinfo{author}{Zhang, X.}, \& \bibinfo{author}{Jin, Y.}
  (\bibinfo{year}{2017}).
\newblock \bibinfo{title}{Platemo: A matlab platform for evolutionary
  multi-objective optimization [educational forum]}.
\newblock {\it \bibinfo{journal}{IEEE Computational Intelligence Magazine}\/},
  {\it \bibinfo{volume}{12}\/}, \bibinfo{pages}{73--87}. \URLprefix
  \url{https://ieeexplore.ieee.org/document/8065138}.
  \DOIprefix\doi{10.1109/MCI.2017.2742868}.
%Type = Inproceedings
\bibitem[{Tian et~al.(2016)Tian, Zhang, Cheng \& Jin}]{tian2016multi}
\bibinfo{author}{Tian, Y.}, \bibinfo{author}{Zhang, X.},
  \bibinfo{author}{Cheng, R.}, \& \bibinfo{author}{Jin, Y.}
  (\bibinfo{year}{2016}).
\newblock \bibinfo{title}{A multi-objective evolutionary algorithm based on an
  enhanced inverted generational distance metric}.
\newblock In {\it \bibinfo{booktitle}{2016 IEEE congress on evolutionary
  computation (CEC)}\/} (pp. \bibinfo{pages}{5222--5229}).
\newblock \bibinfo{organization}{IEEE}.
\newblock \URLprefix \url{https://ieeexplore.ieee.org/document/7748352}.
  \DOIprefix\doi{10.1109/CEC.2016.7748352}.
%Type = Article
\bibitem[{Wang et~al.(2016)Wang, Jin \& Yao}]{wang2016diversity}
\bibinfo{author}{Wang, H.}, \bibinfo{author}{Jin, Y.}, \& \bibinfo{author}{Yao,
  X.} (\bibinfo{year}{2016}).
\newblock \bibinfo{title}{Diversity assessment in many-objective optimization}.
\newblock {\it \bibinfo{journal}{IEEE transactions on cybernetics}\/},  {\it
  \bibinfo{volume}{47}\/}, \bibinfo{pages}{1510--1522}. \URLprefix
  \url{https://ieeexplore.ieee.org/document/7473938}.
  \DOIprefix\doi{10.1109/TCYB.2016.2550502}.
%Type = Article
\bibitem[{Wang et~al.(2010)Wang, Wu \& Yuan}]{wang2010multi}
\bibinfo{author}{Wang, Y.-N.}, \bibinfo{author}{Wu, L.-H.}, \&
  \bibinfo{author}{Yuan, X.-F.} (\bibinfo{year}{2010}).
\newblock \bibinfo{title}{Multi-objective self-adaptive differential evolution
  with elitist archive and crowding entropy-based diversity measure}.
\newblock {\it \bibinfo{journal}{Soft Computing}\/},  {\it
  \bibinfo{volume}{14}\/}, \bibinfo{pages}{193}. \URLprefix
  \url{https://link.springer.com/article/10.1007/s00500-008-0394-9}.
  \DOIprefix\doi{https://doi.org/10.1007/s00500-008-0394-9}.
%Type = Article
\bibitem[{Weinert et~al.(2009)Weinert, Zabel, Kersting, Michelitsch \&
  Wagner}]{weinert2009use}
\bibinfo{author}{Weinert, K.}, \bibinfo{author}{Zabel, A.},
  \bibinfo{author}{Kersting, P.}, \bibinfo{author}{Michelitsch, T.}, \&
  \bibinfo{author}{Wagner, T.} (\bibinfo{year}{2009}).
\newblock \bibinfo{title}{On the use of problem-specific candidate generators
  for the hybrid optimization of multi-objective production engineering
  problems}.
\newblock {\it \bibinfo{journal}{Evolutionary computation}\/},  {\it
  \bibinfo{volume}{17}\/}, \bibinfo{pages}{527--544}. \URLprefix
  \url{https://www.mitpressjournals.org/doi/abs/10.1162/evco.2009.17.4.17405}.
  \DOIprefix\doi{https://doi.org/10.1162/evco.2009.17.4.17405}.
%Type = Article
\bibitem[{While et~al.(2006)While, Hingston, Barone \&
  Huband}]{while2006faster}
\bibinfo{author}{While, L.}, \bibinfo{author}{Hingston, P.},
  \bibinfo{author}{Barone, L.}, \& \bibinfo{author}{Huband, S.}
  (\bibinfo{year}{2006}).
\newblock \bibinfo{title}{A faster algorithm for calculating hypervolume}.
\newblock {\it \bibinfo{journal}{IEEE transactions on evolutionary
  computation}\/},  {\it \bibinfo{volume}{10}\/}, \bibinfo{pages}{29--38}.
  \URLprefix \url{https://ieeexplore.ieee.org/document/1583625}.
  \DOIprefix\doi{10.1109/TEVC.2005.851275}.
%Type = Article
\bibitem[{Wu et~al.(2018)Wu, Chen \& Zhang}]{wu2018multiobjective}
\bibinfo{author}{Wu, C.-F.}, \bibinfo{author}{Chen, B.-S.}, \&
  \bibinfo{author}{Zhang, W.} (\bibinfo{year}{2018}).
\newblock \bibinfo{title}{Multiobjective $h_{2}/h_{\infty}$ control design of
  the nonlinear mean-field stochastic jump-diffusion systems via fuzzy
  approach}.
\newblock {\it \bibinfo{journal}{IEEE Transactions on Fuzzy Systems}\/},  {\it
  \bibinfo{volume}{27}\/}, \bibinfo{pages}{686--700}. \URLprefix
  \url{https://ieeexplore.ieee.org/document/8444730}.
  \DOIprefix\doi{10.1109/TFUZZ.2018.2866823}.
%Type = Inproceedings
\bibitem[{Yuan et~al.(2014)Yuan, Xu \& Wang}]{yuan2014evolutionary}
\bibinfo{author}{Yuan, Y.}, \bibinfo{author}{Xu, H.}, \& \bibinfo{author}{Wang,
  B.} (\bibinfo{year}{2014}).
\newblock \bibinfo{title}{Evolutionary many-objective optimization using
  ensemble fitness ranking}.
\newblock In {\it \bibinfo{booktitle}{Proceedings of the 2014 Annual Conference
  on Genetic and Evolutionary Computation}\/} (pp. \bibinfo{pages}{669--676}).
\newblock \URLprefix \url{https://dl.acm.org/doi/10.1145/2576768.2598345}.
  \DOIprefix\doi{https://doi.org/10.1145/2576768.2598345}.
%Type = Article
\bibitem[{Yuan et~al.(2015{\natexlab{a}})Yuan, Xu, Wang \& Yao}]{yuan2015new}
\bibinfo{author}{Yuan, Y.}, \bibinfo{author}{Xu, H.}, \bibinfo{author}{Wang,
  B.}, \& \bibinfo{author}{Yao, X.} (\bibinfo{year}{2015}{\natexlab{a}}).
\newblock \bibinfo{title}{A new dominance relation-based evolutionary algorithm
  for many-objective optimization}.
\newblock {\it \bibinfo{journal}{IEEE Transactions on Evolutionary
  Computation}\/},  {\it \bibinfo{volume}{20}\/}, \bibinfo{pages}{16--37}.
  \URLprefix \url{https://ieeexplore.ieee.org/document/7080938}.
  \DOIprefix\doi{10.1109/TEVC.2015.2420112}.
%Type = Article
\bibitem[{Yuan et~al.(2015{\natexlab{b}})Yuan, Xu, Wang, Zhang \&
  Yao}]{yuan2015balancing}
\bibinfo{author}{Yuan, Y.}, \bibinfo{author}{Xu, H.}, \bibinfo{author}{Wang,
  B.}, \bibinfo{author}{Zhang, B.}, \& \bibinfo{author}{Yao, X.}
  (\bibinfo{year}{2015}{\natexlab{b}}).
\newblock \bibinfo{title}{Balancing convergence and diversity in
  decomposition-based many-objective optimizers}.
\newblock {\it \bibinfo{journal}{IEEE Transactions on Evolutionary
  Computation}\/},  {\it \bibinfo{volume}{20}\/}, \bibinfo{pages}{180--198}.
  \URLprefix \url{https://ieeexplore.ieee.org/document/7120115}.
  \DOIprefix\doi{10.1109/TEVC.2015.2443001}.
%Type = Article
\bibitem[{Zhang et~al.(2021)Zhang, Zhang, Zhang, Wang \& Chen}]{zhang2021novel}
\bibinfo{author}{Zhang, J.}, \bibinfo{author}{Zhang, B.},
  \bibinfo{author}{Zhang, N.}, \bibinfo{author}{Wang, C.}, \&
  \bibinfo{author}{Chen, Y.} (\bibinfo{year}{2021}).
\newblock \bibinfo{title}{A novel robust event-triggered fault tolerant
  automatic steering control approach of autonomous land vehicles under
  in-vehicle network delay}.
\newblock {\it \bibinfo{journal}{International Journal of Robust and Nonlinear
  Control}\/},  {\it \bibinfo{volume}{31}\/}, \bibinfo{pages}{2436--2464}.
  \URLprefix \url{https://onlinelibrary.wiley.com/doi/abs/10.1002/rnc.5393}.
  \DOIprefix\doi{https://doi.org/10.1002/rnc.5393}.
%Type = Article
\bibitem[{Zhang \& Li(2007)}]{zhang2007moea}
\bibinfo{author}{Zhang, Q.}, \& \bibinfo{author}{Li, H.}
  (\bibinfo{year}{2007}).
\newblock \bibinfo{title}{Moea/d: A multiobjective evolutionary algorithm based
  on decomposition}.
\newblock {\it \bibinfo{journal}{IEEE Transactions on evolutionary
  computation}\/},  {\it \bibinfo{volume}{11}\/}, \bibinfo{pages}{712--731}.
  \URLprefix \url{https://ieeexplore.ieee.org/abstract/document/4358754}.
  \DOIprefix\doi{10.1109/TEVC.2007.892759}.
%Type = Article
\bibitem[{Zhang \& Rockett(2011)}]{zhang2011generic}
\bibinfo{author}{Zhang, Y.}, \& \bibinfo{author}{Rockett, P.~I.}
  (\bibinfo{year}{2011}).
\newblock \bibinfo{title}{A generic optimising feature extraction method using
  multiobjective genetic programming}.
\newblock {\it \bibinfo{journal}{Applied Soft Computing}\/},  {\it
  \bibinfo{volume}{11}\/}, \bibinfo{pages}{1087--1097}. \URLprefix
  \url{https://www.sciencedirect.com/science/article/abs/pii/S1568494610000372}.
  \DOIprefix\doi{https://doi.org/10.1016/j.asoc.2010.02.008}.
%Type = Article
\bibitem[{Zhao et~al.(2011)Zhao, Iruthayarajan, Baskar \&
  Suganthan}]{zhao2011multi}
\bibinfo{author}{Zhao, S.-Z.}, \bibinfo{author}{Iruthayarajan, M.~W.},
  \bibinfo{author}{Baskar, S.}, \& \bibinfo{author}{Suganthan, P.~N.}
  (\bibinfo{year}{2011}).
\newblock \bibinfo{title}{Multi-objective robust pid controller tuning using
  two lbests multi-objective particle swarm optimization}.
\newblock {\it \bibinfo{journal}{Information Sciences}\/},  {\it
  \bibinfo{volume}{181}\/}, \bibinfo{pages}{3323--3335}. \URLprefix
  \url{https://www.sciencedirect.com/science/article/pii/S002002551100171X}.
  \DOIprefix\doi{https://doi.org/10.1016/j.ins.2011.04.003}.
%Type = Article
\bibitem[{Zhao et~al.(2018)Zhao, Zhang \& Li}]{zhao2018displacement}
\bibinfo{author}{Zhao, W.}, \bibinfo{author}{Zhang, H.}, \&
  \bibinfo{author}{Li, Y.} (\bibinfo{year}{2018}).
\newblock \bibinfo{title}{Displacement and force coupling control design for
  automotive active front steering system}.
\newblock {\it \bibinfo{journal}{Mechanical Systems and Signal Processing}\/},
  {\it \bibinfo{volume}{106}\/}, \bibinfo{pages}{76--93}. \URLprefix
  \url{https://www.sciencedirect.com/science/article/abs/pii/S0888327017306751}.
  \DOIprefix\doi{https://doi.org/10.1016/j.ymssp.2017.12.037}.
%Type = Article
\bibitem[{Zhou et~al.(2011)Zhou, Qu, Li, Zhao, Suganthan \&
  Zhang}]{zhou2011multiobjective}
\bibinfo{author}{Zhou, A.}, \bibinfo{author}{Qu, B.-Y.}, \bibinfo{author}{Li,
  H.}, \bibinfo{author}{Zhao, S.-Z.}, \bibinfo{author}{Suganthan, P.~N.}, \&
  \bibinfo{author}{Zhang, Q.} (\bibinfo{year}{2011}).
\newblock \bibinfo{title}{Multiobjective evolutionary algorithms: A survey of
  the state of the art}.
\newblock {\it \bibinfo{journal}{Swarm and Evolutionary Computation}\/},  {\it
  \bibinfo{volume}{1}\/}, \bibinfo{pages}{32--49}. \URLprefix
  \url{https://www.sciencedirect.com/science/article/abs/pii/S2210650211000058}.
  \DOIprefix\doi{https://doi.org/10.1016/j.swevo.2011.03.001}.
%Type = Article
\bibitem[{Zhou et~al.(2019)Zhou, Gao, Yao, Chan, Zhang, Li \&
  Lin}]{zhou2019decomposition}
\bibinfo{author}{Zhou, J.}, \bibinfo{author}{Gao, L.}, \bibinfo{author}{Yao,
  X.}, \bibinfo{author}{Chan, F.~T.}, \bibinfo{author}{Zhang, J.},
  \bibinfo{author}{Li, X.}, \& \bibinfo{author}{Lin, Y.}
  (\bibinfo{year}{2019}).
\newblock \bibinfo{title}{A decomposition and statistical learning based
  many-objective artificial bee colony optimizer}.
\newblock {\it \bibinfo{journal}{Information Sciences}\/},  {\it
  \bibinfo{volume}{496}\/}, \bibinfo{pages}{82--108}. \URLprefix
  \url{https://www.sciencedirect.com/science/article/pii/S0020025519304001}.
  \DOIprefix\doi{https://doi.org/10.1016/j.ins.2019.05.014}.
%Type = Techreport
\bibitem[{Zitzler et~al.(1999)Zitzler, Deb \& Thiele}]{zitzler1999comparison}
\bibinfo{author}{Zitzler, E.}, \bibinfo{author}{Deb, K.}, \&
  \bibinfo{author}{Thiele, L.} (\bibinfo{year}{1999}).
\newblock {\it \bibinfo{title}{Comparison of multiobjective evolutionary
  algorithms: Empirical results (revised version)}\/}.
\newblock \bibinfo{type}{Technical Report} Technical Report 70, Computer
  Engineering and Networks Laboratory (TIK~….
\newblock \URLprefix
  \url{https://www.mitpressjournals.org/doi/10.1162/106365600568202}.
  \DOIprefix\doi{10.1162/106365600568202}.
%Type = Article
\bibitem[{Zitzler et~al.(2001)Zitzler, Laumanns \& Thiele}]{zitzler2001spea2}
\bibinfo{author}{Zitzler, E.}, \bibinfo{author}{Laumanns, M.}, \&
  \bibinfo{author}{Thiele, L.} (\bibinfo{year}{2001}).
\newblock \bibinfo{title}{Spea2: Improving the strength pareto evolutionary
  algorithm}.
\newblock {\it \bibinfo{journal}{TIK-report}\/},  {\it
  \bibinfo{volume}{103}\/}. \URLprefix
  \url{https://www.research-collection.ethz.ch/handle/20.500.11850/145755}.
  \DOIprefix\doi{https://doi.org/10.3929/ethz-a-004284029}.

\end{thebibliography}

\end{document}